\documentclass[prb,
superscriptaddress,
onecolumn,
floatfix,
nofootinbib,
notitlepage,
]{revtex4-1}

\usepackage{xcolor}
\usepackage{graphicx}
\usepackage{dcolumn}
\usepackage{bm}
\usepackage{color}
\usepackage[colorlinks=true]{hyperref}
\usepackage{units}
\usepackage{multirow}
\usepackage{hhline}
\usepackage{amsmath,amssymb}
\usepackage{longtable}
\usepackage{dcolumn}

\newcolumntype{d}{D{.}{.}{2}}
\newcolumntype{f}{D{.}{.}{4}}
\newcolumntype{h}{D{.}{.}{3}}

\newcommand{\rewrite}[1]{\textcolor{black}{#1}}
\newcommand{\review}[1]{\textcolor{black}{#1}}

\newcommand{\EoS}{equation of state}
\newcommand{\EoSs}{equations of state}
\newcommand{\EL}{Euler--Lagrange}
\newcommand{\GL}{Ginzburg--Landau}

\newcommand{\WS}{Wigner--Seitz}
\newcommand{\gs}{ground state}
\newcommand{\ghs}{ground-state}
\newcommand{\nablab}{\boldsymbol{\nabla}}
\renewcommand\Re{\text{Re}}
\renewcommand\Im{\text{Im}}

\newcommand{\dd}{\mathrm{d}}
\newcommand{\ee}{\mathrm{e}}
\newcommand{\ii}{\mathrm{i}}
\newcommand{\bb}{\text{b}} % Abbreviation of baryon
\newcommand{\cc}{\text{c}} % Abbreviation of critical
\newcommand{\pp}{\text{p}} % Abbreviation of proton
\newcommand{\nn}{\text{n}} % Abbreviation of neutron
\newcommand{\pn}{\text{pn}} % Abbreviation of proton-neutron
\newcommand{\FF}{\text{F}} % Abbreviation of Fermi

\newcommand{\Vp}{\mathbf{V}_\text{p}}
\newcommand{\Vn}{\mathbf{V}_\text{n}}
\newcommand{\gpp}{g_\text{pp}}
\newcommand{\gnn}{g_\text{nn}}
\newcommand{\gpn}{g_\text{pn}}
\newcommand{\mpp}{m_\text{p}}
\newcommand{\mnn}{m_\text{n}}
\newcommand{\mau}{m_\text{u}}
\newcommand{\psip}{\psi_\text{p}}
\newcommand{\psin}{\psi_\text{n}}
\newcommand{\rhop}{\rho_\text{p}}
\newcommand{\rhon}{\rho_\text{n}}

\newcommand{\gr}{\alpha}
\newcommand{\Fg}{\mathcal{F}_\text{g}}

\begin{document}

%%%%%%%%%%%%%%%%%%%%%%%%%%%%%%%%%%%%%%%%%%%%%%%%%%%%%%%%%%%%%%

\title{Superconducting phases in a two-component microscale model of neutron star cores}

\author{Toby S.~Wood}
\email{toby.wood@newcastle.ac.uk}
\affiliation{School of Mathematics, Statistics and Physics, Newcastle University, NE1 7RU, United Kingdom}

\author{Vanessa Graber}
\email{graber@ice.csic.es}
\affiliation{Institute of Space Sciences (ICE, CSIC), Campus UAB, 08193 Barcelona, Spain}
\affiliation{Institut d'Estudis Espacials de Catalunya (IEEC), 08034 Barcelona, Spain}
\affiliation{Department of Physics and McGill Space Institute, McGill University, Montreal QC H3A 2T8, Canada}

\author{William G.~Newton}
\affiliation{Department of Physics and Astronomy, Texas A\&M University-Commerce, Commerce, TX 75429-3011, USA}

%%%%%%%%%%%%%%%%%%%%%%%%%%%%%%%%%%%%%%%%%%%%%%%%%%%%%%%%%%%%%%

\date{\today}

\begin{abstract}
\rewrite{We identify the possible \gs s for a mixture of two superfluid condensates}
(one neutral, the other electrically charged)
\rewrite{using a phenomenological} \GL\ model.
While this framework is applicable to any interacting condensed-matter mixture of a charged and a neutral component,
we focus on nuclear matter in neutron star cores,
where proton and neutron condensates are coupled via non-dissipative entrainment.
\review{We employ the Skyrme interaction to determine the neutron star's equilibrium composition,
and hence obtain realistic coefficients for our \GL\ model
at each depth within the star's core.
We then use the \GL\ model to determine the \gs\ in the presence of a magnetic field.}
\review{In this way,}
we obtain superconducting phase diagrams for six representative Skyrme models,
revealing the microphysical magnetic flux distribution throughout the neutron star core.
The phase diagrams are rather complex
and the locations of most of the phase transitions can only be determined through numerical calculations.
Nonetheless, we find that for all \EoSs\ considered in this work,
much of the outer core exhibits type-1.5 superconductivity,
rather than type-II superconductivity as is generally assumed.
For local magnetic field strengths
$\lesssim \unit[10^{14}]{G}$,
the magnetic flux is distributed inhomogeneously,
with bundles of magnetic fluxtubes separated by flux-free Meissner regions.
We provide \rewrite{an approximate} criterion to determine the transition
between this type-1.5 phase and the type-I region in the inner core.
\end{abstract}

\keywords{Suggested keywords}

\maketitle

%%%%%%%%%%%%%%%%%%%%%%%%%%%%%%%%%%%%%%%%%%%%%%%%%%%%%%%%%%%%%%
%%%%%%%%%%%%%%%%%%%%%%%%%%%%%%%%%%%%%%%%%%%%%%%%%%%%%%%%%%%%%%

\section{Introduction}
\label{sec:intro}

Macroscopic quantum behavior is prominent in many physical systems,
ranging from superfluid phases in ultra-cold atomic gases and heavy-ion collisions
to superconducting transitions in metals
and exotic quantum phases in dense nuclear matter and quark matter.
Here, we consider the scenario of two coexisting superfluid condensates,
one of which is electrically charged,
that are coupled via density and density-gradient interactions.
We are particularly interested in the effects of entrainment:
the non-dissipative coupling between two quantum fluids
first discussed in the context of superfluid helium-3 and helium-4.\citep{AndreevBashkin76}
To this end,
we study the equilibrium phases of the superconducting condensate and determine
how its \gs\ is influenced by the neutral component.
Our results are generic for any interacting condensed-matter mixture of a charged and a neutral component,
but our main focus is on nuclear matter in the cores of neutron stars.
Within these compact objects, the mass density exceeds $\unitfrac[10^{14}]{g}{cm^3}$
(potentially reaching up to $\unitfrac[10^{15}]{g}{cm^3}$)
and protons and neutrons both form Cooper pairs,
resulting in the formation of two inter-penetrating quantum condensates.

The neutron star interior is a complex system,
and the nature of the magnetic field on large as well as on small scales is poorly understood.
The conventional picture was established in the seminal work of \citet{Baym-etal69},
who argued that the time for magnetic flux to be expelled from the stellar interior,
as the star cools below the critical temperature for superconductivity,
is much longer than the characteristic ages of neutron stars,
owing to the large conductivity of normal matter.
Because of this,
the transition to superconductivity occurs with an imposed magnetic flux,
and the resulting microphysical state depends on the characteristic length scales involved.
In a simple single-component superconductor,
the behavior is dictated by the \GL\ parameter, $\kappa$,
defined as the ratio of the London penetration length, $\lambda$, to the proton coherence length, $\xi_\pp$.
For $\kappa > 1/\sqrt{2}$,
the magnetic field resolves into an array of discrete fluxtubes,
each of which carries a quantum of magnetic flux.
The fluxtubes are mutually repulsive and therefore stable, resulting in type-II superconductivity.
In the type-I regime with $\kappa < 1/\sqrt{2}$,
fluxtubes are mutually attractive
and merge into macroscopic non-superconducting
domains that exist alongside flux-free Meissner regions;
this is known as an intermediate type-I state.
Using expressions for $\lambda$ and $\xi_\pp$ well-known from standard superconductor theory,
$\kappa$ can be estimated for the neutron star interior.
Owing to the density dependence of the underlying parameters,
$\kappa$ decreases with depth,
and in most models attains the value $1/\sqrt{2}$ at some depth within the star's core.
This suggests a transition between type-II superconductivity in the outer core
and type-I superconductivity in the inner core
(provided that nuclear matter does not turn into quark matter at such high densities).

Most hydrodynamic models of the star's large-scale behavior focus on the outer core
and assume the existence of a regular fluxtube array,
employing fluid equations obtained by averaging over many individual fluxtubes.\citep{Glampedakis-etal11, Graber-etal15}
While this simplification is convenient,
there are several reasons why such a model may be incorrect.
\citet{Link2003} has argued
that if a regular fluxtube array coexists with a neutron vortex array,
resulting from the rotation of the superfluid,
relative motion between the two arrays will be strongly damped due to Kelvin wave excitations.
This would make long-period precession in neutron stars impossible,
yet observational evidence for such behavior exists,\citep{Stairs-etal00,Shabanova-etal01,Haberl-etal06, Ashton-etal16}
casting doubt on the presence of a large-scale type-II superconductor in the outer core.
These discrepancies could be resolved if the entire core were in a type-I regime.\citep{Link2003,Sedrakian05}
\citet{Buckley-etal04-a,Buckley-etal04-b} have suggested that a type-I regime is favored
because of strong density interactions between the neutron and proton condensates,
but the coupling strengths they considered \rewrite{could be} unrealistically large
for the nuclear matter in the neutron star interior.\citep{Alford-etal05}
\citet{CharbonneauZhitnitsky07} have argued
that the fluxtube array in the outer core could become unstable due to a helical instability
if an induced longitudinal current were present,
effectively destroying the regularity of the lattice.
Although not providing a rigorous analysis of the resulting superconducting phase,
the authors suggest that an intermediate type-I state could be formed.

These considerations notwithstanding,
it is clear that the original analysis by \citet{Baym-etal69} misses an important piece of physics:
a realistic treatment of the coupling of the proton superconductor to the neutron condensate.
An improved formalism was first presented by \citet{AlfordGood08};
by incorporating density and density-gradient coupling terms into a \rewrite{phenomenological} \GL\ description of the condensates,
they found not only that the critical value of the \GL\ parameter $\kappa$ is changed,
but also that in some cases the transition is mediated by domains of ``type-II(n)'' superconductivity,
wherein each fluxtube carries $n$ magnetic flux quanta.
A more in-depth analysis by \citet{HaberSchmitt17}
found that these type-II(n) fluxtubes are generally unstable,
and instead there is a regime of ``type-1.5'' superconductivity,
in which the fluxtubes form bunches with a preferred separation.
We will discuss both studies in more detail below.

The possibility of multi-component systems having highly inhomogeneous magnetic field configurations
is not just an abstract concept applicable to exotic neutron star matter.
Similar ideas have been invoked to explain the behavior of unconventional superconductors,
such as the type-1.5 behavior observed in terrestrial experiments with MgB$_2$.\citep{Moshchalkov-etal09, Gutierrez-etal12}
Diverse field distributions in multi-band systems
generally result from the existence of three or more characteristic length scales
satisfying a specific hierarchy.\citep{BabaevSpeith05, Babaev-etal10, Babaev-etal17}
In a two-component superconductor, this corresponds to two coherence lengths,
connected to two energy gaps in a microphysical picture~\citep{SilaevBabaev11, SilaevBabaev12},
that satisfy $\xi_1 < \sqrt{2} \lambda < \xi_2$.
Under these conditions
fluxtubes are mutually attractive for large separations,
but repulsive for short separations,
resulting in the formation of fluxtube bundles that characterizes type-1.5 superconductivity.
We show later that entrainment between the two condensates can cause a similar ordering of length scales,
and may lead to type-1.5 superconductivity throughout much of the neutron star core.
This may have consequences not only for the magnetic field dynamics
but also for the star's rotational evolution,
because of expected interactions between fluxtubes and neutron vortices.

We follow previous studies~\citep{Alpar-etal84, AlfordGood08, Sinha2015, HaberSchmitt17, KobyakovPethick17}
and employ an effective \GL\ model \rewrite{for the condensates.
Although the \GL\ description is only formally valid close to the condensation temperature,
it provides the simplest phenomenological framework in which to study superfluid--superconductor interactions.}
\rewrite{Our goal in the present work is to identify the type of superconductivity for a given set of model parameters,
and so we are concerned only with the \gs\ for the condensates.
This means that we do not consider interactions between the condensates and the ``normal'' components of the core,
which include electrons, thermal excitations and normal protons and neutrons.
Neutron stars cool very efficiently by neutrino emission,\citep{Yakovlev-etal01, YakovlevPethick04}
and after $\sim \unit[10^{4}]{yr}$ lie far below
the critical temperatures for superconductivity and superfluidity,\citep{Yakovlev-etal99,Kaminker-etal01, Kaminker-etal02}
so the \emph{local} \gs\ is actually a good approximation of the real microphysical state in mature neutron stars.}

We will work with the most general \GL\ functional
that permits a consistent treatment of entrainment
while also \rewrite{correctly} satisfying Galilean\citep{DobaczewskiDudek95, ChamelHaensel06} invariance \rewrite{on small scales}.
Our resulting energy functional is connected to the Skyrme interaction potential,\cite{Bender-etal03, Dutra-etal12}
commonly used to describe the \ghs\ characteristics of finite nuclei and nuclear matter at high densities.
\review{After determining the neutron star's equilibrium composition
based on the standard Skyrme interaction,
we take advantage of this connection
to deduce realistic coefficients for our \GL\ model at each depth within the stellar core.}
Based on this description,
we subsequently determine the characteristics of the superconductor by minimizing the free energy of the coupled two-component system
to obtain the \gs\ in the presence of a magnetic field.
Hence, we construct phase diagrams
(using the nuclear matter density as the control parameter)
that indicate the domains of the different types of superconductivity in the neutron star core.
To explore how the phase diagram is affected by the underlying superfluid parameters and \EoS ,
we study the superconducting phase for density-dependent energy gaps~\citep{Kaminker-etal01, Andersson-etal05, Ho-etal15}
and a set of representative Skyrme models.\cite{Bender-etal03, Dutra-etal12}

The paper is organized as follows:
In Sec.~\ref{sec:formalism} we will present \rewrite{our \GL\ model,}
including a review of entrainment, Galilean invariance, and the connection to the Skyrme functional.
Section~\ref{sec:parameter-ranges} introduces parameters of six representative Skyrme models
\review{that we use to construct one-dimensional neutron star structure models,}
and the characteristic properties of the superfluids.
Sec.~\ref{sec:superconductivity} discusses our analytical and numerical approaches to construct the phase diagrams;
in particular, we examine two distinct types of ``experiments'' to study the superconductor's magnetic response.
Corresponding results are presented in Sec.~\ref{sec:results}.
Finally, we provide a conclusion and outlook into future work in Sec.~\ref{sec:conclusion}.
\rewrite{Details of some of the calculations} are provided in Appendices~\ref{sec:code}, \ref{sec:Kramer} and~\ref{sec:Abrikosov}.

%%%%%%%%%%%%%%%%%%%%%%%%%%%%%%%%%%%%%%%%%%%%%%%%%%%%%%%%%%%%%%
%%%%%%%%%%%%%%%%%%%%%%%%%%%%%%%%%%%%%%%%%%%%%%%%%%%%%%%%%%%%%%

\section{Basic Formalism}
\label{sec:formalism}

Our goal is to formulate a simple, phenomenological model of the neutron and proton condensates that includes
(a) their mutual entrainment,
and (b) the coupling to the magnetic field.
The simplest such model~\citep{Alpar-etal84,AlfordGood08,DrummondMelatos17}
is a \rewrite{phenomenological} \GL\ model,
in which the free energy is expressed in terms of complex scalar order parameters for the condensates, $\psip$ and $\psin$,
and the magnetic vector potential, $\mathbf{A}$.
We define the order parameters such that
$|\psip|^2$, for example, is the number density of proton Cooper pairs, and so $\rhop = 2\mpp|\psip|^2$ is the mass density of the proton condensate,
where $\mpp$ is the mass of a proton, and
the order parameters are therefore two-particle mean-field wave functions
for the condensates.
\review{As noted earlier, the core temperature in mature neutron stars lies far below the critical temperature for superconductivity,
so in the absence of magnetic flux practically all of the proton matter would reside in the condensed state.
In the presence of magnetic flux, however,
there will be normal proton matter present in the cores of fluxtubes and in any non-superconducting regions,
where the proton condensate is absent.
We are concerned here only with the \gs\ for the condensates, wherein interactions with any normal matter are suppressed,
and so we can safely disregard the normal matter in what follows.}

%%%%%%%%%%%%%%%%%%%%%%%%%%%%%%%%%%%%%%%%%%%%%%%%%%%%%%%%%%%%%%

\subsection{Entrainment and local phase invariance}
\label{subsec:entrainment}

To illustrate the effect of entrainment between the condensates,
we temporarily neglect any coupling to the magnetic field by setting $\mathbf{A}=\mathbf{0}$;
the magnetic field will be reintroduced later by invoking gauge invariance.
The hydrodynamical momenta of the condensates
are then proportional to the gradient of the
phases of the order parameters, i.e.,
\begin{align}
    \hslash\nablab\arg\psip = 2\mpp\Vp \,,
        \qquad
    \hslash\nablab\arg\psin = 2\mnn\Vn\,,
    \label{eq:velocity}
\end{align}
where $\Vp$ and $\Vn$ are the superfluid velocities.
In the presence of entrainment,
the velocity-dependent terms in the free-energy density,
$F_\text{vel}$,
\review{of the condensates}
must take the form
\begin{equation}
    F_\text{vel} =
        \tfrac{1}{2}\rhop|\Vp|^2 + \tfrac{1}{2}\rhon|\Vn|^2 - \tfrac{1}{2}\rho^\pn|\Vp-\Vn|^2\,,
    \label{eq:Galilean}
\end{equation}
where $\rhop$ and $\rhon$ are the true mass densities
\review{of the condensates}
and the coefficient $\rho^\pn$, which determines the strength of entrainment,
is generally negative.\citep{AndreevBashkin76}
This form of the free energy \review{--- with an interaction term that depends only on the relative velocity ---} is necessary to ensure Galilean invariance \citep{ChamelHaensel06}
\review{(as well as the more general constraint of ``local phase invariance'' \citep{DobaczewskiDudek95}).}

Equation~\eqref{eq:Galilean} demonstrates that
the primary effect of entrainment is to disfavor any relative flow between the two condensates
\review{by imposing an energetic penalty wherever $\Vp \neq \Vn$.}
A more subtle but equally important consequence is that
the condensates' hydrodynamical momenta, given in Eq.~\eqref{eq:velocity},
are no longer proportional to their mass fluxes,
defined as $\partial F_\text{vel}/\partial\mathbf{V}_x$ for $x \in \{\pp,\nn\}$.
This has significant consequences for the structure of fluxtubes and vortices,
and for their mutual interactions,\citep{Alpar-etal84}
but in the present work we are concerned only with fluxtubes,
i.e., the response of the proton condensate.
For the same reason, we will also neglect rotation.
While ignoring defects in the neutron condensate is certainly a simplification,
it is justified when deriving a microscale model of the neutron star interior,
where the proton fluxtube density is many orders of magnitude larger than
that of the neutron vortices.\citep{Graber-etal17}

Within the \GL\ mean-field framework,
entrainment first enters the free-energy density at fourth order in the order parameters,
and at second order in their derivatives.\citep{Alpar-etal84}
The most general such term
that satisfies global U(1) symmetry in each condensate
is a linear combination of the quantities
\begin{equation}
    |\psi_x|^2|\nablab\psi_y|^2,\
    \psi_x\psi_y\nablab\psi_x^\star\cdot\nablab\psi_y^\star,\
    \psi_x\psi_y^\star\nablab\psi_x^\star\cdot\nablab\psi_y,\
    \psi_x^\star\psi_y^\star\nablab\psi_x\cdot\nablab\psi_y,
\end{equation}
where $x,y \in \{\pp,\nn\}$.
With the additional constraint of Galilean invariance~\eqref{eq:Galilean},
the most general form of the entrainment term is found to be
\begin{align}
    F_\textrm{ent}
        &= \tfrac{1}{2}(h_1+h_2)\left|\left(\tfrac{\mnn}{\mpp}\right)^{1/2}\psin^\star\nablab\psip +
        \left(\tfrac{\mpp}{\mnn}\right)^{1/2}\psip\nablab\psin^\star\right|^2 \nonumber \\
    &+ \tfrac{1}{2}(h_1-h_2)\left|\left(\tfrac{\mnn}{\mpp}\right)^{1/2}\psin\nablab\psip -
        \left(\tfrac{\mpp}{\mnn}\right)^{1/2}\psip\nablab\psin\right|^2 \nonumber \\
    &+ \tfrac{1}{4}h_3\bigl|\nablab(\psip\psip^\star)\bigr|^2 + \tfrac{1}{4}h_4\bigl|\nablab(\psin\psin^\star)\bigr|^2\,,
  \label{eq:entrainment}
\end{align}
which includes four real, independent parameters $h_1$, \textellipsis, $h_4$.
In terms of the superfluid densities and velocities, we can write this as
\begin{align}
    F_\textrm{ent}
        &= h_1\left[\frac{\rhon}{4\mpp^2}\left|\nablab\rhop^{1/2}\right|^2
        + \frac{\rhop}{4\mnn^2}\left|\nablab\rhon^{1/2}\right|^2
        + \frac{\rhop\rhon}{\hslash^2}|\Vp-\Vn|^2\right] \nonumber \\
    &+ \frac{h_2}{8\mpp\mnn}\nablab\rhop\cdot\nablab\rhon
        + \frac{h_3}{16\mpp^2}|\nablab\rhop|^2
        + \frac{h_4}{16\mnn^2}|\nablab\rhon|^2\,,
    \label{eq:entrainment_fluid}
\end{align}
and so by comparison with Eq.~\eqref{eq:Galilean} the entrainment coefficient is
\begin{equation}
    \rho^\pn = -\dfrac{2}{\hslash^2}h_1\rhop\rhon.
    \label{eq:entrainment_coefficient}
\end{equation}
The remaining $h_i$ parameters only provide density-gradient coupling,
and the simplest model of entrainment would therefore set these parameters to zero.
On scales much larger than the fluxtube cores
the superfluid densities are approximately constant,
and so these terms will have negligible effect.
However, we will show later that the density-gradient terms
play a significant role in the transition between type-I and type-II superconductivity,
and therefore must be included in the construction of phase diagrams.

%%%%%%%%%%%%%%%%%%%%%%%%%%%%%%%%%%%%%%%%%%%%%%%%%%%%%%%%%%%%%%

\subsection{Connection with previous work, and with the Skyrme model}
\label{subsec:Skyrme-connection}

The velocity contributions to the free-energy density,
given by Eq.~(\ref{eq:Galilean}),
can equivalently be expressed as
\begin{equation}
    F_\text{vel} = \tfrac{1}{2}\rho^\text{pp}|\Vp|^2
        + \tfrac{1}{2}\rho^\text{nn}|\Vn|^2
        + \rho^\pn\Vp\cdot\Vn\,,
\end{equation}
where
$\rho^\text{pp}\equiv \rhop - \rho^\pn$
and $\rho^\text{nn} \equiv \rhon - \rho^\pn$
represent ``effective'' proton and neutron mass densities.
Therefore, on scales much larger than the vortex and fluxtube cores,
for which
the superfluid densities are approximately constant,
entrainment can be described by including in the free-energy density
a term proportional to $\Vp\cdot\Vn$,
and renormalizing the proton and neutron masses accordingly.
However, in a microscale model that \rewrite{correctly} includes density variations in the fluxtube cores,
the dependence of the coefficients $\rho^{xy}$
on the condensate densities
must be chosen carefully to preserve Galilean invariance\citep{ChamelHaensel06,Kobyakov20}
and additional density-gradient coupling terms
have to be included.
A number of previous studies~\citep{Alpar-etal84,DrummondMelatos17} \rewrite{do not treat entrainment on small scales consistently}
because their entrainment interactions are incompatible with
Eqs.~\eqref{eq:entrainment} and~\eqref{eq:entrainment_fluid}.
\citet{HaberSchmitt17} have introduced a relativistic model of \rewrite{density and derivative couplings}
\rewrite{that in the non-relativistic limit is also incompatible with our Eq.~\eqref{eq:entrainment_fluid},
cf.~their Eq.~(5).}
The model of \citet{AlfordGood08} \emph{is}
compatible with Eqs.~\eqref{eq:entrainment} and~\eqref{eq:entrainment_fluid},
but it only includes the $h_2$ term.
Hence their model actually has no entrainment at all, i.e., $\rho^\pn = 0$.
This appears to be an oversight on their part,
because they chose the value for $h_2$ based on prior estimates of $\rho^\pn$.
\rewrite{Finally, the model of \citet{Kobyakov20} has a similar but subtly different form to Eq.~\eqref{eq:entrainment},
because it takes the entrainment term to be}
\begin{equation}
  \frac{\left(\Im\{|\psin|^2\psip^\star\nablab\psip - |\psip|^2\psin^\star\nablab\psin\}\right)^2}{|\psip|^2|\psin|^2}\,.
\end{equation}
\rewrite{Although this quantity is Galilean invariant,
it cannot be obtained from products of the order parameters, their conjugates and derivatives,
and therefore cannot arise in our mean-field formalism.}

Given \rewrite{these inconsistencies,}
it is perhaps worth independently verifying
that the form of the entrainment energy given by Eq.~\eqref{eq:entrainment}
can be obtained as a suitable limit of the more realistic Skyrme energy density functional,
which is a mean-field model for the many-body interactions in
finite nuclei and nuclear matter.\citep{Bender-etal03}
Neglecting for simplicity the spin-orbit terms,
which cannot be described by scalar order parameters,
the entrainment and density-gradient terms in the Skyrme model are typically expressed in the form
\begin{equation}
    F_\text{Skyrme} = \sum_{t=0,1}\left[ C_t^{\Delta \rho} \rho_t \nabla^2 \rho_t
        + C_t^\tau(\rho_t\tau_t - j_t^2)\right]\,,
    \label{eq:Skyrme}
\end{equation}
where the local densities can be related to our two-particle scalar order parameters as follows:
\begin{align*}
    \rho_0 &= |\psin|^2 + |\psip|^2 \, , \\
    \tau_0 &= |\nablab\psin|^2 + |\nablab\psip|^2  \, , \\
    \mathbf{j}_0 &= \Im\{\psin^\star\nablab\psin + \psip^\star\nablab\psip\} \, , \\
    \rho_1 &= |\psin|^2 - |\psip|^2 \, , \\
    \tau_1 &= |\nablab\psin|^2 - |\nablab\psip|^2  \, , \\
    \mathbf{j}_1 &= \Im\{\psin^\star\nablab\psin - \psip^\star\nablab\psip\}\,.
\end{align*}
After applying integration by parts to the Laplacian terms
and neglecting surface contributions,
Eq.~\eqref{eq:Skyrme} can be written in the form of Eq.~\eqref{eq:entrainment},
by assuming that $\mpp = \mnn = \mau$,
where $\mau$ is the atomic mass unit,
and defining
\begin{align}
    h_1 &= C_0^\tau - C_1^\tau \, , \label{eq:h_1} \\
    h_2 &= - 4 C_0^{\Delta \rho} + 4  C_1^{\Delta \rho} \, , \\
    h_3 &= h_4 = C_0^\tau + C_1^\tau - 4  C_0^{\Delta \rho} - 4 C_1^{\Delta \rho}\,.
    \label{eq:h_3}
\end{align}
We will use this analogy between the Skyrme model and the Galilean-invariant \GL\ model
to identify physically meaningful choices for our parameters $h_i$.
In the next section,
we discuss realistic ranges for these and other model parameters.
We do however caution
that it is difficult to establish a direct one-to-one correspondence between the two descriptions,
because the Skyrme model contains many additional parameters and degrees of freedom
that are not reproducible in our \GL\ framework, which only considers scalar order parameters.
Nonetheless, the value that we obtain for the entrainment parameter in Eq.~\eqref{eq:h_1}
does at least match that used by \citet{ChamelHaensel06},
which can also be derived more rigorously in
the continuum limit of the Hartre--Fock--Bogoliubov theory.\cite{Chamel20}

%%%%%%%%%%%%%%%%%%%%%%%%%%%%%%%%%%%%%%%%%%%%%%%%%%%%%%%%%%%%%%
%%%%%%%%%%%%%%%%%%%%%%%%%%%%%%%%%%%%%%%%%%%%%%%%%%%%%%%%%%%%%%

\section{Parameter ranges}
\label{sec:parameter-ranges}

As outlined above,
our considerations are generic for interacting condensed-matter mixtures with a charged and a neutral constituent.
However, we are specifically interested in the neutron star core,
where neutrons and protons form two interpenetrating condensates.
For this system, we aim to construct phase diagrams of the resulting superconducting state.
The natural parameter controlling the physics in the stellar interior is the nuclear density,
which increases from $\sim \unitfrac[10^{14}]{g}{cm^3}$ at the crust-core interface
to $\sim \unitfrac[10^{15}]{g}{cm^3}$ at the star's center.
Such mass densities correspond to baryon number densities in the range of $\sim \unitfrac[0.06 - 0.6]{1}{fm^3}$.
\review{The \GL\ model that we introduced in the previous section cannot be applied over such a range of densities.
We will therefore construct realistic one-dimensional
stellar models using the Skyrme interaction,
and then apply the \GL\ model at each depth within the star's core,
using local parameter values deduced from the one-dimensional model.}

To examine the interacting two-component condensate,
we require not only the coupling parameters $h_i$,
assumed to be constant throughout the star,
but also the composition of nuclear matter and the energy gaps of the superfluid and superconductor
at any given density.
For the purpose of our initial analysis,
we present parameter ranges for several explicit examples,
but note that the method introduced in the following subsections
could be readily extended to other Skyrme models and energy gap parametrizations.
Software to reproduce the results of this section can be found at \url{https://github.com/vanessagraber/NS_EoS}.

\begin{table}[t]
\begin{ruledtabular}
\begin{tabular}{c ccdcccfffccc}
    Model & $t_0$  & $t_1$  & \multicolumn{1}{c}{$t_2$} & $t_3$ & $t_4$ & $x_0$ & \multicolumn{1}{c}{$x_1$} & \multicolumn{1}{c}{$x_2$} & \multicolumn{1}{c}{$x_3$} & $x_4$ & $\sigma$ & $\sigma_2$\\
    \hline
    LNS\citep{Cao-etal06} & $-2484.97$ & $266.74$ & -337.14 & $14588.20$ & - & $0.0623$ & 0.6585 & -0.9538 & -0.0341 & - & $1/6$ & - \\
    NRAPR\citep{Steiner-etal05} & $-2719.70$ & $417.64$ & -66.69 & $15042.00$ & - & $0.1615$ & -0.0480 & 0.0272 & 0.1361 & - & $0.1442$ & - \\
    Sk$\chi$450 \citep{LimHolt17} & $-1803.29$ & $301.82$ & -273.28 & $12783.86$ & $564.10$ & $0.4430$ & -0.3622 & -0.4105 & 0.6545 & $-11.3160$ & $1/3$ & $1$\\
    SLy4\citep{Chabanat-etal98} & $-2488.91$ & $486.82$ & -546.39 & $13777.00$ & - & $0.8340$ & -0.3440 & -1.0000 & 1.3540 & - & $1/6$ & - \\
    SQMC700\citep{Guichon-etal06} & $-2429.10$ & $370.97$ & -96.69 & $13773.43$ & - & $0.1000$ & 0.0000 & 0.0000 & 0.0000 & - & $1/6$ & - \\
    Ska35s20\citep{Dutra-etal12} & $-1768.80$ & $263.90$ & -158.30 & $12904.80$ & - & $0.1300$ & -0.8000 & 0.0000 & 0.0100 & - & $0.3500$ & -\\
\end{tabular}
\end{ruledtabular}
\caption{Parameters of six representative Skyrme models studied in this paper.
The parameters have units such that the energy density is given in $\unitfrac[]{MeV}{fm^3}$,
implying that $t_0$ is in $\unit[]{MeV \, fm^3}$, $t_1$ and $t_2$ are in $\unit[]{MeV \, fm^5}$,
$t_3$ is in $\unit[]{MeV \, fm^{3 + 3 \sigma}}$, and $t_4$ is in $\unit[]{MeV \, fm^{3 + 3 \sigma_2}}$,
while $x_0, x_1, x_2, x_3$, $x_4$, $\sigma$, and $\sigma_2$ are dimensionless.
Note that only for model Sk$\chi$450 the parameters $t_4, x_4$ and $\sigma_2$ are fitted.
For details see \citet{LimHolt17}.}
\label{tbl:Skyrme}
\end{table}

%%%%%%%%%%%%%%%%%%%%%%%%%%%%%%%%%%%%%%%%%%%%%%%%%%%%%%%%%%%%%%

\subsection{Skyrme models}
\label{subsec:Skyrme-parameters}

The development of Skyrme models has been driven by the idea
to employ an effective density-dependent many-body interaction for the description of finite nuclei
as well as nuclear matter.
The advantage of this approach is that all relevant quantities can be calculated analytically,
as model parameters are fitted to match certain data of finite nuclei.
While generally performing well at saturation density $\rho_0 \sim \unitfrac[2.82 \times 10^{14}]{g}{cm^3}$,
or equivalently $n_0 \sim \unitfrac[0.17]{1}{fm^3}$,
these models can predict characteristics of nuclear matter that differ significantly at high densities.
As a result,
it is generally unclear up to which density the Skyrme prescription remains valid.
To proceed,
we make the assumption that nuclear matter in the density range $\sim \unitfrac[10^{14} - 10^{15}]{g}{cm^3}$
is well characterized by the Skyrme functional.

The literature on the subject provides a wide range of possibilities
(see \citealt{Dutra-etal12} for a recent review)
with different models exhibiting different degrees of success in satisfying macroscopic constraints as,
e.g., deduced from heavy-ion collisions.
In the following, we choose six Skyrme models that are relatively successful in doing so\citep{Dutra-etal12}
and have been previously applied to model neutron stars,
namely LNS,\citep{Cao-etal06} NRAPR,\citep{Steiner-etal05} Sk$\chi$450,\citep{LimHolt17}
SLy4,\citep{Chabanat-etal98} SQMC700,\citep{Guichon-etal06} and Ska35s20.\citep{Dutra-etal12}
Each of these models can be represented by a set of parameters
$x_0, x_1, x_2, x_3, x_4, t_0, t_1, t_2, t_3, t_4, \sigma, \sigma_2$,\citep{Dutra-etal12}
summarized in Table~\ref{tbl:Skyrme}.
These parameters are directly related to the coefficients in Eq.~\eqref{eq:Skyrme} via\citep{Bender-etal03}
\begin{align}
    C_0^\tau &= \frac{3}{16} \, t_1 + \frac{1}{4} t_2 \left( \frac{5}{4} + x_2 \right) \, ,\\
    C_1^\tau &= - \frac{1}{8} \, t_1 \left( \frac{1}{2} + x_1 \right) + \frac{1}{8} t_2 \left( \frac{1}{2} + x_2 \right) \, , \\
    C_0^{\Delta \rho} &= -\frac{9}{64} \, t_1 + \frac{1}{16} t_2 \left( \frac{5}{4} + x_2 \right) \, , \\
    C_1^{\Delta \rho} &= \frac{3}{32} \, t_1 \left( \frac{1}{2} + x_1 \right) + \frac{1}{32} t_2 \left( \frac{1}{2} + x_2 \right) \, .
    \label{eq:C_coefficients}
\end{align}
We can then use
Eqs.~\eqref{eq:h_1}--\eqref{eq:h_3},
to calculate the coupling coefficients $h_i$,
which for convenience are presented in Table~\ref{tbl:h_i}.
We point out that
although we do not probe a continuous range of Skyrme parameters and their influence on the superconducting phase,
we have chosen models that cover a typical parameter range.
Moreover, we have included the model Ska35s20,
which results in an $h_1$ value (directly related to the entrainment parameter $\rho^\pn$ via Eq.~\eqref{eq:entrainment})
that is several orders of magnitude smaller than the $h_1$ values of the other five models,
as a comparison, in order to illustrate the influence of weak versus strong entrainment on the superconducting state.

\begin{table}[b]
\begin{minipage}[c]{0.4\textwidth}
\begin{ruledtabular}
\begin{tabular}{c hhh}
    Model & \multicolumn{1}{c}{$h_1$} & \multicolumn{1}{c}{$h_2$} & \multicolumn{1}{c}{$h_3$}\\
    \hline
    LNS & 44.552 & 310.013 & 45.546 \\
    NRAPR & 85.007 & 322.613 & 218.840 \\
    Sk$\chi$450 & 7.493 & 239.668 & 205.570 \\
    SLy4 & 32.473 & 370.614 & 327.143 \\
    SQMC700 & 68.570 & 302.400 & 185.485 \\
    Ska35s20 & 0.010 & 158.330 & 237.510 \\
\end{tabular}
\end{ruledtabular}
\end{minipage}
\caption{Coefficients of the \GL\ model for six Skyrme models. $h_i$ parameters are given in $\unit[]{MeV \, fm^5}$.}
\label{tbl:h_i}
\end{table}

\begin{figure}[t]
\centering
    \includegraphics[width=0.48\textwidth]{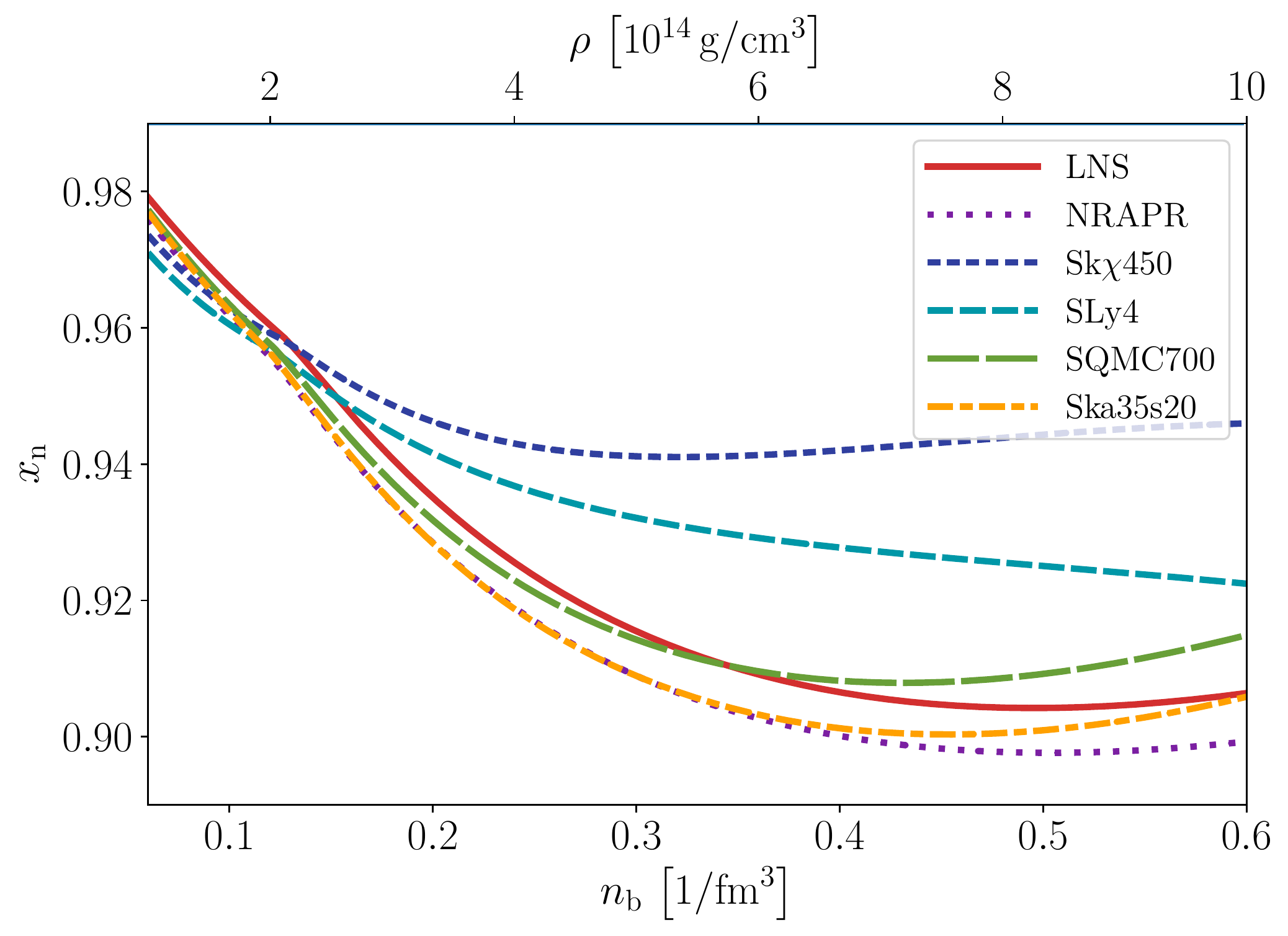}
    \hspace{0.1cm}
    \includegraphics[width=0.48\textwidth]{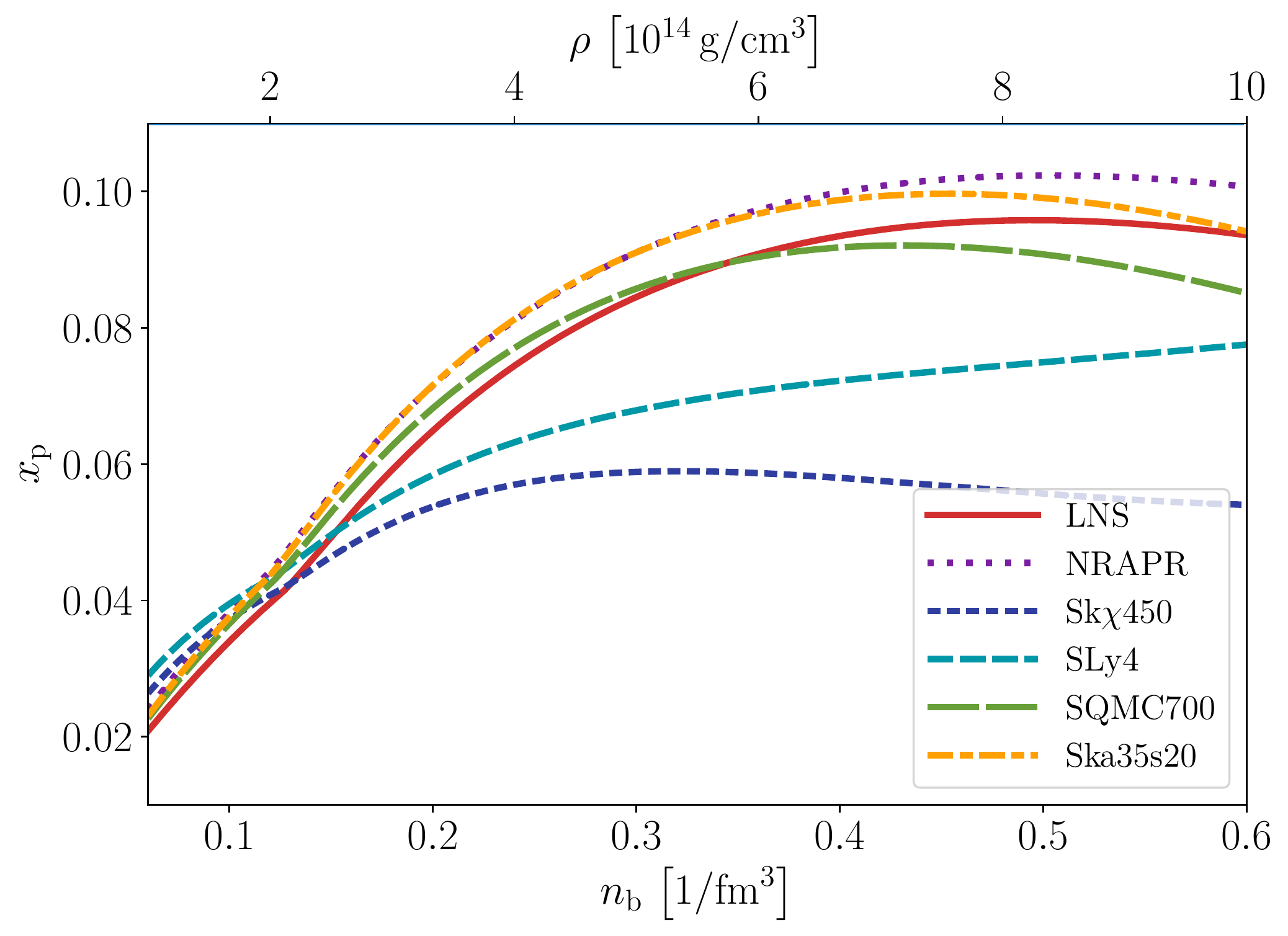}
\caption{Neutron (left) and proton (right) particle number fractions, $x_x \equiv n_x / n_\bb$,
in the neutron star core as a function of baryon number density $n_\bb$ (lower x-axis)
and mass density $\rho$ (upper x-axis) for six Skyrme models.
Note that $x_\nn + x_\pp = 1$, due to baryon number conservation.}
\label{fig:particle_fractions}
\end{figure}

In addition to calculating suitable estimates for the coefficients of the \GL\ functional,
we also employ the Skyrme models to determine the neutron star composition.
Following \citet{Chamel08},
we simultaneously solve equations for baryon conservation, charge neutrality, beta equilibrium and
\review{the equilibrium due to weak processes} for a given set of Skyrme parameters.
This allows us to deduce number densities and particle fractions of the neutrons, protons, electrons and muons
(which appear once the \review{electron chemical potential} exceeds the muon rest-mass energy)
for any given baryon density;
neutron and proton fractions for all Skyrme models are shown in Fig.~\ref{fig:particle_fractions}.
The approach also provides dynamical effective masses related to the condensates' entrainment
as well as Landau effective masses,
characterizing the static \gs s,
as a function of stellar density.
For an explanation of the different effective masses,
we refer the reader to the detailed discussion in \citet{ChamelHaensel06}.

%%%%%%%%%%%%%%%%%%%%%%%%%%%%%%%%%%%%%%%%%%%%%%%%%%%%%%%%%%%%%%

\subsection{Energy gaps and coherence lengths}
\label{subsec:gaps-coherence-length}

The Skyrme models is designed to provide a mean-field description of interacting particles,
and a separate microscopic pairing force is usually specified to calculate the pairing gaps,
leading to an independent macroscopic description of the pairing properties of the condensates.
We thus adopt a separate formalism for the behavior of the pairing gap as a function of density.
Protons are expected to pair in a spin-singlet state, while neutrons pair in a triplet state.
Superfluidity is present if the formation of Cooper pairs results in a lowering of the \ghs\ energy.
The parameter characterizing this process is the energy gap, $\Delta$,
which corresponds to the energy needed to create a quasi-particle of momentum $k$.
The energy gap at the Fermi level $\Delta(k_{\FF x})$,
where $k_{\FF x} \equiv (3 \pi^2 n_{x})^{1/3}$ is the Fermi number,
is therefore half the energy required to break a Cooper pair.
We remind the reader
that we relate the true number densities to the Cooper pair densities through $n_x = 2 |\review{\psi}_x|^2$.
Gap computations are difficult,
because theoretical models have to go beyond the bare nucleon-nucleon interaction and include additional physics
such as in-medium effects.
For the proton condensate, one of the difficulties is to correctly account for the neutron background
and current calculations arrive at maximum gaps of $\sim \unit[0.4 - 1]{MeV}$
in the range $k_{\FF\pp} \sim \unitfrac[0.4 - 0.7]{1}{fm}$.
For the triplet-paired neutrons details are even more unclear,
since the state is anisotropic and requires solution of the anisotropic gap equation.
Current models typically predict maximum gaps on the order of $\sim \unit[0.1 - 0.6]{MeV}$
at $k_{\FF\nn} \sim \unitfrac[1.4 - 2.5]{1}{fm}$.

\begin{figure}[t]
\centering
    \includegraphics[width=0.48\textwidth]{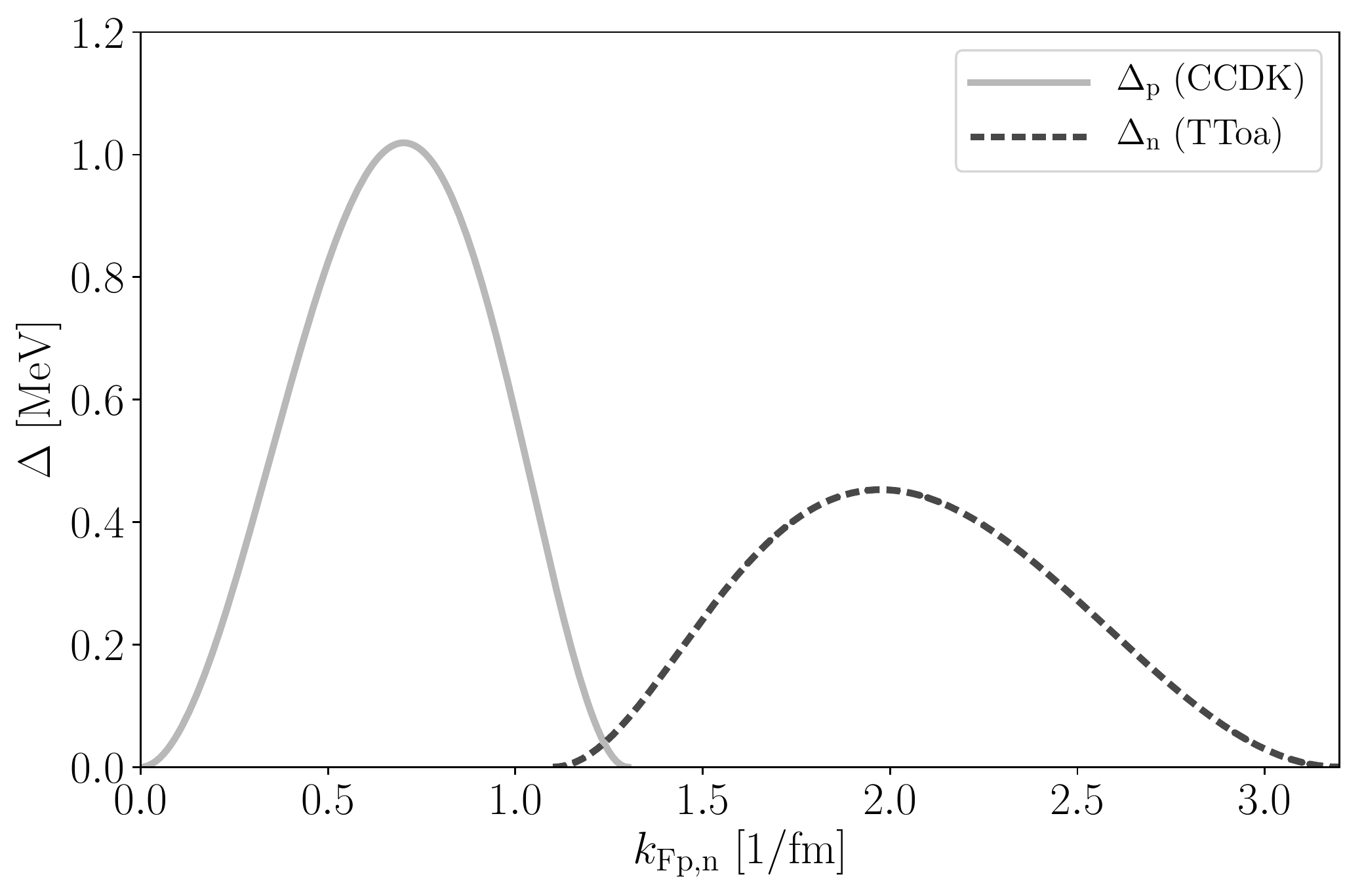}
\caption{Singlet proton (CCDK) and triplet neutron (TToa) energy gaps $\Delta_x$
as a function of Fermi wave number $k_{\FF x}$.}
\label{fig:gapskF}
\end{figure}

\begin{figure}[t]
\centering
    \includegraphics[width=0.48\textwidth]{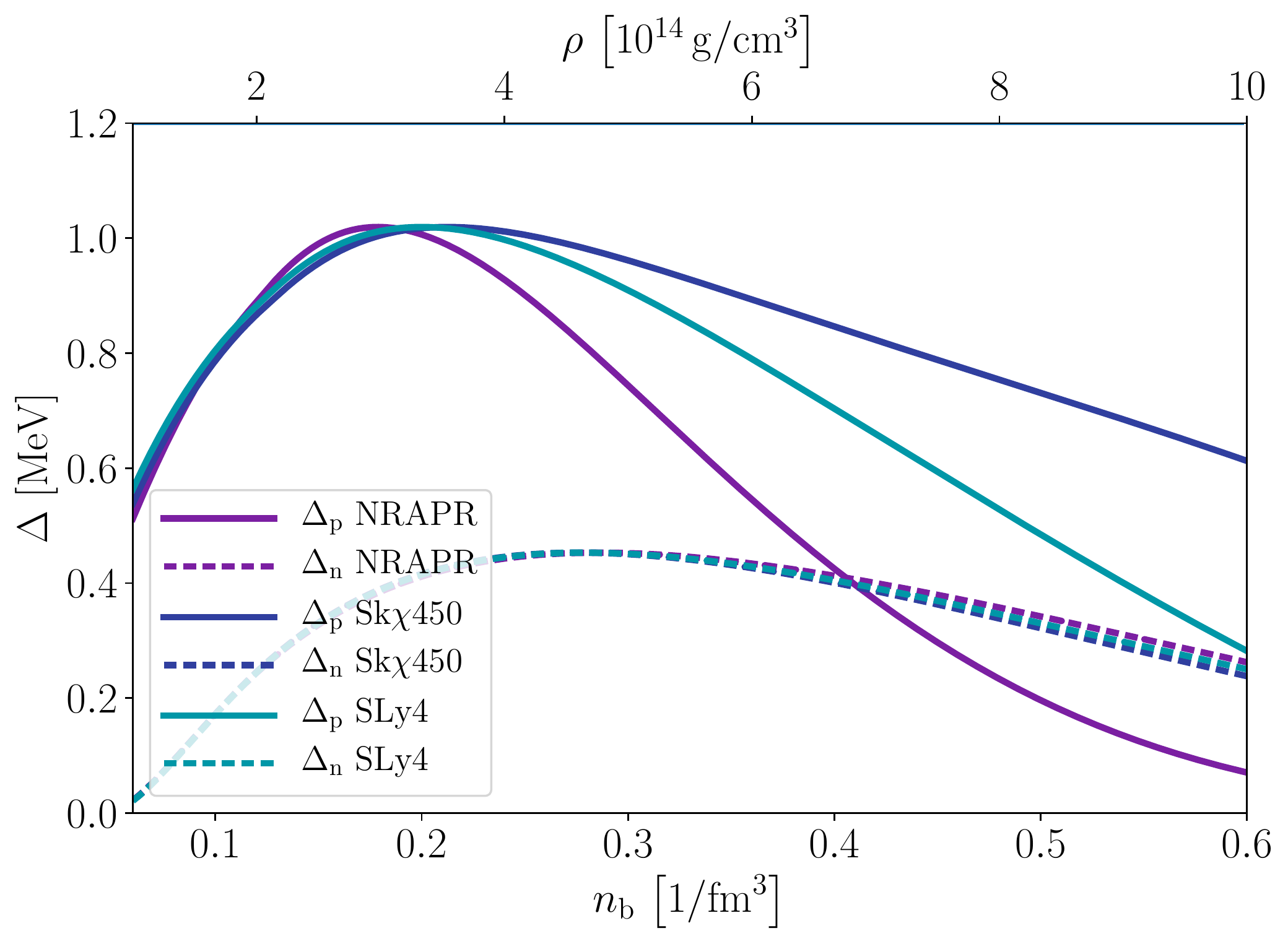}
    \hspace{0.1cm}
    \includegraphics[width=0.48\textwidth]{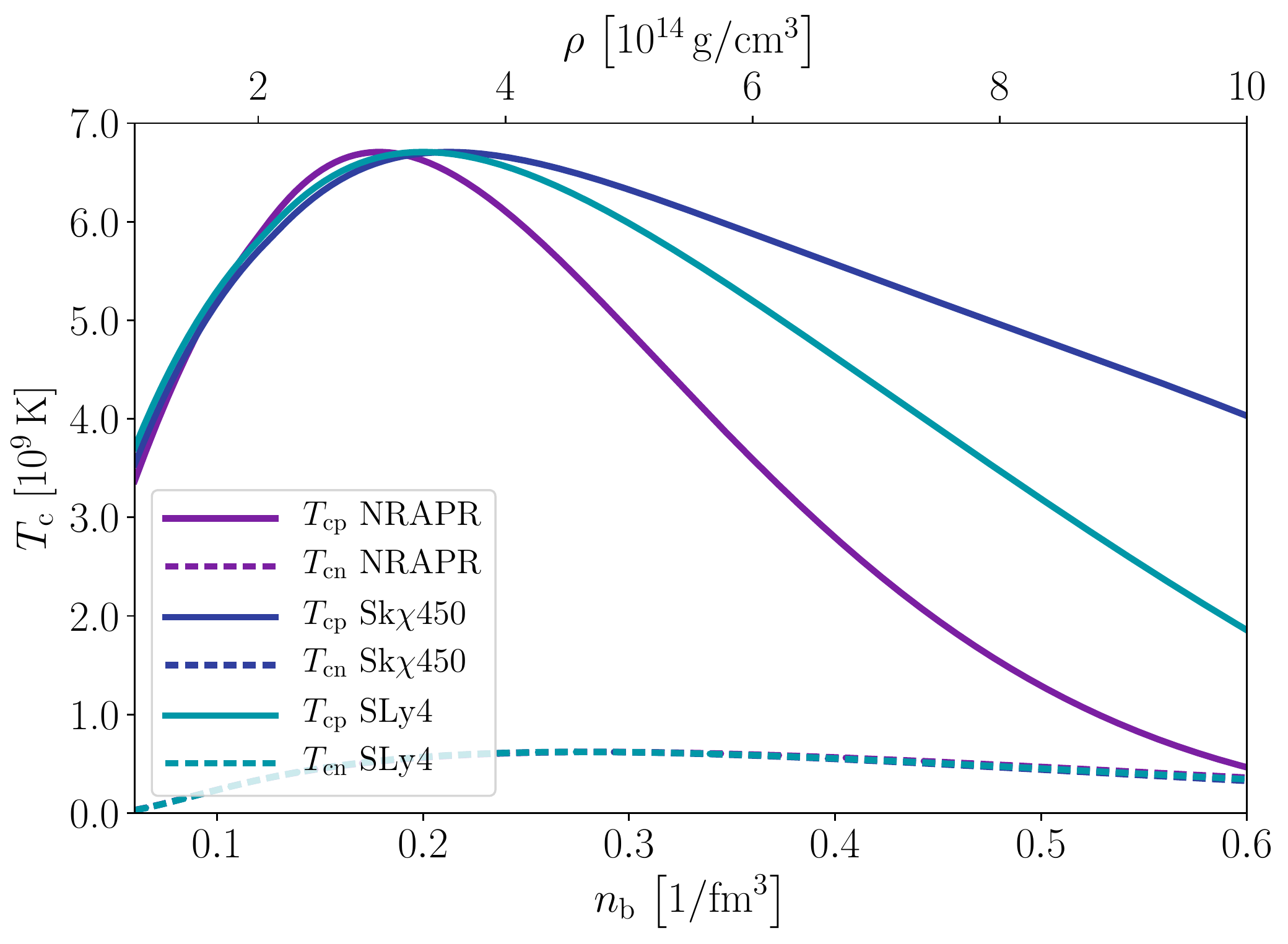}
\caption{Energy gaps $\Delta_x$ (left) and
transition temperatures $T_{\cc x}$ (right)
as a function of baryon number density $n_\bb$ (lower x-axis)
and mass density $\rho$ (upper x-axis),
calculated for three of the six Skyrme models to illustrate the quantities' typical spread.}
\label{fig:gapsTc}
\end{figure}

To provide an approach that can easily be repeated for different gap models,
we follow \citet{Andersson-etal05} (see also \citealt{Kaminker-etal01})
and represent the energy gaps at the Fermi surface by a phenomenological formula:
\begin{equation}
    \Delta_{x} (k_{\FF x}) = \Delta_0 \, \frac{(k_{\FF x} - k_1)^2} {(k_{\FF x} - k_1)^2 + k_2} \,
        \frac{(k_{\FF x} - k_3)^2} {(k_{\FF x} - k_3)^2 + k_4}\, .
\end{equation}
The fit parameters $\Delta_0$ and $k_i$ allow one to adapt the shape of specific gaps available in the literature.
As a study of different gaps is beyond the scope of this paper and will be reserved for future work,
we focus on two of the models presented in \citet{Ho-etal15},
namely the $^1S_0$ CCDK \citep{Chen-etal93} gap for the protons
and the $^3P_2$ TToa \citep{TakatsukaTamagaki04} model for the neutrons.
Both gaps find tentative support in explaining the observed cooling behavior
of the neutron star in the Cassiopeia A supernova remnant.\citep{Page-etal11, Ho-etal15, Wijngaarden-etal19}
The corresponding fit parameters are
\begin{align}
    \text{CCDK proton gap:} \hspace{1cm} & \Delta_0 = \rewrite{\unit[102.0]{MeV}}, k_1 = \unitfrac[0.0]{1}{fm},
        k_2 = \unitfrac[9.0]{1}{fm^2}, k_3 = \unitfrac[1.3]{1}{fm}, k_4 = \unitfrac[1.5]{1}{fm^2} \, , \\[1.1ex]
    \text{TToa neutron gap:} \hspace{1cm} & \Delta_0 = \unit[2.1]{MeV}, k_1 = \unitfrac[1.1]{1}{fm},
        k_2= \unitfrac[0.6]{1}{fm^2}, k_3 = \unitfrac[3.2]{1}{fm}, k_4 = \unitfrac[2.4]{1}{fm^2} \, .
\end{align}
Both gaps as a function of $k_{\FF x}$ are shown in Fig.~\ref{fig:gapskF}.
Having calculated the composition of the star for a given Skyrme model,
we already have information about the Fermi wave numbers as a function of density
and can, hence, calculate $\Delta_{x}$ for any depth inside the star;
gaps as a function of mass and number density are shown
in the left panel of Fig.~\ref{fig:gapsTc} for three of the six Skyrme parametrization.

We further note that the energy gap also provides information on the critical temperature of each condensate.
In the zero-temperature limit,
the gap takes its maximum value and we have
$k_B T_{\cc\pp} \approx 0.567 \Delta_{\pp}$ for the isotropic singlet proton gap
and $k_B T_{\cc\nn} \approx 0.118 \Delta_{\nn}$ for the anisotropic triplet neutron gap.\citep{TakatsukaTamagaki71}
Transition temperatures as a function of density are shown in the right panel of Fig.~\ref{fig:gapsTc},
indicating that for equilibrium neutron stars with $T \sim \unit[10^7]{K}$,
the neutrons and the protons in the entire core are paired,
validating a crucial assumption of our \GL\ model.

In the following,
we also require the coherence lengths of both condensates.
In the zero-temperature \review{and single-component} limit, a rough estimate for \review{these length scales} can be obtained from Pippard's expression
\begin{equation}
    \xi_{x} \equiv \frac{\hbar^2 k_{\FF x}}{\pi m_{x}^{\oplus} \Delta_{x}}\, ,
    \label{eq:coherencelength}
\end{equation}
where we follow \citet{ChamelHaensel06} in denoting the Landau effective masses with $m_{x}^{\oplus}$.
Note that expression~\eqref{eq:coherencelength} is derived under the assumption of weak coupling
\review{and neglects in-medium effects.
However, accounting for these does not significantly change the respective values \citep{deBlasio-etal97, Matsuo06}
and we will therefore determine the ``bare'' coherence lengths $\xi_{x}$ in the absence of coupling between the neutrons and the protons} using the above equation.

%%%%%%%%%%%%%%%%%%%%%%%%%%%%%%%%%%%%%%%%%%%%%%%%%%%%%%%%%%%%%%
%%%%%%%%%%%%%%%%%%%%%%%%%%%%%%%%%%%%%%%%%%%%%%%%%%%%%%%%%%%%%%

\section{Superconductivity}
\label{sec:superconductivity}

\subsection{The Helmholtz and Gibbs free energies}

Our goal now is to determine the type of superconductivity at each depth within the neutron star,
for each of the six equations of state.
To do so, we employ a zero-temperature \GL\ model,
with parameters that are chosen to match
the one-dimensional structure models described in Sec.~\ref{sec:parameter-ranges}.

The total free-energy density in our model is obtained by adding the entrainment terms~\eqref{eq:entrainment} \review{(with $\mpp = \mnn = \mau$ as well as $h_3 = h_4$ for consistency with the Skyrme interaction)}
to the usual free energy of a two-component superfluid,
and introducing the magnetic vector potential $\mathbf{A}$ by minimal coupling.
In Gaussian cgs units, the result \review{in its most compact form} can be expressed as
\begin{align}
    F[\psip,\psin,\mathbf{A}]
        &= F_0 - \mu_\pp|\psip|^2 - \mu_\nn|\psin|^2 + \frac{\gpp}{2}|\psip|^4
        + \frac{\gnn}{2}|\psin|^4 + g_\pn|\psip|^2|\psin|^2 \nonumber \\
    &+ \frac{\hslash^2}{4\mau}\left|\left(\nablab - \frac{2\ii e}{\hslash c}\mathbf{A}\right)\psip\right|^2
        + \frac{\hslash^2}{4\mau}\left|\nablab\psin\right|^2
        + \frac{1}{8\pi}|\nablab\times\mathbf{A}|^2
        \nonumber \\
    &+ h_1\left|\left(\nablab - \frac{2\ii e}{\hslash c}\mathbf{A}\right)(\psin^\star\psip)\right|^2
        + \dfrac{1}{2}(h_2-h_1)\nablab(|\psip|^2)\cdot\nablab(|\psin|^2) \nonumber \\
    &+ \dfrac{1}{4}h_3\left(\bigl|\nablab(|\psip|^2)\bigr|^2 + \bigl|\nablab(|\psin|^2)\bigr|^2\right)\,,
\label{eqn:Ffull}
\end{align}
where $F_0$ is an arbitrary reference level,
and where we have assumed that the proton Cooper pairs have charge $2\ee$.
The coefficients $\mu_\pp$ and $\mu_\nn$ are the chemical potentials of the
\review{proton and neutron Cooper pairs},
$\gpp$ and $\gnn$ define the self-repulsion of the condensates,
and $\gpn$ defines their mutual repulsion.

In the absence of magnetic fields,
we expect the \gs\ for this system to be a uniform mixture of proton and neutron condensates
with position-independent densities $|\psip|^2$ and $|\psin|^2$,
whose values depend on the chemical potentials $\mu_\pp$ and $\mu_\nn$.
Following \citet{AlfordGood08},
we choose $\mu_\pp$ and $\mu_\nn$ such that these densities match those obtained
in the one-dimensional structure models of Sec.~\ref{sec:parameter-ranges},
i.e., $\mu_\nn = (\gnn n_\nn + \gpn n_\pp)/2$ and $\mu_\pp = (\gpp n_\pp + \gpn n_\nn)/2$.
However, if the mutual attraction/repulsion between the condensates is too strong,
such that $\gpn^2 > \gpp\gnn$,
then this two-component system becomes unstable.
Such behavior is not expected in neutron stars,
where the two condensates are believed to be only weakly attractive,\citep{Alford-etal05}
and so in what follows we will always assume that $\gpn^2 < \gpp\gnn$.
Moreover, we will assume that the neutron chemical potential, $\mu_\nn$,
is positive, meaning that a neutron condensate is present even in non-superconducting regions, where $\psi_\pp = 0$.
This implies a further restriction $\gpn > - \gnn n_\nn/n_\pp$.
The consequences of violating these restrictions have been discussed in detail by \citet{HaberSchmitt17},
for instance.
In most studies of two-component condensates the coefficient $\gpn$ represents the principle interaction
between the two components,\citep{Esry-etal97,Law-etal97,BashkinVagov97,RiboliModugno02}
\review{and its effect on superconductivity in the neutron star core has been studied extensively.\citep{AlfordGood08,HaberSchmitt17,Kobyakov20}
In the present work, however, our main focus is on the effect of entrainment and other higher-order coupling terms.}
In the numerical results we present later, we therefore generally \review{take} $\gpn=0$,
and \review{study} the effect of the $h_i$ parameters \review{on the superconductor}.
\review{For completeness, and to facilitate comparison with earlier studies, we retain $\gpn$ in our analytical results.}

In the absence of coupling between the condensates (i.e.~for $\gpn = 0$ and $h_i = 0$)
the ``bare'' coherence lengths \review{(equivalent to those given in Eq.~\eqref{eq:coherencelength})} are defined as
\begin{equation}
    \xi_\pp \equiv \frac{\hslash}{\sqrt{2\mau\gpp n_\pp}}
        \qquad\mbox{and}\qquad
    \xi_\nn \equiv \frac{\hslash}{\sqrt{2\mau\gnn n_\nn}}\,,
\label{eq:bare_coherence}
\end{equation}
and the \rewrite{``bare''} London length is defined as
\begin{equation}
    \lambda \equiv \sqrt{\frac{\mau c^2}{4\pi\ee^2n_\pp}}\,.
\label{eq:bare_London}
\end{equation}
We will show below
how the effective coherence lengths and London length are modified
by the coupling between the condensates,
including their mutual entrainment.

\begin{figure}[t]
\centering
    \includegraphics[width=0.48\textwidth]{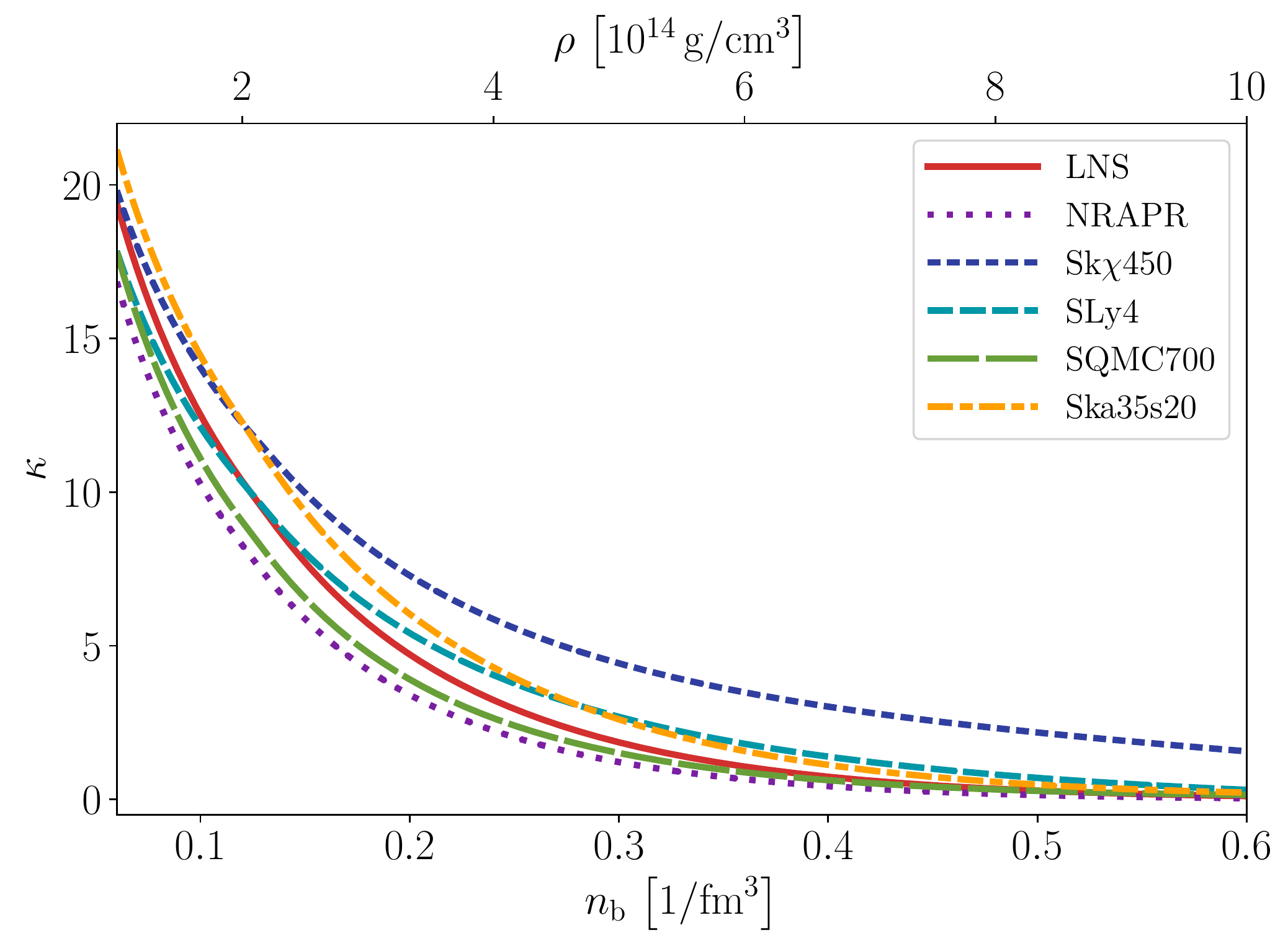}
    \hspace{0.1cm}
    \includegraphics[width=0.48\textwidth]{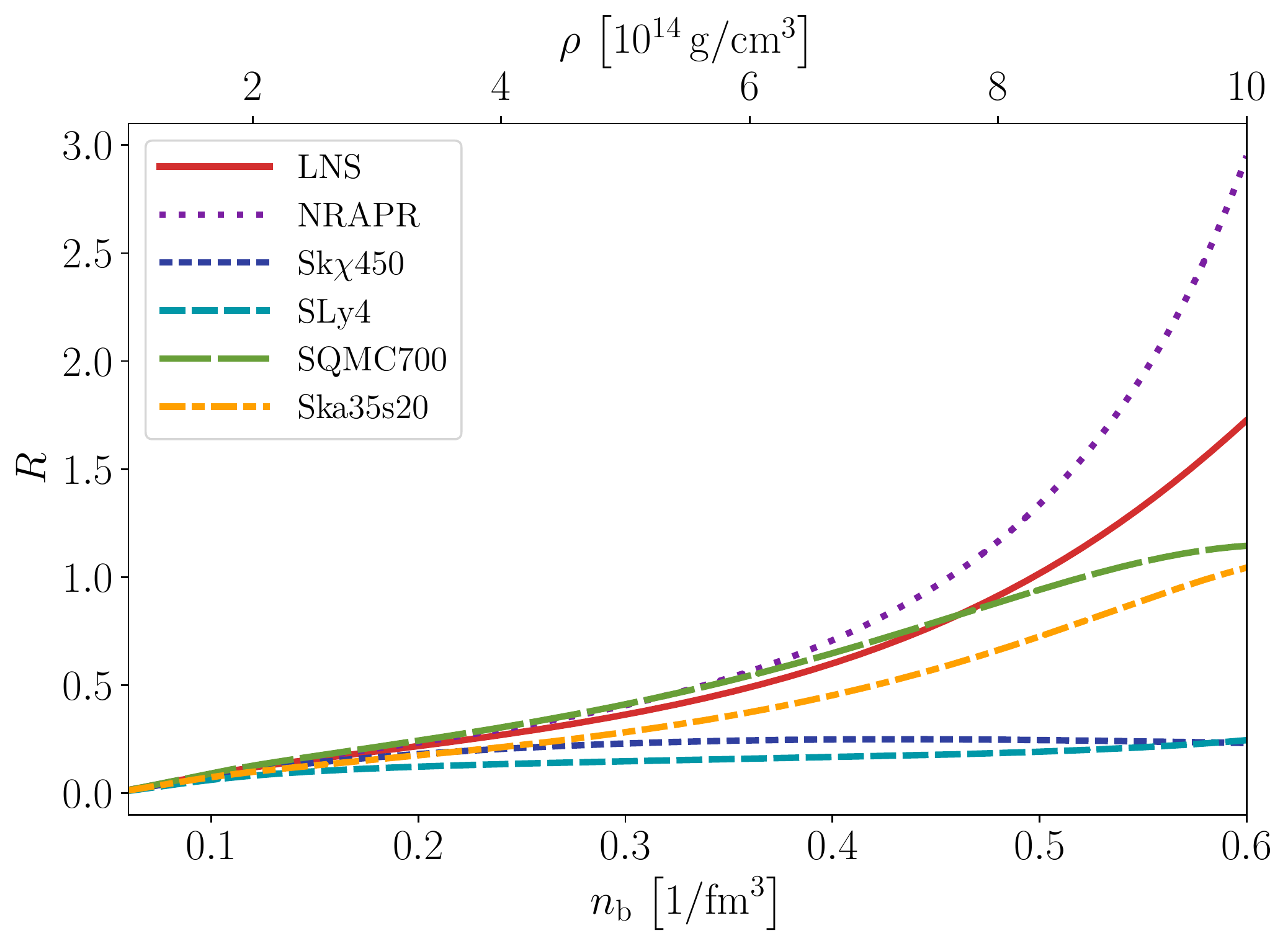}
\caption{Dimensionless \GL\ parameter $\kappa$ (left) and
$R$ (right)
as a function of baryon number density $n_\bb$ (lower x-axis)
and mass density $\rho$ (upper x-axis)
for the six Skyrme models studied in this paper.}
\label{fig:kappa_R}
\end{figure}

In order to simplify the mathematical model,
we now nondimensionalize the free-energy density~\eqref{eqn:Ffull} by measuring $\psip$ and $\psin$
in units of $\sqrt{n_\pp/2}$ and $\sqrt{n_\nn/2}$, respectively, lengths in units of $\xi_\pp$,
$\mathbf{A}$ in units of $\hslash c /2\ee\xi_\pp$,
and the coupling coefficients $h_i$ in units of $\gpp\xi_\pp^2$.
Note that while the dimensional $h_i$
coupling coefficients
are independent of density,
their dimensionless counterparts vary with depth inside the star.
To improve the readability of our equations,
we avoid introducing specific notation for dimensionless quantities and instead point out that,
from now on,
all parameters refer to dimensionless quantities.
The dimensionless free-energy density is then,
in units of $\gpp n_\pp^2/4$,
\begin{align}
    F[\psip,\psin,\mathbf{A}]
        &= \frac{1}{2}(1-|\psip|^2)^2
        + \frac{R^2}{2\epsilon}(1-|\psin|^2)^2
        + \frac{\gr}{\epsilon}(1-|\psip|^2)(1-|\psin|^2)
        \nonumber \\
    &+ \left|\left(\nablab - \ii\mathbf{A}\right)\psip\right|^2
        + \frac{1}{\epsilon}\left|\nablab\psin\right|^2
        + \kappa^2|\nablab\times\mathbf{A}|^2
        \nonumber \\
    &+ \frac{h_1}{\epsilon}\left|\left(\nablab - \ii\mathbf{A}\right)(\psin^\star\psip)\right|^2
        + \frac{(h_2-h_1)}{2\epsilon}\nablab(|\psip|^2)\cdot\nablab(|\psin|^2)
        \nonumber \\
    &+ \frac{h_3}{4}\left(\bigl|\nablab(|\psip|^2)\bigr|^2
        + \frac{1}{\epsilon^2}\bigl|\nablab(|\psin|^2)\bigr|^2\right)\,,
\label{eq:F_dimless}
\end{align}
where we have chosen the reference level to be $F_0 = 1/2 + R^2 / (2 \epsilon) + \alpha/ \epsilon$,
such that the free-energy density vanishes in the absence of magnetic fields,
with $|\psip| = 1$ and $|\psin| = 1$,
and we have defined the following parameters:
\begin{gather}
    \kappa \equiv \frac{\lambda}{\xi_\pp}\,, \quad
    R \equiv \frac{\xi_\pp}{\xi_\nn} \,, \quad
    \epsilon \equiv \frac{n_\pp}{n_\nn}\,, \quad
    \gr \equiv \frac{\gpn}{\gpp}\,.
\end{gather}
Note that $\kappa$ is equivalent to our dimensionless \rewrite{``bare'' London length}.
To illustrate their variation within the neutron star interior,
the first two dimensionless quantities are plotted in Fig.~\ref{fig:kappa_R}
as a function of baryon density for our six Skyrme models.
We do not show a separate plot for $\epsilon$,
which would closely resemble the proton fraction $x_\pp$ in the right panel of Fig.~\ref{fig:particle_fractions},
because the large neutron fraction inside the core
(see left panel of Fig.~\ref{fig:particle_fractions})
dictates $n_\bb \simeq n_\nn$ and thus $\epsilon \simeq x_\pp$.

\begin{figure}[t]
\centering
    \includegraphics[width=0.48\textwidth]{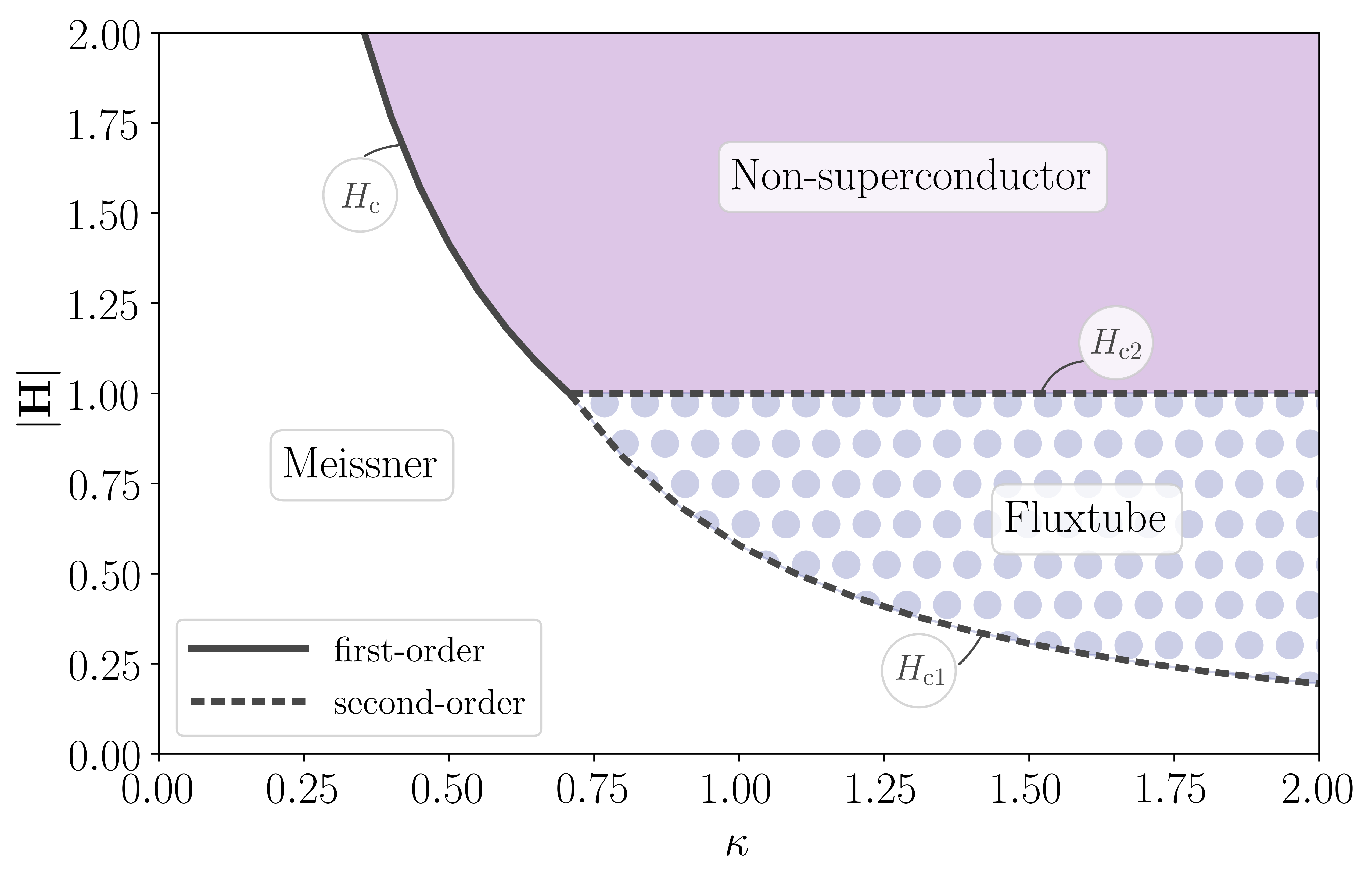}
    \hspace{0.1cm}
    \includegraphics[width=0.48\textwidth]{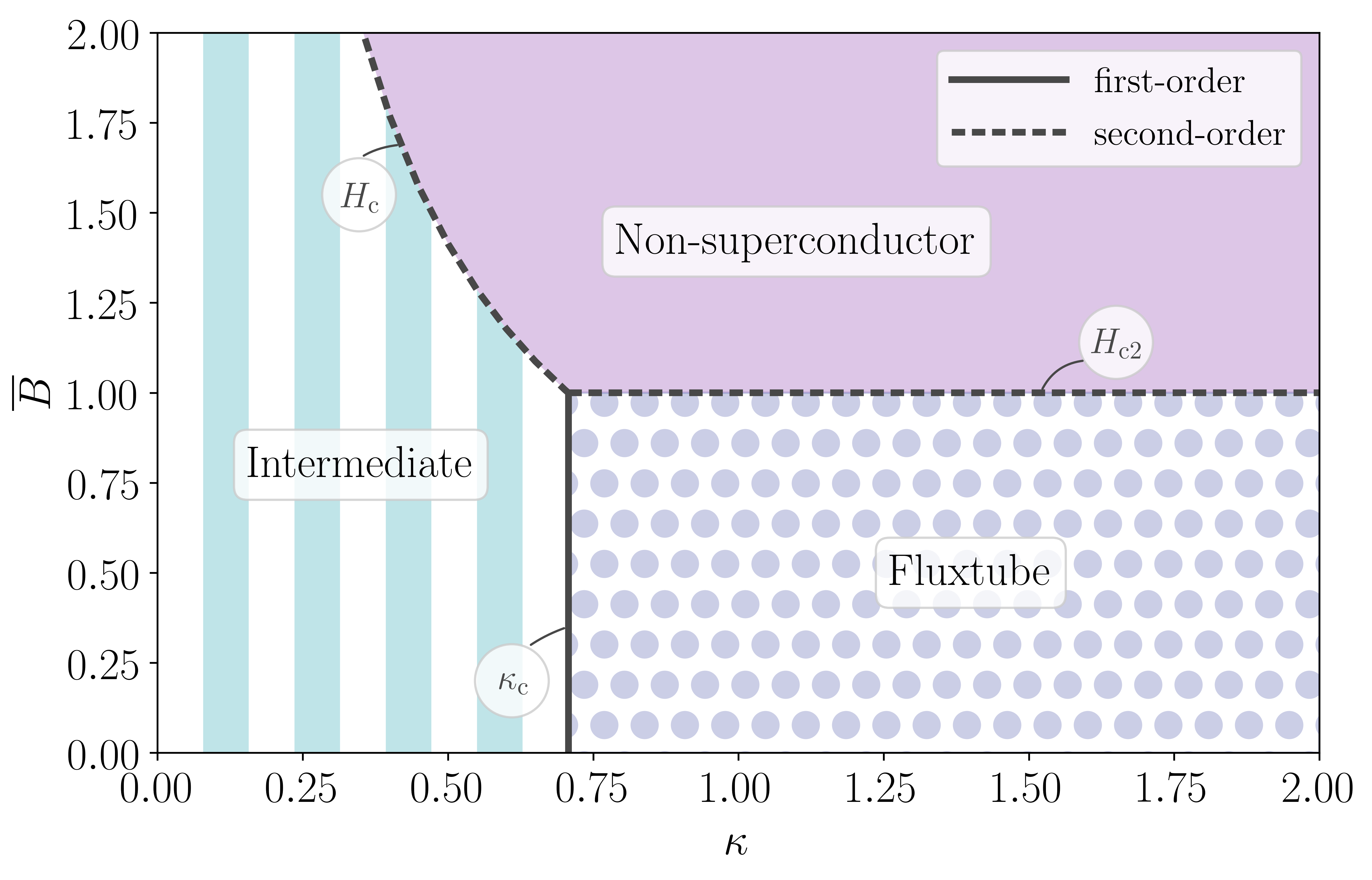}
\caption{Phase diagrams for a one-component \GL\ superconductor,
for different values of the \GL\ parameter, $\kappa$.
The left panel shows the experiment with an imposed external field,
$|\mathbf{H}|$, in our nondimensional units.
The first-order and second-order transitions at the different critical fields are indicated by
solid and dashed black lines, respectively,
and the resulting phases labelled accordingly.
Shading of the respective regions is indicative of the magnetic flux distribution.
The right panel shows the phase transitions in the experiment
with an imposed mean flux, $\overline{B}$.
For more details see the text.}
\label{fig:GL}
\end{figure}

We now seek the \gs\ for this system in the presence of an imposed magnetic field.
There are two distinct thought-experiments that can be considered.
In the first experiment, we control the magnetic flux density,
$\mathbf{B} = \nablab\times\mathbf{A}$,
by imposing a mean or net magnetic flux,
and minimize the Helmholtz free energy,
\begin{align}
    \mathcal{F} = \langle F\rangle\,,
\end{align}
where the angled brackets represent some kind of integral over our physical domain,
which could be finite or infinite.
This experiment closely approximates the conditions in the core of a neutron star,
which becomes superconducting as the star cools in the presence of a pre-existing magnetic flux.
However, as we will discuss below,
the \gs\ under these conditions can be inhomogeneous,
i.e., macroscopic domains of distinct physical behavior can appear.
For conceptual convenience, we can consider an alternative experiment
in which the system is coupled to a thermodynamic external magnetic field, $\mathbf{H}$,
by minimising the dimensionless Gibbs free energy,
\begin{align}
    \mathcal{G}
        &= \langle F - 2\kappa^2\mathbf{H}\cdot\nablab\times\mathbf{A}\rangle
    \nonumber \\
        &= \mathcal{F} - 2\kappa^2\mathbf{H}\cdot\langle\mathbf{B}\rangle\,.
\end{align}
In an unbounded domain, the \gs\ in this experiment is guaranteed to be homogeneous,
and hence the phase diagram is generally simpler.
For later reference, we present in Fig.~\ref{fig:GL}
the phase diagrams for a single-component \GL\ superconductor,
i.e., we discuss its state as a function of the \GL\ parameter, $\kappa$.
(For more details,
we refer the reader to standard textbooks on superconductivity,
e.g., \citet{Tinkham04}).
For $\kappa < 1/\sqrt{2}$,
we have a type-I superconductor;
when $\mathbf{H}$ is used as the control parameter,
there is a first-order transition between the Meissner state (with $\mathbf{B} = \mathbf{0}$)
and the non-superconducting state (with $\mathbf{B} = \mathbf{H}$)
at the critical value $|\mathbf{H}| = H_\cc = 1/(\sqrt{2} \kappa)$ in our dimensionless units.
When the mean magnetic flux, $\overline{B}$,
is used as the control parameter, this discontinuity resolves into an intermediate phase for $0 < \overline{B} < H_\cc$,
in which Meissner regions alternate with non-superconducting ones.
For $\kappa > 1/\sqrt{2}$, on the other hand,
we have a type-II superconductor;
for $H_{\cc1} < |\mathbf{H}| < H_{\cc2}$ the magnetic flux organizes into a hexagonal lattice of discrete fluxtubes.
The transitions at the lower critical field, $H_{\cc1}$, and the upper critical field, $H_{\cc2}$,
are both second-order,
because fluxtubes appear with infinite separation at $|\mathbf{H}| = H_{\cc1}$,
and the superconductor density becomes vanishingly small at $|\mathbf{H}| = H_{\cc2}$.
In our dimensionless units, $H_{\cc2} = 1$ and
$H_{\cc1} = \mathcal{F}_{\infty}/(4 \pi \kappa^2)$,
where $\mathcal{F}_{\infty}$ is the energy per unit length of a single fluxtube,
which can be determined numerically by solving the single-component \GL\ equations.
When $\overline{B}$ is the control parameter,
there is a similar second-order transition at $\overline{B}=H_{\cc2}$,
and a first-order transition between the intermediate and fluxtube states at $\kappa = 1/\sqrt{2}$.

For our two-component system,
we anticipate that the phase diagram will be more complicated
than that shown in Fig.~\ref{fig:GL}.
In particular, \citet{HaberSchmitt17} have argued that the upper and lower transitions to and from the fluxtube state
can become first-order in some cases,
occurring at $|\mathbf{H}| = H_{\cc1'} < H_{\cc1}$ and $|\mathbf{H}| = H_{\cc2'} > H_{\cc2}$, respectively.
In that case, in the experiment with an imposed mean magnetic flux, $\overline{B}$,
the \gs\ can feature an irregular array of fluxtubes, even in an unbounded domain.
Some aspects of this phase space can be determined analytically,
as we describe in Sec.~\ref{sec:transitions}.
However, in order to produce a complete phase diagram,
it is necessary to solve the \EL\ equations arising from one of the functionals $\mathcal{F}$ or $\mathcal{G}$ numerically,
as we describe in the next section.

%%%%%%%%%%%%%%%%%%%%%%%%%%%%%%%%%%%%%%%%%%%%%%%%%%%%%%%%%%%%%%

\subsection{The numerical model}
\label{sec:numerical}

Whether we choose to work with the Helmholtz free energy, $\mathcal{F}$,
or with the Gibbs free energy, $\mathcal{G}$,
we obtain the same system of \EL\ equations:
\begin{align}
    \kappa^2\nablab\times(\nablab\times\mathbf{A})
        &= \Im\left\{\psip^\star(\nablab-\ii\mathbf{A})\psip
        + \frac{h_1}{\epsilon}\psin\psip^\star(\nablab-\ii\mathbf{A})
        (\psin^\star\psip)\right\}
        \label{eq:ground_A} \, , \\[1.2ex]
    \nabla^2\psin
        &= R^2(|\psin|^2-1)\psin
        + \gr (|\psip|^2 - 1)\psin
        - h_1\psip(\nablab+\ii\mathbf{A})^2(\psip^\star\psin)
        - \psin\nabla^2\left(\frac{h_2-h_1}{2}|\psip|^2
        + \frac{h_3}{2\epsilon}|\psin|^2\right)
        \label{eq:ground_n} \, , \\
    (\nablab-\ii\mathbf{A})^2\psip
        &= (|\psip|^2-1)\psip
        + \frac{\gr}{\epsilon} (|\psin|^2 - 1)\psip
        - \frac{h_1}{\epsilon}\psin(\nablab-\ii\mathbf{A})^2(\psin^\star\psip)
        - \psip\nabla^2\left(\frac{h_2-h_1}{2\epsilon}|\psin|^2 + \frac{h_3}{2}|\psip|^2\right) \, .
        \label{eq:ground_p}
\end{align}
However, the appropriate boundary conditions for these two experiments are different,
and also depend on the particular size and shape chosen for the domain.
Without loss of generality, we will assume from here on
that the magnetic field is oriented in the $z$-direction,
and that all variables are independent of $z$;
so our domain will be some region within the $xy$-plane.
We solve a discretized version of Eqs.~\eqref{eq:ground_A}--\eqref{eq:ground_p},
which are obtained by minimising a discrete approximation to the free energy on a regular grid in $x$ and $y$.
The gauge field is included via a Peierls substitution,
in order to maintain gauge invariance.
The equations are solved using a simple relaxation method,
and the grid resolution is repeatedly refined until a sufficient level of accuracy has been obtained.
Additional details on the numerical algorithm can be found in Appendix~\ref{sec:code}.

\begin{figure}[t]
\centering
    \includegraphics[width=10cm]{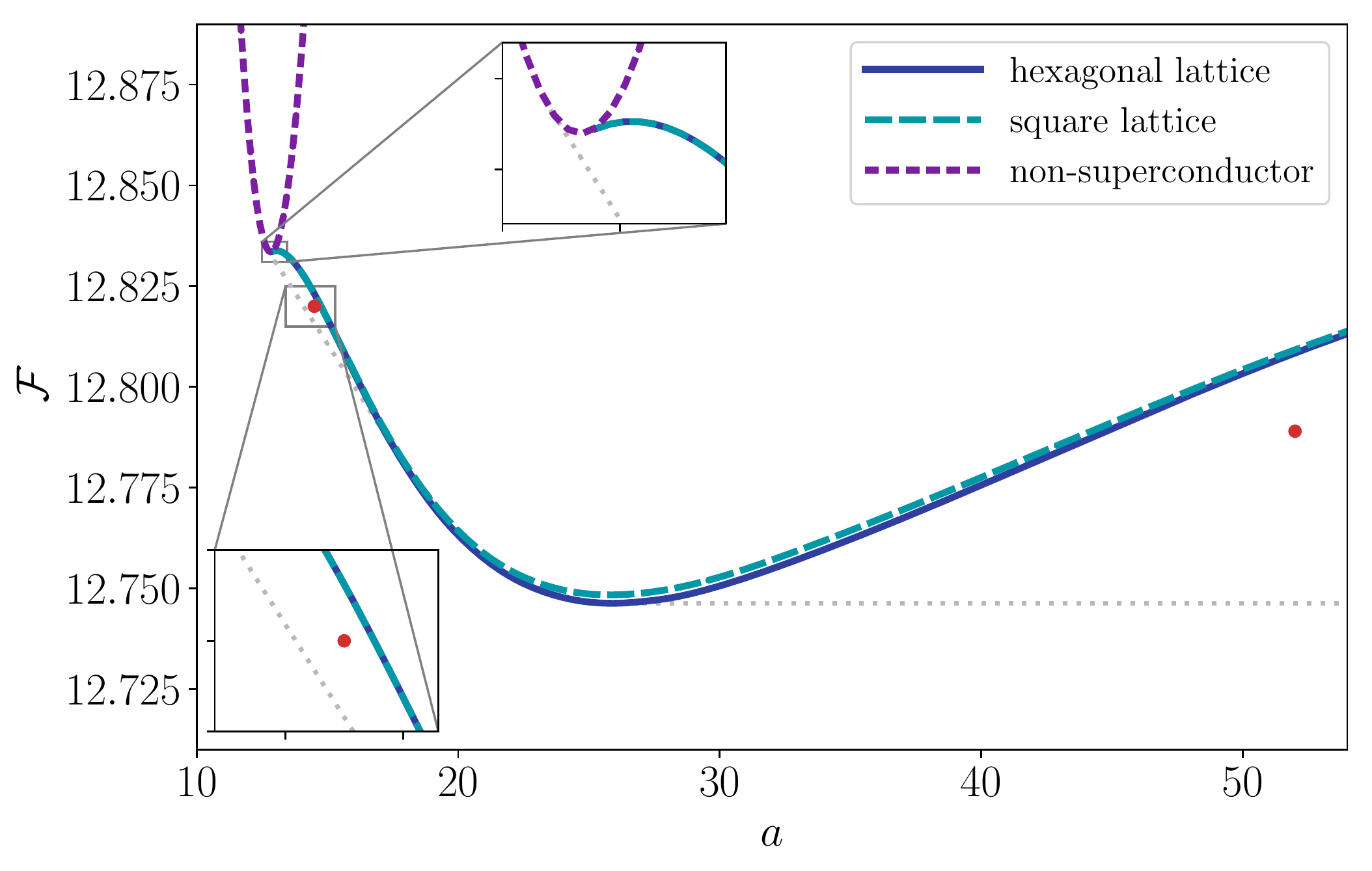}
\caption{The Helmholtz free energy per flux quantum per unit length, $\mathcal{F}$,
as a function of the area per magnetic flux quantum, $a$,
for the NRAPR \EoS\ at a baryon density of $n_\bb = \unitfrac[0.283]{1}{fm^3}$.
The energy in both the square (long-dashed, cyan) and hexagonal (solid, blue) lattice states matches smoothly
onto the energy of the non-superconducting state (short-dashed, purple) at $a \simeq 12.9$
(region enlarged in the upper inset),
and both converge to the same finite value as $a\to\infty$.
The dotted gray lines indicate the lower convex envelope,
which is the true \gs\ in an unbounded domain.
The two red dots indicate the values for the two simulations in Fig.~\ref{fig:mixed}.
We show an enlarged view of the left point in the lower inset.}
\label{fig:NRAPR_0.283}
\end{figure}

In the present study,
we are not interested in the effect of physical boundaries on the phase diagram,
and so we would ideally use an infinite domain,
but for numerical calculations the domain must of course be finite.
Moreover we cannot use periodic boundary conditions,
because in the presence of fluxtubes neither $\mathbf{A}$ nor $\psip$ is spatially periodic.
Instead, we must use quasi-periodic boundary conditions,\citep{Wood-etal19}
which involves specifying not only the size of the domain,
$L_x \times L_y$, say,
but also the number, $N$, of magnetic flux quanta within the domain.
Working in the symmetric gauge,
the quasi-periodic boundary conditions for our dimensionless variables take the form
\begin{align}
    \mathbf{A}(\mathbf{x}+\mathbf{L})
        &= \mathbf{A}(\mathbf{x}) + \frac{N\pi}{L_xL_y} \mathbf{e}_z\times\mathbf{L} \, , \\
    \psip(\mathbf{x}+\mathbf{L})
        &= \psip(\mathbf{x})\exp\left(\ii\frac{N\pi}{L_xL_y}\mathbf{e}_z
        \times\mathbf{L}\cdot\mathbf{x}\right) \, , \\
    \psin(\mathbf{x}+\mathbf{L}) &= \psin(\mathbf{x}) \, ,
\end{align}
where $\mathbf{L}$ represents either of the translational symmetries $(L_x,0)$ or $(0,L_y)$.
These boundary conditions impose a mean magnetic flux through the domain,
$\overline{B} = 2\pi N / (L_xL_y)$
(in our dimensionless units, the quantum of magnetic flux is $2\pi$),
and therefore they are appropriate only for the experiment involving the Helmholtz free energy, $\mathcal{F}$.
Moreover, the choice of domain aspect ratio
affects the configuration of any fluxtube array that forms.
In particular, we can impose either a square or a hexagonal lattice symmetry
by using the following domain shapes:
\begin{itemize}
    \item for a square lattice, we take $N=1$ and $L_x/L_y = 1$;
    \item for a hexagonal lattice, we take $N=2$ and $L_x/L_y = \sqrt{3}$.
\end{itemize}
In order to directly compare these two cases,
we calculate the Helmholtz free energy per magnetic flux quantum per unit length:
\begin{equation}
    \mathcal{F} \equiv \frac{1}{N}\int_{x=0}^{L_x}\int_{y=0}^{L_y} F\,\dd x\,\dd y\,.
\end{equation}
As an example,
in Fig.~\ref{fig:NRAPR_0.283}
we plot $\mathcal{F}$ as a function of the area per magnetic flux quantum,
\begin{equation}
    a \equiv \frac{L_xL_y}{N} = \frac{2\pi}{\overline{B}}\,,
\end{equation}
for the NRAPR \EoS\ at the baryon density $n_\bb = \unitfrac[0.283]{1}{fm^3}$.
Note that in the case of a fluxtube lattice, $a$ corresponds to the area of a single \WS\ cell.
This plot was produced by computing the \gs\ for both square
and hexagonal lattices for domains of various sizes.
We also plot the energy in the non-superconducting state,
which has $\psip=0$ and a uniform magnetic field $\mathbf{B} = (0,0,2\pi/a)$,
and is known analytically (see Sec.~\ref{sec:H_c}).

However, as discussed in the previous section,
in some cases the true \gs\ might be inhomogeneous,
if this allows the average energy to be lower than that of any homogeneous state.
Suppose, for example,
that a fraction, $\phi$, of the total magnetic flux is contained in regions with $a = a_1$ and $\mathcal{F} = \mathcal{F}_1$,
while the rest is in regions with $a = a_2$ and $\mathcal{F} = \mathcal{F}_2$.
In that case, the overall values of $a$ and $\mathcal{F}$ are given by the lever rule,
\begin{align}
    \overline{a} &= \phi a_1 + (1-\phi)a_2 \, , \\
    \overline{\mathcal{F}} &= \phi\mathcal{F}_1 + (1-\phi)\mathcal{F}_2\,.
\end{align}
In this way,
an energy $\overline{\mathcal{F}}$ that is lower than $\mathcal{F}(\overline{a})$ can be achieved in any range of $a$
for which the function $\mathcal{F}(a)$ is not convex.
In fact,
the true \gs\ in an unbounded domain is given by the lower convex envelope of all the homogeneous states,
which is indicated by the dotted gray lines in Fig.~\ref{fig:NRAPR_0.283}.
We will use the notation $\Fg(a)$
to refer to the true \ghs\ energy as a function of the area $a$.
For this particular case,
there are four distinct behaviors seen across the full range of $a$:
\begin{itemize}
    \item for $0 < a \lesssim 12.7$ the \gs\ is non-superconducting;
    \item for $12.7 \lesssim a \lesssim 18.5$ the \gs\ is a mixture of non-superconductor and a hexagonal fluxtube lattice;
    \item for $18.5 \lesssim a \lesssim 26$ the \gs\ is a hexagonal fluxtube lattice;
    \item for $a \gtrsim 26$ the \gs\ is a mixture of a hexagonal fluxtube lattice and the Meissner state.
\end{itemize}

\begin{figure}[t]
\centering
    \includegraphics[width=10cm]{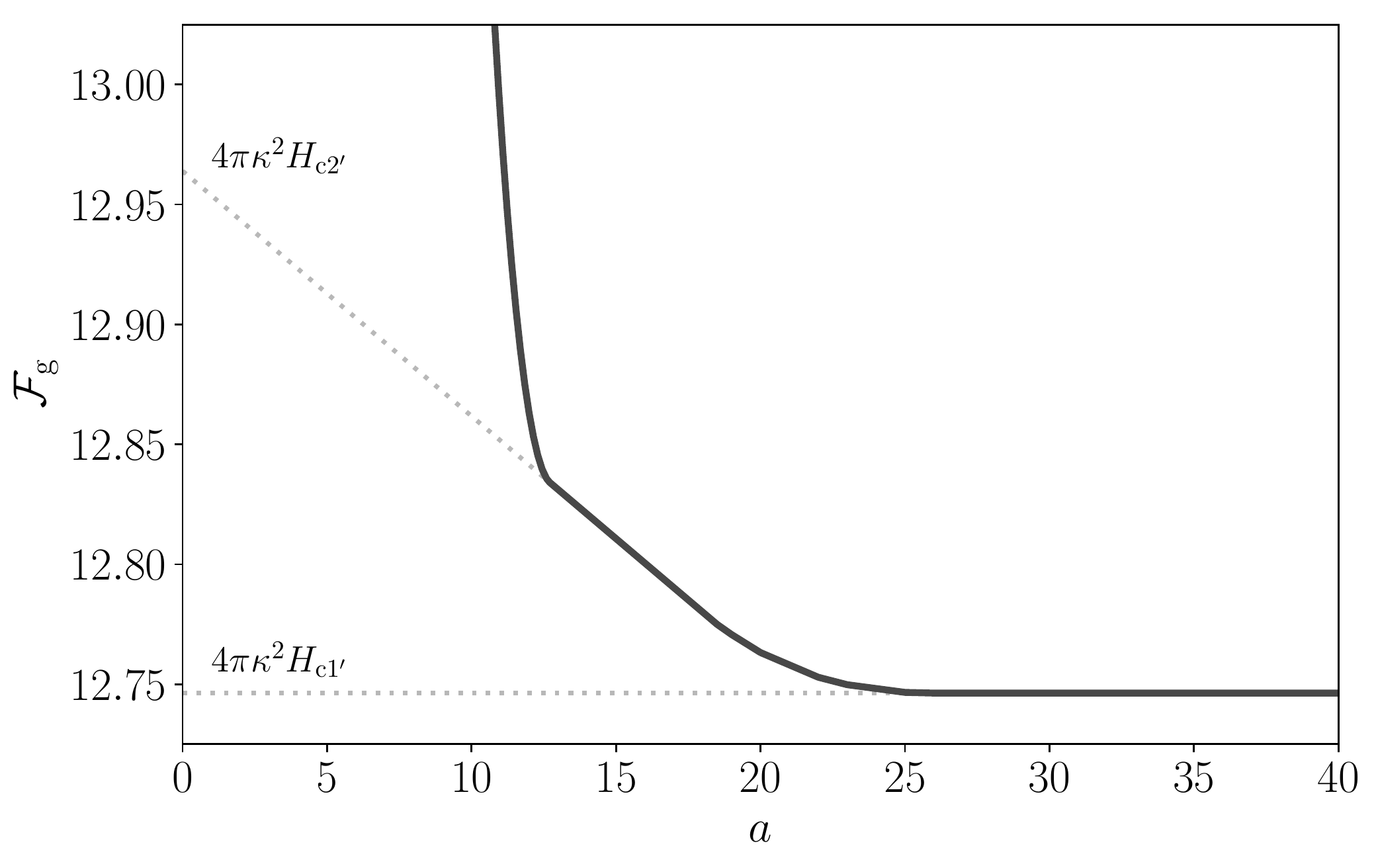}
\caption{The \ghs\ Helmholtz free energy per unit length, $\Fg(a)$,
can be used to infer the minimum Gibbs free-energy density, $\overline{G}$,
as a function of the external field, $\mathbf{H}$.
The tangent to each point on the curve $\Fg(a)$
has slope $\overline{G}$,
and intersects the vertical axis at the point $\mathcal{F} = 4\pi\kappa^2|\mathbf{H}|$.
The dotted lines indicate the transitions
at $H_{\cc1'}$ and $H_{\cc2'}$,
both of which are first-order in this example.}
\label{fig:NRAPR_0.283_2}
\end{figure}

Once the function $\Fg(a)$ is known,
it is straightforward to also determine the minimum Gibbs energy
as a function of $\mathbf{H}$.
In fact, the mean Gibbs energy density is
\begin{equation}
    \overline{G} = \mathcal{G}/a
        = \frac{\Fg(a)}{a} - \frac{4\pi\kappa^2}{a}|\mathbf{H}|\,,
        \label{eq:Gbar}
\end{equation}
and the minimum of $\overline{G}$ over all $a$ can be interpreted graphically from the plots in Fig.~\ref{fig:NRAPR_0.283}.
Since $\Fg(a)$ is a convex and monotonically decreasing function,
each point on the curve $\Fg(a)$
corresponds to a \gs\ with energy
$\overline{G} = \Fg'(a)$,
and the corresponding value of $|\mathbf{H}|$
can be found by extrapolating the tangent line up to the $\mathcal{F}$-axis,
as shown in Fig.~\ref{fig:NRAPR_0.283_2}.
The two ranges of $a$ for which the function $\mathcal{F}_g(a)$ is linear
give rise to two critical values of $|\mathbf{H}|$ (i.e.~$H_{\cc1'}$ and $H_{\cc2'}$)
at which the \gs\ changes discontinuously.
At these values there are first-order transitions
between a hexagonal fluxtube lattice with finite mean field and either the Meissner state (at $H_{\cc1'}$) or the non-superconducting state (at $H_{\cc2'}$).

\begin{figure}[t]
\centering
    \includegraphics[width=0.43\textwidth]{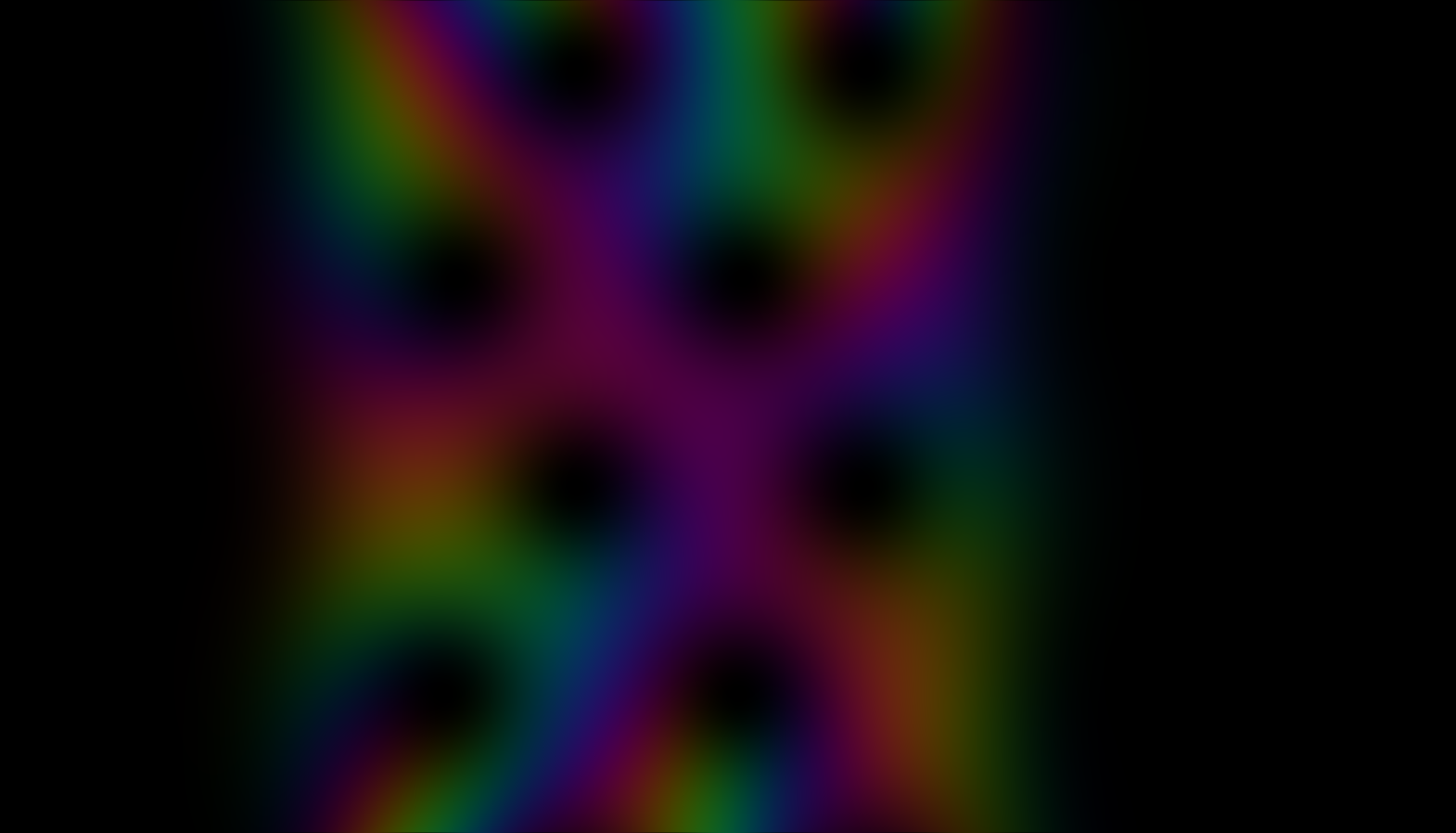}
    \hspace{1cm}
    \includegraphics[width=0.43\textwidth]{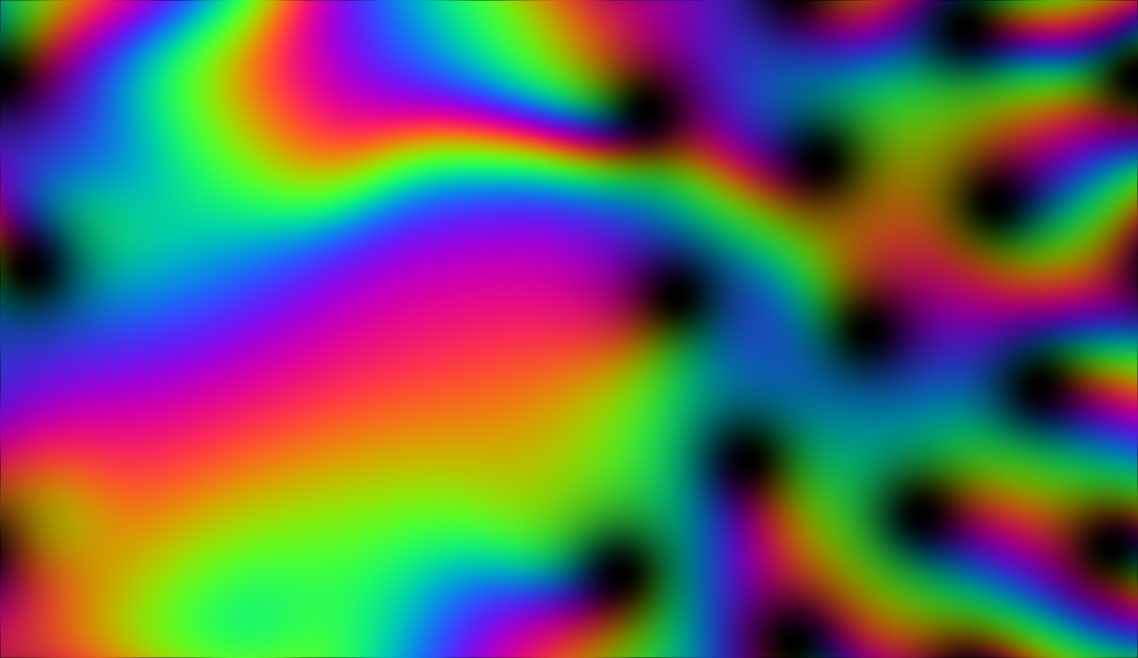}
\caption{Inhomogeneous \gs s for the NRAPR \EoS\
at a baryon density of $n_\bb = \unitfrac[0.283]{1}{fm^3}$.
The brightness and hue indicate the density and phase of the proton order parameter, $\psip$, respectively.
The left panel shows a case with $N=24$ magnetic flux quanta,
and a (dimensionless) area of $aN = 14.5\times24$,
corresponding to a mixture of non-superconducting protons and hexagonal fluxtube lattice.
Approximately 2/3 of the magnetic flux is contained in the non-superconducting domain,
visible as dark bands on both sides of the image,
and hence only 8 fluxtubes are visible.
The low brightness of the lattice domain
indicates the low density of the proton condensate there, i.e., $|\psi_\pp|^2 \ll 1$.
The right panel shows a case with $N=14$ magnetic flux quanta,
and a (dimensionless) area of $aN = 52\times14$;
this is a mixture of Meissner state and hexagonal fluxtube lattice.
In both cases the aspect ratio is $\sqrt{3}$,
which means that a pure hexagonal lattice is a possible state of the system,
but is not the \gs.}
\label{fig:mixed}
\end{figure}

We emphasize that the function $\Fg$ represents the minimum free energy
only for a hypothetical unbounded domain,
free from any geometrical constraints.
In any simulation with a finite domain size, the free energy in the \gs\ will certainly exceed this value.
Nevertheless, by using a large enough computational domain,
and choosing values of $a$ within the appropriate ranges,
we can obtain examples of the inhomogeneous \gs s described above.
Fig.~\ref{fig:mixed} shows two such examples, with $a=14.5$ and $a=52$.
These values of $a$ were chosen so that
in each case approximately half of the domain contains a hexagonal lattice.
As shown in Fig.~\ref{fig:NRAPR_0.283},
in each case the free energy is lower than that of a pure lattice,
but still significantly higher
than for the true \gs\ in an unbounded domain.

%%%%%%%%%%%%%%%%%%%%%%%%%%%%%%%%%%%%%%%%%%%%%%%%%%%%%%%%%%%%%%

\subsection{Phase transitions with $\mathbf{H}$}
\label{sec:transitions}

In this section we will analyse in detail
the phase transitions in the experiment involving the external magnetic field, $\mathbf{H}$.
As is clear from the previous section,
the existence of first-order transitions at the lower and upper critical fields,
$H_{\cc1'}$ and $H_{\cc2'}$,
results from the non-convexity of the free energy $\mathcal{F}(a)$ in the pure lattice state.
To describe these phase transitions in general
requires quite detailed knowledge of the function $\Fg(a)$,
which can only be determined with a 2D numerical model.
However,
some important features of the superconducting phase diagram can be determined either analytically,
or from knowledge of the structure of a single fluxtube.
This allows the phase diagram to be constructed more efficiently,
because it reduces reliance on the 2D code,
and it provides some physical insight into the origins of these first-order phase transitions.
In the following sections we describe those features of the function $\mathcal{F}(a)$ that can
be determined analytically or semi-analytically.
Readers who are only interested in the final phase diagrams
can proceed directly to Sec.~\ref{sec:results}.

%%%%%%%%%%%%%%%%%%%%%%%%%%%%%%%%%%%%%%%%%%%%%%%%%%%%%%%%%%%%%%

\subsubsection{The critical field, $H_\cc$}
\label{sec:H_c}

The two simplest solutions of the \EL\ equations~(\ref{eq:ground_A})--(\ref{eq:ground_p})
are the Meissner (flux-free) state,
which has $|\psip|=|\psin|=1$ and $\mathbf{B}=\mathbf{0}$,
and the non-superconducting state,
which has $|\psip|=0$, $|\psin|= \sqrt{1 + \gr/R^2}$
and a uniform magnetic flux density $\mathbf{B}$ with $|\mathbf{B}|= 2\pi/a$.
As discussed earlier,
these two states can only be realized if the mutual repulsion/attraction $g_\pn$, and thus $\alpha$,
between the two condensates is sufficiently weak.
Expressed in terms of the dimensionless parameter $\alpha$,
this implies the conditions $\alpha^2 < R^2\epsilon$
and $\alpha > -R^2$.
We will assume from here on that both of these are satisfied.

In order to determine the thermodynamical critical field, $H_\cc$,
we equate the energy of the Meissner state with that of the non-superconducting one.
By construction, the free-energy density (\ref{eq:F_dimless}) in the Meissner state is $F=0$,
and, hence, its mean Gibbs energy is $\overline{G}=0$.
For a type-I superconductor,
the transition to the non-superconducting state therefore occurs
when the corresponding energy density $\overline{G}$ becomes negative.

The purple short-dashed curve in Fig.~\ref{fig:NRAPR_0.283} represents the free energy per unit length
per quantum of flux in the non-superconducting state.
From Eq.~\eqref{eq:F_dimless}, we obtain
\begin{align}
    \mathcal{F}
        &= \frac{a}{2}\left(1 - \frac{\gr^2}{\epsilon R^2}\right) + \frac{(2\pi\kappa)^2}{a}\,.
    \label{eq:non-super}
\end{align}
Substituting into Eq.~\eqref{eq:Gbar},
we then find that the minimum mean Gibbs energy density, $\overline{G}$,
is achieved for $a = 2\pi/|\mathbf{H}|$, as expected,
and that this minimum is
\begin{align}
    \overline{G}
        &= \frac{1}{2}\left(1 - \frac{\gr^2}{\epsilon R^2}\right) - \kappa^2|\mathbf{H}|^2\,.
\end{align}
Therefore, assuming that we have a type-I superconductor,
there is a first-order transition between the Meissner and the non-superconducting state at
\begin{equation}
    |\mathbf{H}| = H_\cc \equiv \frac{1}{\sqrt{2}\kappa}
        \sqrt{1 - \frac{\gr^2}{\epsilon R^2}}\,.
    \label{eq:H_c}
\end{equation}
This defines the critical field, $H_\cc$,
\rewrite{in agreement with previous studies. \citep{HaberSchmitt17,Kobyakov20}}

%%%%%%%%%%%%%%%%%%%%%%%%%%%%%%%%%%%%%%%%%%%%%%%%%%%%%%%%%%%%%%

\subsubsection{The lower critical field, $H_{\cc1}$ vs.~$H_{\cc1'}$}
\label{sec:lower}

As shown in Sec.~\ref{sec:numerical},
the lower critical field is a first-order phase transition
if the Helmholtz free energy per unit length per flux quantum $\mathcal{F}(a)$ in the pure lattice state
has a minimum at some finite value of $a$,
since then a fluxtube lattice forms with finite separation.
If this minimum is $\mathcal{F}_\text{min}$, say,
then we have $H_{\cc1'} = \mathcal{F}_\text{min}/(4\pi\kappa^2)$.
If, on the other hand,
$\mathcal{F}$ is a monotonically decreasing function of $a$ in the lattice state,
then there is a second-order transition at $H_{\cc1} = \mathcal{F}_\infty/(4\pi\kappa^2)$,
as in the case of a single-component superconductor,
where
\begin{equation}
    \mathcal{F}_\infty = \lim_{a\to\infty}\mathcal{F}(a)\,.
\end{equation}
In this limit, interactions between the fluxtubes vanish,
and $\mathcal{F}_\infty$ is equivalent to the energy per unit length of a single fluxtube in an infinite domain.
This energy can be computed efficiently by using polar coordinates $r,\theta$,
centered on the fluxtube core,
and seeking solutions of Eqs.~\eqref{eq:ground_A}--\eqref{eq:ground_p} in the form
\citep{AlfordGood08}
\begin{align}
    \psip &= f(r) \, \ee^{\ii\theta}\,, \\
    \psin &= g(r)\,, \\
    \mathbf{A} &= A_\theta(r)\,\mathbf{e}_\theta\,.
\end{align}
This ansatz assumes that the fluxtube carries a single quantum of magnetic flux,
and that there is no corresponding phase defect in the neutron condensate.
It further results in a system of ordinary differential equations that can be solved numerically,
yielding the value of $\mathcal{F}_\infty$ to high accuracy.
However,
to determine whether the lower transition is second-order or first-order,
we need to know whether the function $\mathcal{F}(a)$ tends to $\mathcal{F}_\infty$
from above or from below,
which is equivalent to asking whether the long-range interaction between fluxtubes is repulsive or attractive.
This can be derived rigorously using a method introduced by \citet{Kramer71},
which we describe in detail in Appendix~\ref{sec:Kramer}.
However, the result can be anticipated heuristically
by considering the perturbations produced by a fluxtube in the far-field,
i.e., at a large distance from its core.
It is convenient to work with the real variables
\begin{align}
    f &\equiv |\psip|\,, \\
    g &\equiv |\psin|\,, \\
    \chi &\equiv \arg\psin\,, \\
    \mathbf{V} &\equiv \nablab(\arg\psip) - \mathbf{A}\,.
\end{align}
The gauge invariance of the free energy density \eqref{eq:F_dimless} guarantees that
it can be rewritten in terms of these variables without loss of generality;
for the full expression see Eq.~\eqref{eq:F2}.
The corresponding \EL\ equations are then
\begin{align}
    0 &= f(f^2-1) + \frac{\gr}{\epsilon}f(g^2-1) - \nabla^2f + f|\mathbf{V}|^2 \nonumber \\
        &+ \frac{h_1}{\epsilon}[f|\nablab g|^2 + fg^2|\mathbf{V}-\nablab\chi|^2 - \nablab\cdot(g^2\nablab f)]
        - \frac{h_2}{2\epsilon} f\nabla^2(g^2) - \frac{h_3}{2}f\nabla^2(f^2)\,,
            \label{eq:EL1} \\
    0 &= \frac{R^2}{\epsilon}g(g^2-1) + \frac{\gr}{\epsilon}g(f^2-1) - \frac{1}{\epsilon}\nabla^2g
        + \frac{1}{\epsilon}g|\nablab\chi|^2 \nonumber \\
        &+ \frac{h_1}{\epsilon}[g|\nablab f|^2 + gf^2|\mathbf{V}-\nablab\chi|^2 - \nablab\cdot(f^2\nablab g)]
        - \frac{h_2}{2\epsilon}g\nabla^2(f^2) - \frac{h_3}{2\epsilon^2}g\nabla^2(g^2) \,, \\
    0 &= \nablab\cdot[g^2\nablab\chi + h_1f^2g^2(\nablab\chi-\mathbf{V})] \,, \\
    0 &= f^2\mathbf{V} + \frac{h_1}{\epsilon}f^2g^2(\mathbf{V}-\nablab\chi)
        + \kappa^2\nablab\times(\nablab\times\mathbf{V}) \,.
            \label{eq:EL4}
\end{align}
The far-field structure of the fluxtube can be determined by
linearizing these about the uniform state with $f=g=1$, and $\mathbf{V} = \nablab \chi = \mathbf{0}$.
This leads to the following system:
\begin{align}
    \left(1+\frac{h_1}{\epsilon}+h_3\right)\nabla^2\delta f + \frac{h_2}{\epsilon}\nabla^2\delta g
        &= 2\delta f + \frac{2\gr}{\epsilon}\delta g \,,
            \label{eq:lin_f} \\
    \left(1+h_1+\frac{h_3}{\epsilon}\right)\nabla^2\delta g + h_2\nabla^2\delta f
        &= 2R^2\delta g + 2\gr\delta f \,,
            \label{eq:lin_g} \\
    \kappa^2\nablab\times(\nablab\times\delta\mathbf{V})
        &= \frac{h_1}{\epsilon}\nablab\delta\chi - \left(1 + \frac{h_1}{\epsilon}\right)\delta\mathbf{V} \,,
            \label{eq:lin_V} \\
    \left(1 + h_1\right)\nabla^2\delta\chi
        &= h_1\nablab\cdot\delta\mathbf{V}\,,
            \label{eq:lin_chi}
\end{align}
where $\delta f$, $\delta g$, $\delta\mathbf{V}$, and $\delta\chi$ denote the linear perturbations.
In the case of a single fluxtube,
we are interested in the solution that is axisymmetric and decays at large distance from the origin.
We deduce from Eqs.~\eqref{eq:lin_V} and \eqref{eq:lin_chi}
that this solution has $\delta\chi = 0$ and
\begin{equation}
    \delta\mathbf{V} = V_0 \, K _1 \left( \frac{r}{\lambda_\star} \right) \, \mathbf{e}_\theta \, ,
        \label{eq:V0}
\end{equation}
for some constant coefficient $V_0$,
where $K_1$ is a modified Bessel function and $\lambda_\star$
the effective London length,
\begin{equation}
  \lambda_\star = \kappa\left(1 + \frac{h_1}{\epsilon}\right)^{-1/2}\,.
  \label{eq:lambda_star}
\end{equation}
Note that, compared to the London length $\lambda$ in absence of coupling,
$\lambda_\star$ is made smaller by the parameter $h_1$,
essentially because the effective mass of the protons is made smaller
by the entrainment of neutrons.\citep{Alpar-etal84}
From Eqs.~\eqref{eq:lin_f} and \eqref{eq:lin_g} we find that,
owing to the coupling between the condensates,
the fluxtube in the far-field has a double-coherence-length structure, i.e.,
\begin{align}
    \delta f = f_1 \, K_0 \left( \frac{\sqrt{2}r}{\xi_1} \right) + f_2 \, K_0 \left(\frac{\sqrt{2}r}{\xi_2} \right)
        \qquad\mbox{and}\qquad
    \delta g = g_1 \, K_0 \left(\frac{\sqrt{2}r}{\xi_1} \right) + g_2 \, K_0 \left(\frac{\sqrt{2}r}{\xi_2} \right)\,,
        \label{eq:fg0}
\end{align}
where $K_0$ is a zeroth-order modified Bessel function, and the coefficients $f_i,g_i$ and the effective coherence lengths $\xi_i$ satisfy the equations
\begin{align}
    \left(1+\frac{h_1}{\epsilon}+h_3\right) \frac{f_i}{\xi_i^2} + \frac{h_2}{\epsilon} \frac{g_i}{\xi_i^2}
        &= f_i + \frac{\gr}{\epsilon}g_i\,,
        \label{eq:fi} \\
    \left(1+h_1+\frac{h_3}{\epsilon}\right) \frac{g_i}{\xi_i^2} + h_2 \frac{f_i}{\xi_i^2}
        &= R^2g_i + \gr f_i\,.
        \label{eq:gi}
\end{align}
This is reminiscent of the fluxtube structure found
in a two-component superconductor,
although in that case the two coherence lengths arise because
a fluxtube is a phase defect in \emph{both} condensates.\citep{Svistunov-etal15}
In our model the fluxtubes are phase defects only in the proton condensate,
but the coupling coefficients $\gr$ and $h_2$ ultimately achieve a similar effect.
We might therefore expect our model to exhibit type-1.5 superconductivity in some parameter regimes,
which is common in two-component superconductors,
and generally occurs when the London length lies between the two coherence lengths.\citep{Babaev-etal17}
The physical reason is that the electromagnetic interaction between fluxtubes,
which decays on the London length, is generally repulsive,
whereas the density interactions are generally attractive.
Thus type-1.5 superconductivity arises because fluxtubes are mutually attractive at large separation distances,
but become mutually repulsive at shorter separations, leading to a state of fluxtube bunches
at a preferred mean separation, as indicated by the minimum in the $\mathcal{F}(a)$ curve.

As mentioned earlier, this heuristic argument can be put on a more rigorous basis by calculating the
long-range interaction energy between fluxtubes in a lattice configuration,
using the method first described by \citet{Kramer71}.
A similar method was used by \citet{HaberSchmitt17} to demonstrate the existence of type-1.5 superconductivity
resulting from \rewrite{density and derivative couplings}.
\rewrite{As we evaluate the interaction energy for a more general case than \citet{HaberSchmitt17}
and thus obtain a different result,
we present our analysis in full in Appendix~\ref{sec:Kramer},}
although the steps closely follow those of \citet{Kramer71}
\rewrite{for a single-component superconductor}.
Our final result for the interaction energy is
\begin{align}
    \mathcal{F} - \mathcal{F}_\infty
        &\simeq 2\pi\kappa^2 V_0^2\sum_{i\neq0}K_0\left( \frac{|\mathbf{x}_i|}{\lambda_\star}\right)
            - 2\pi \sum_{j=1,2} \left[f_j^2 + 2\dfrac{\gr}{\epsilon} f_jg_j
            + \dfrac{R^2}{\epsilon}g_j^2\right]
        \xi_j^2\sum_{i\neq0}K_0\left( \frac{\sqrt{2}|\mathbf{x}_i|}{\xi_j}\right)\,,
    \label{eq:result}
\end{align}
where $\mathbf{x}_i$ is the location of the $i$-th lattice point,
assuming that $\mathbf{x}_0 = \mathbf{0}$.
In the absence of any coupling between the two fluids ($\gr=h_1=h_2=h_3=0$)
this reduces exactly to the result of \citet{Kramer71}.

In principle, we can use this result to estimate the free energy of a particular lattice state
using just the values of the coefficients $f_i,g_i,V_0$,
which can themselves be inferred from the nonlinear solution for a single fluxtube.
However, this result is only strictly valid if the fluxtubes are very widely separated,
and it becomes inaccurate once the fluxtubes are close enough to interact nonlinearly.
Nevertheless, we can deduce that, in the asymptotic limit $a\to\infty$,
\begin{equation}
    \mathcal{F} - \mathcal{F}_\infty \propto
        \kappa^2 V_0^2 \sqrt{\frac{\lambda_\star}{d}} \, \ee^{-d/\lambda_\star}
        - \left[f_+^2 + 2\dfrac{\gr}{\epsilon}f_+g_+ + \dfrac{R^2}{\epsilon}g_+^2\right]
            \xi_+^2\sqrt{\frac{\xi_+}{\sqrt{2}d}} \, \ee^{-\sqrt{2}d/\xi_+}
  \label{eq:long-range}
\end{equation}
where $d \propto \sqrt{a}$ is the lattice constant,
and $\xi_+$ represents the larger of the two \rewrite{effective} coherence lengths,
\rewrite{which in practice is always larger than both of the bare coherence lengths.
If $\lambda_\star > \xi_+ / \sqrt{2}$ then,
according to Eq.~\eqref{eq:long-range},}
the long-range interaction energy is positive,
implying that there is a second-order transition at the lower critical field $H_{\cc1}$,
as for a type-II superconductor.
Conversely, if $\lambda_\star < \xi_+ / \sqrt{2}$ then
the interaction energy is negative,
implying that the $\mathcal{F}(a)$ curve has a minimum at a finite value of $a$.
In that case, there is a first-order transition
at the lower critical field $H_{\cc1'}$,
and we have a type-1.5 superconductor,
which confirms the heuristic argument given above.

%%%%%%%%%%%%%%%%%%%%%%%%%%%%%%%%%%%%%%%%%%%%%%%%%%%%%%%%%%%%%%

\begin{figure}[t]
\centering
  \includegraphics[width=0.48\textwidth]{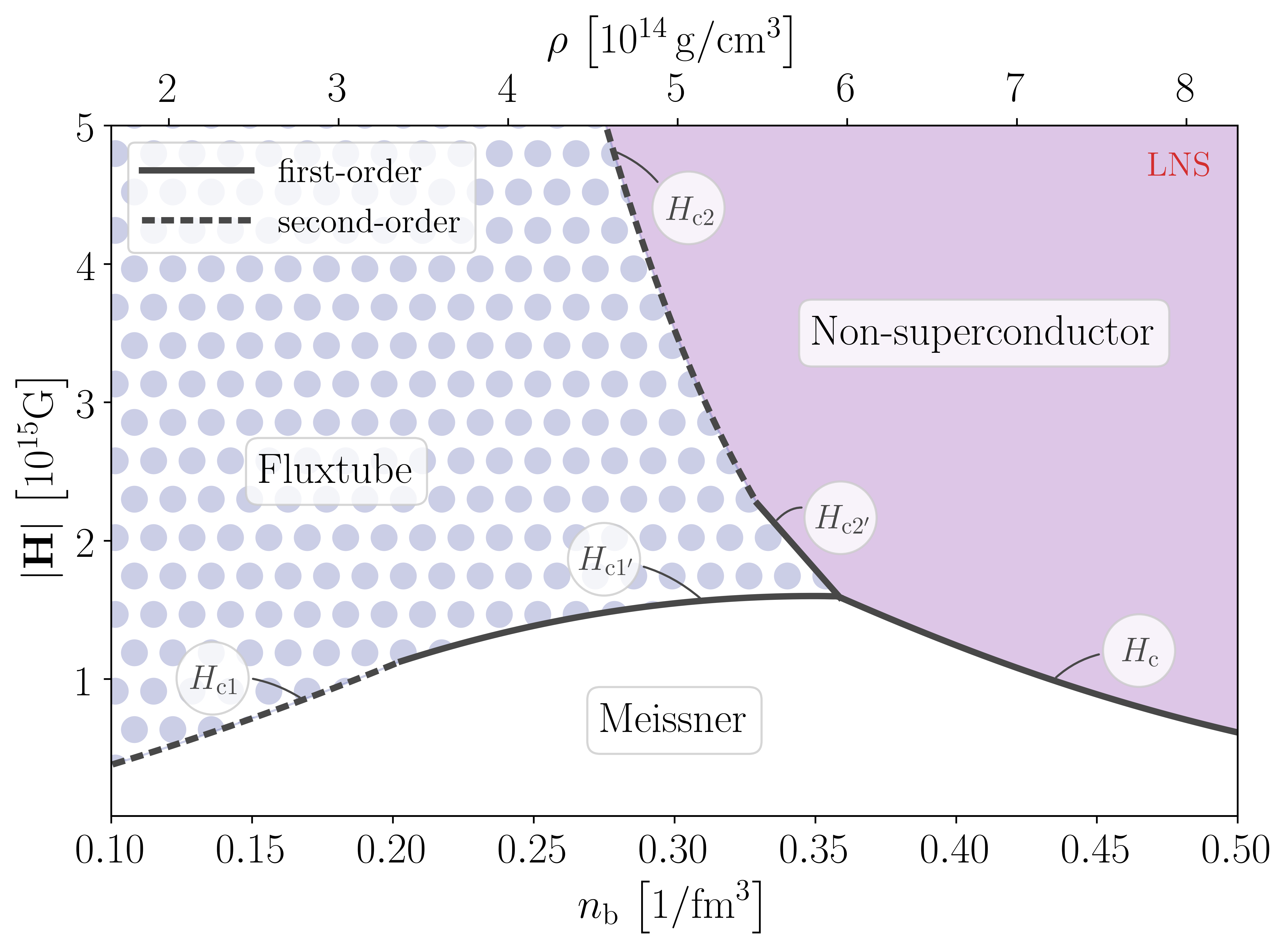}
  \hspace{0.1cm}
  \includegraphics[width=0.48\textwidth]{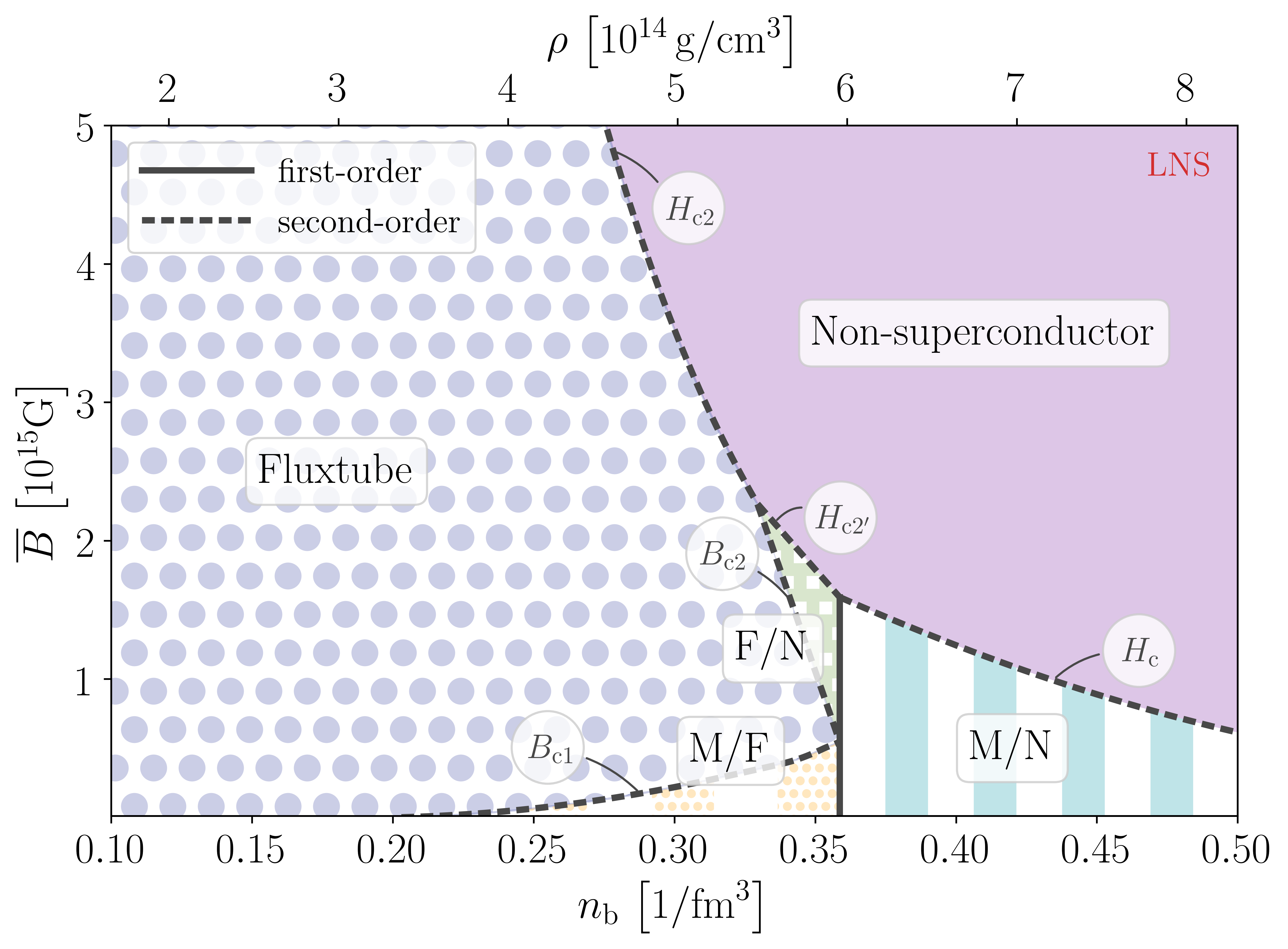}
\caption{Phase diagrams for the LNS \EoS\
with imposed external field $|\mathbf{H}|$ (left panel)
and imposed mean flux $\overline{B}$ (right panel).
Both figures are plotted in the same style as Fig.~\ref{fig:GL},
except that here the entire core cross-section of the neutron star is shown,
with the lower abscissa representing the baryon number density $n_\bb$,
and the upper abscissa the mass density $\rho$.
The first-order and second-order transitions at the different critical fields
are indicated by solid and dashed black lines, respectively,
and the resulting phases labeled accordingly.
Shading of the respective regions is indicative of the magnetic flux distribution.
In addition to the phases observed for the single-component case
(we label the intermediate type-I phase as M/N),
we also obtain inhomogeneous regimes
where Meissner and fluxtube regions (M/F)
as well as fluxtube and non-superconducting regions (F/N) alternate.
These are associated with the appearance of the critical fields
$H_{\cc1'}$ and $H_{\cc2'}$, respectively.}
\label{fig:LNS}
\end{figure}

\subsubsection{The upper critical field, $H_{\cc2}$ vs.~$H_{\cc2'}$}
\label{sec:upper}

As shown in Sec.~\ref{sec:numerical},
if the free energy $\mathcal{F}(a)$ is convex
(and monotonically decreasing),
then the transition between the fluxtube lattice state and the non-superconducting state is second-order.
This means that the order parameter $\psip$
vanishes smoothly at the transition point,
$H_{\cc2}$,
which can therefore be determined analytically
by considering linear perturbations to the non-superconducting state.
Working in the symmetric gauge,
the non-superconducting state is given by
$|\psip| = 0$, $|\psin| = \sqrt{1 + \gr/R^2}$,
and $\mathbf{A} = \tfrac{1}{2}\overline{B}\mathbf{e}_z\times\mathbf{x}$,
where $\overline{B} = 2\pi/a$ is the (uniform) mean magnetic flux.
From the linearized version of Eq.~\eqref{eq:ground_p},
we find that the perturbation $\delta\psip$ must satisfy the equation
\begin{equation}
    \left[1 + \frac{h_1}{\epsilon} \left(1 + \frac{\gr}{R^2}\right)\right]
        \left(\nablab - \frac{\ii}{2} \overline{B} \mathbf{e}_z \times\mathbf{x} \right)^2 \delta\psip
        = - \left(1 - \frac{\gr^2}{\epsilon R^2}\right)\delta\psip\,.
\end{equation}
As known from single-component systems,\cite{Tinkham04}
bounded solutions of this equation first appear when
\begin{equation}
    \overline{B} = H_{\cc2} \equiv \frac{1 - \frac{\gr^2}{\epsilon R^2}}
        {1 + \frac{h_1}{\epsilon}\left(1 + \frac{\gr}{R^2}\right)}\,.
        \label{eq:H_c2}
\end{equation}
Therefore the solid lines in Fig.~\ref{fig:NRAPR_0.283},
as highlighted in the upper inset,
meet the dashed line at the point where
$a = 2\pi/H_{\cc2}$.
However, if the function $\mathcal{F}(a)$ is not convex at this point
then the upper transition will occur not at $|\mathbf{H}| = H_{\cc2}$,
but at a higher value $|\mathbf{H}| = H_{\cc2'}$, and will be first-order.
To determine whether the function $\mathcal{F}(a)$ is convex at this point,
we can seek weakly nonlinear solutions in the vicinity of $a = 2\pi/H_{\cc2}$,
following the method of \citet{Abrikosov57}.
\rewrite{We present the details of this calculation in Appendix~\ref{sec:Abrikosov}.}
The main result is that
the function $\mathcal{F}(a)$ becomes non-convex when
\begin{align}
    \left[\frac{2}{H_{\cc2}} + h_3 - \left(1 - \frac{\gr^2}{\epsilon R^2}\right)^2
        \frac{1}{\kappa^2H_{\cc2}^3}\right]\pi\beta\epsilon R^2
    = \sum_{i}\exp(-\tfrac{1}{2}H_{\cc2} |\mathbf{x}_i|^2) \,
        \frac{\left[(h_2-h_1)\tfrac{1}{2}H_{\cc2} |\mathbf{x}_i|^2 + h_1
        + \gr\frac{1}{H_{\cc2}}\right]^2}
        {\left(\frac{1}{R^2 + \gr} + \frac{h_3}{\epsilon R^2}\right)
        \tfrac{1}{2}H_{\cc2}|\mathbf{x}_i|^2 + \frac{1}{H_{\cc2}}}\,,
            \label{eq:result2}
\end{align}
where $\mathbf{x}_i$ is the location of the $i$-th lattice point, as in Sec.~\ref{sec:lower},
and $\beta$ is the kurtosis of the proton order parameter,
\begin{equation}
    \beta \equiv \frac{\overline{|\psip|^4}}{(\overline{|\psip|^2})^2}
        = \sum_{i}\exp(-\tfrac{1}{2}H_{\cc2} |\mathbf{x}_i|^2)\,.
            \label{eq:beta}
\end{equation}
Equation~\eqref{eq:result2} is valid for both hexagonal and square lattices of singly-charged fluxtubes.
(Singly-charged here means that each fluxtube carries a single quantum of magnetic flux.)
Although there are cases in which other fluxtube configurations have a lower free energy
(see Appendix~\ref{sec:Abrikosov}),
the point of transition between $H_{\cc2}$ and $H_{\cc2'}$
is always determined by the singly-charged hexagonal lattice
(for which $\beta \simeq 1.1596$ \citep{Kleiner-etal64}).

%%%%%%%%%%%%%%%%%%%%%%%%%%%%%%%%%%%%%%%%%%%%%%%%%%%%%%%%%%%%%%
%%%%%%%%%%%%%%%%%%%%%%%%%%%%%%%%%%%%%%%%%%%%%%%%%%%%%%%%%%%%%%

\section{Results}
\label{sec:results}

Having derived some features of the superconducting phase transitions analytically
in the previous section,
we now present the full phase diagrams for each of the six equations of state listed in Table~\ref{tbl:Skyrme},
indicating the \gs\ as a function of the nuclear density and the magnetic field strength
measured either in terms of the external field $|\mathbf{H}|$
or the mean flux $\overline{B}$.

\begin{figure}[t]
\begin{minipage}[b]{0.98\linewidth}
  \centering
    % \includegraphics[width=0.48\textwidth, height=0.35\textwidth]{example-image-a}
    % \hspace{0.1cm}
    % \includegraphics[width=0.48\textwidth, height=0.35\textwidth]{example-image-a}
  \includegraphics[width=0.48\textwidth]{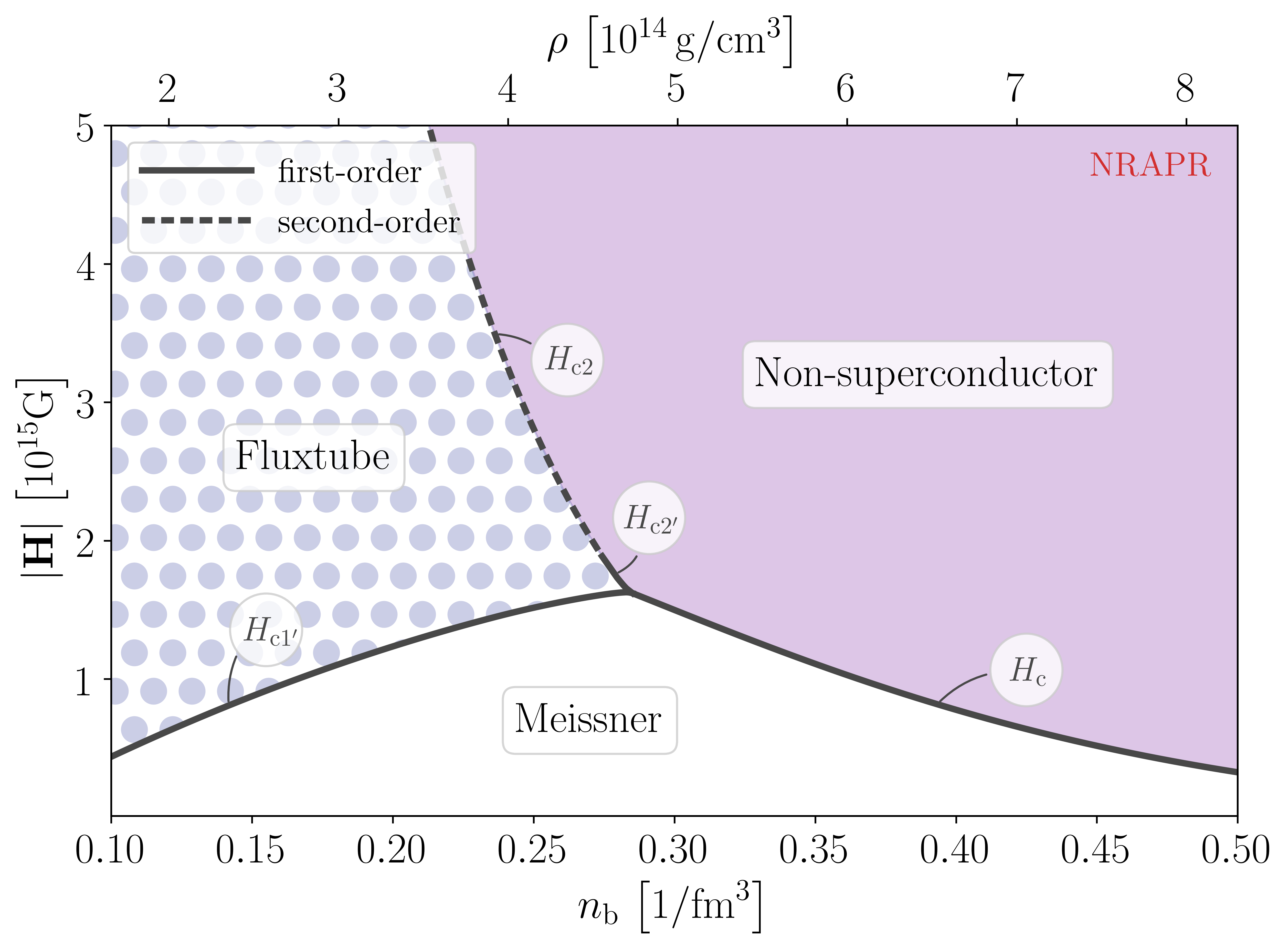}
  \hspace{0.1cm}
  \includegraphics[width=0.48\textwidth]{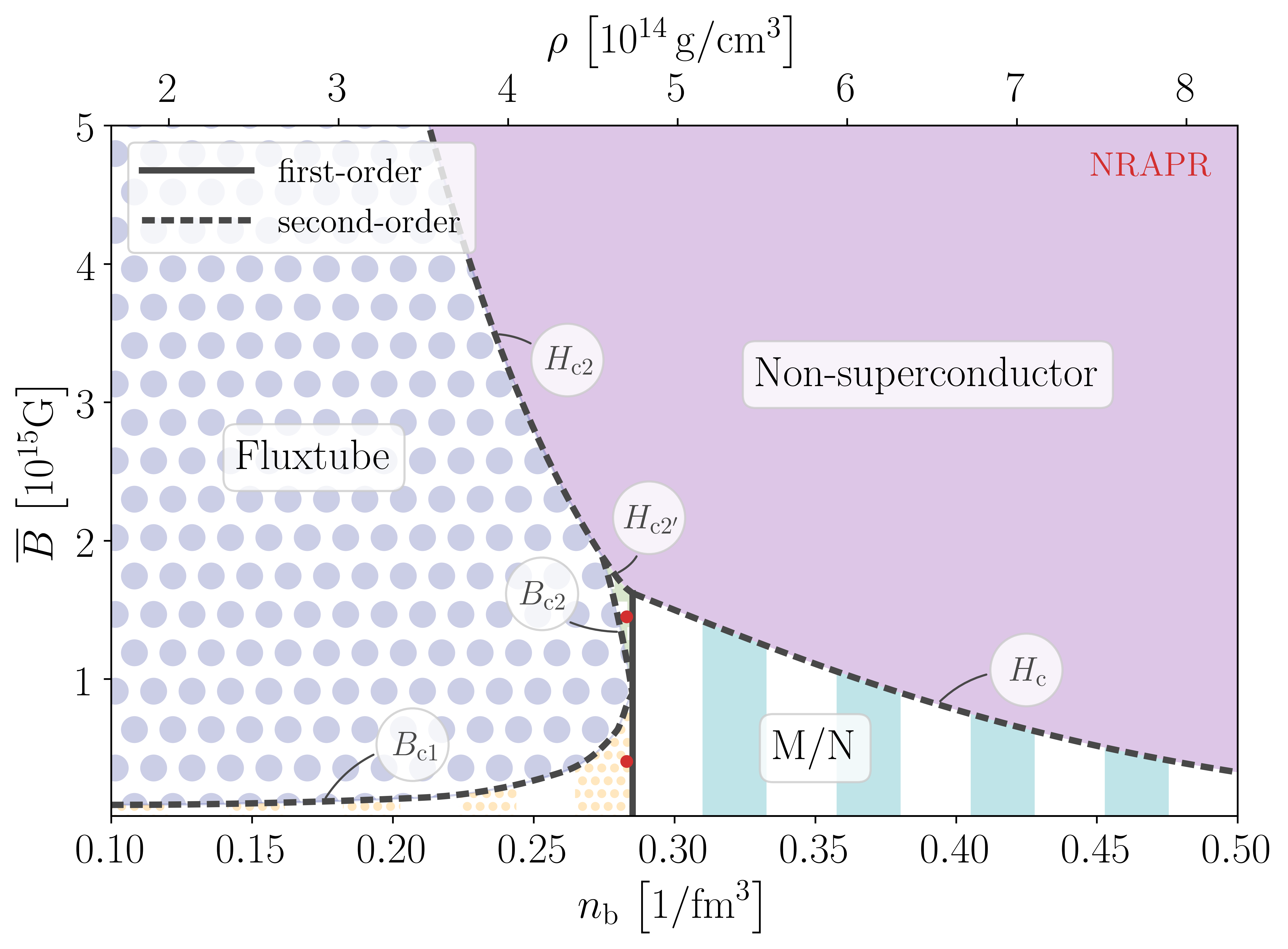}
\end{minipage}
\hspace{0.5cm}
\begin{minipage}[b]{0.98\linewidth}
  \centering
    % \includegraphics[width=0.48\textwidth, height=0.35\textwidth]{example-image-a}
    % \hspace{0.1cm}
    % \includegraphics[width=0.48\textwidth, height=0.35\textwidth]{example-image-a}
  \includegraphics[width=0.48\textwidth]{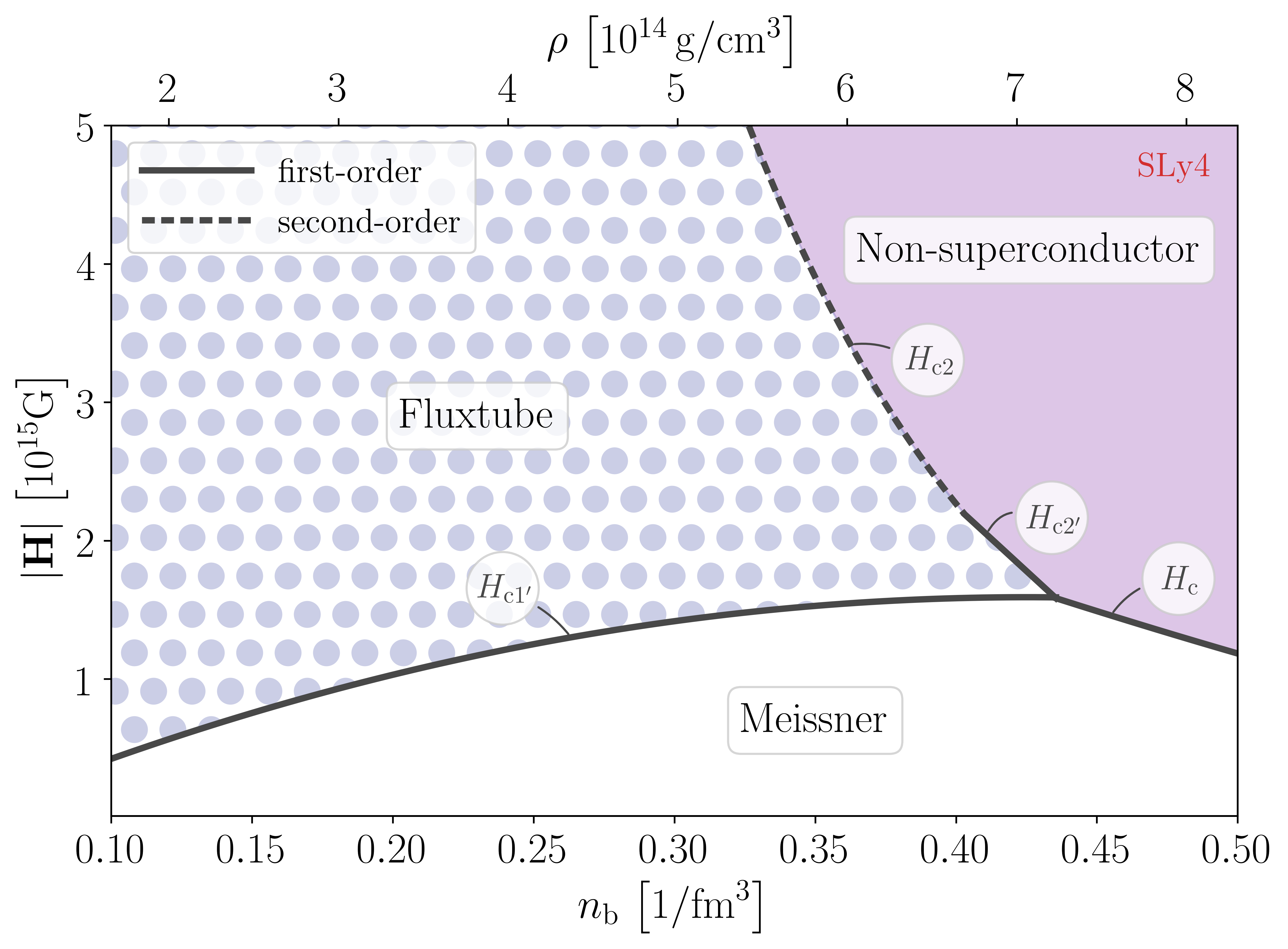}
  \hspace{0.1cm}
  \includegraphics[width=0.48\textwidth]{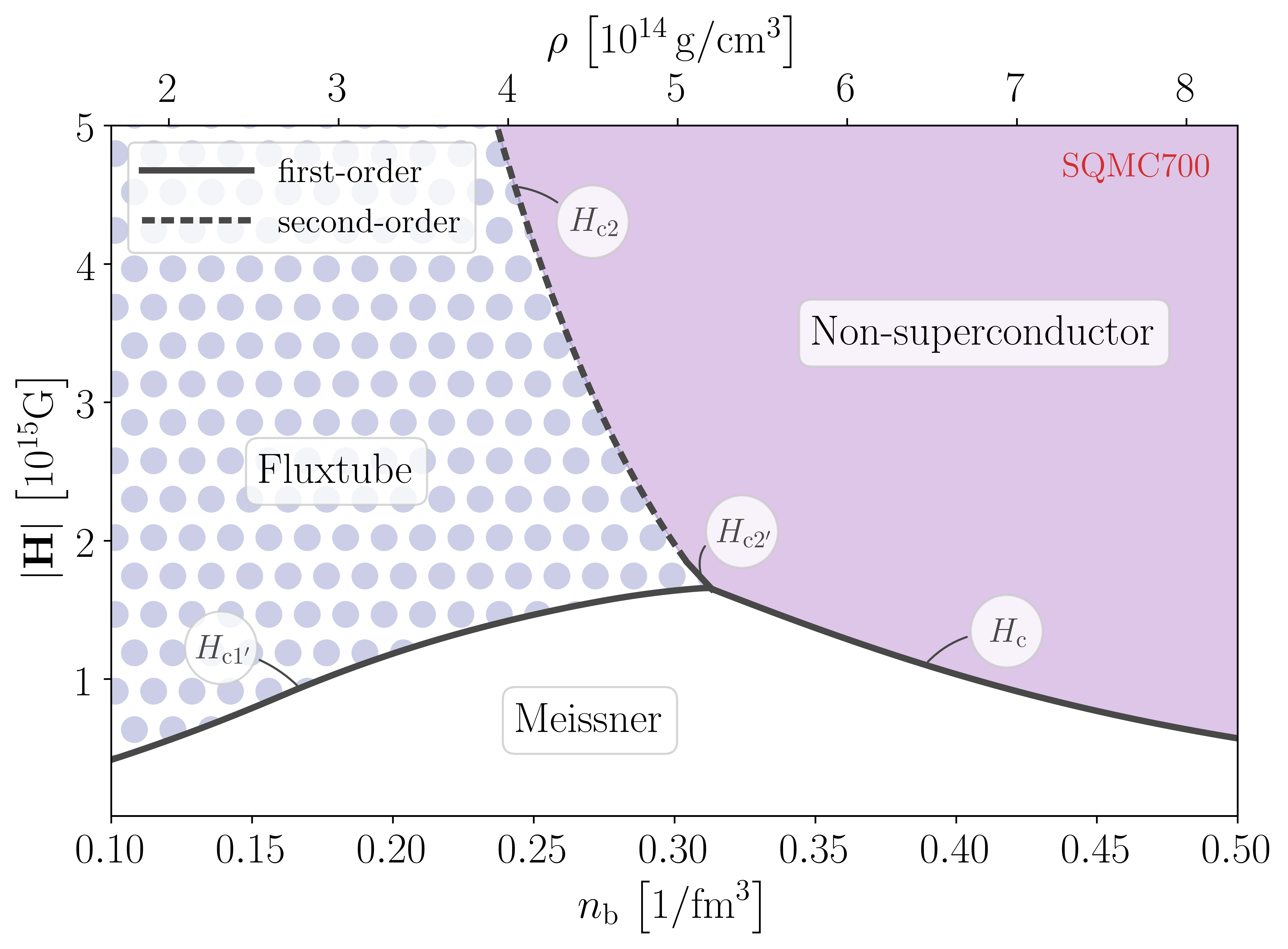}
\end{minipage}
\caption{Phase diagrams for the NRAPR, SLy4, and SQMC700 \EoSs\
(labelled in red in the top right corner of each figure)
with imposed external field $|\mathbf{H}|$ and imposed mean flux $\overline{B}$.
The figures are plotted in the same style as Fig.~\ref{fig:LNS}.
In the $\overline{B}$ experiment for NRAPR (top row, right panel),
two red dots mark the parameter values
corresponding to the simulation of the inhomogeneous \gs s shown in Fig.~\ref{fig:mixed}.
Note that we have omitted the $\overline{B}$ plots for SLy4 and SQMC700,
because they closely resemble those of LNS and NRAPR,
albeit with critical points located at different densities within the star.}
\label{fig:EoS_combined}
\end{figure}

We begin with the LNS \EoS,
in Fig.~\ref{fig:LNS},
which illustrates all of the phase transitions
that we described in Sec.~\ref{sec:superconductivity}.
Note that to clarify the nature of the different regions in the phase diagrams,
we choose the shading to be indicative of the respective magnetic flux distribution.
At lower densities, $n_\bb \lesssim \unitfrac[0.20]{1}{fm^3}$,
the proton superconductor behaves like a classic single-component, type-II superconductor,
with second-order phase transitions from the Meissner to the fluxtube state
at the lower critical field $H_{\cc1}$,
and from the fluxtube to the non-superconducting state
at the upper critical field $H_{\cc2}$, respectively.
Throughout the density range $0.20 \lesssim n_\bb \lesssim \unitfrac[0.36]{1}{fm^3}$,
when fluxtubes are present,
they arrange themselves in a hexagonal lattice with a preferred (finite) separation,
characteristic of a type-1.5 superconductor.
In the experiment, where we control $|\mathbf{H}|$,
this implies the existence of a first-order phase transition
from the Meissner state at the critical field $|\mathbf{H}| = H_{\cc1'} (< H_{\cc1})$,
i.e., fluxtubes appear with finite separation as $|\mathbf{H}|$ is increased above this value.
In the $\overline{B}$ picture,
this corresponds to a critical value $\overline{B} = B_{\cc1}$,
marking a second-order phase transition,
below which the \gs\ is a mixture of the Meissner state and a hexagonal lattice of fluxtubes.
Over a smaller density range, $0.33 \lesssim n_\bb \lesssim \unitfrac[0.36]{1}{fm^3}$,
the upper transition to the non-superconducting state is also first order,
provided that we impose the external field $|\mathbf{H}|$;
i.e., when $|\mathbf{H}|$ exceeds the critical value $H_{\cc2'} (> H_{\cc2})$
the superconductor abruptly breaks down,
causing an abrupt increase in the mean magnetic flux.
In the experiment where we impose the mean magnetic flux $\overline{B}$,
there is instead a range, $B_{\cc2} < \overline{B} < H_{\cc2'}$,
for which the \gs\ is a mixture of a non-superconductor and a hexagonal lattice,
with second-order transitions at either end of this range.
For high densities, $n_\bb \gtrsim \unitfrac[0.36]{1}{fm^3}$,
we recover the behavior of a single-component type-I superconductor,
with a first-order transition at the critical field $|\mathbf{H}| = H_{\cc}$,
and an intermediate state of Meissner and non-superconducting regions
for $\overline{B} < H_{\cc}$.

The situation is similar for the NRAPR, SLy4, and SQMC700 \EoSs,
which are presented in Fig.~\ref{fig:EoS_combined},
except that the entire outer core is now in a type-1.5 state and no type-II region is present.
The inner core remains a type-I superconductor, and there is a thin layer outside this
with an upper transition at $H_{\cc2'}$.
So the phase diagram again features three inhomogeneous phases,
in the case where we control $\overline{B}$,
although the mixture of fluxtubes and non-superconductor
occupies only a small part of the parameter space.

\begin{figure}[t]
  \centering
    % \includegraphics[width=0.48\textwidth, height=0.35\textwidth]{example-image-a}
    % \hspace{0.1cm}
    %  \includegraphics[width=0.48\textwidth, height=0.35\textwidth]{example-image-a}
  \includegraphics[width=0.48\textwidth]{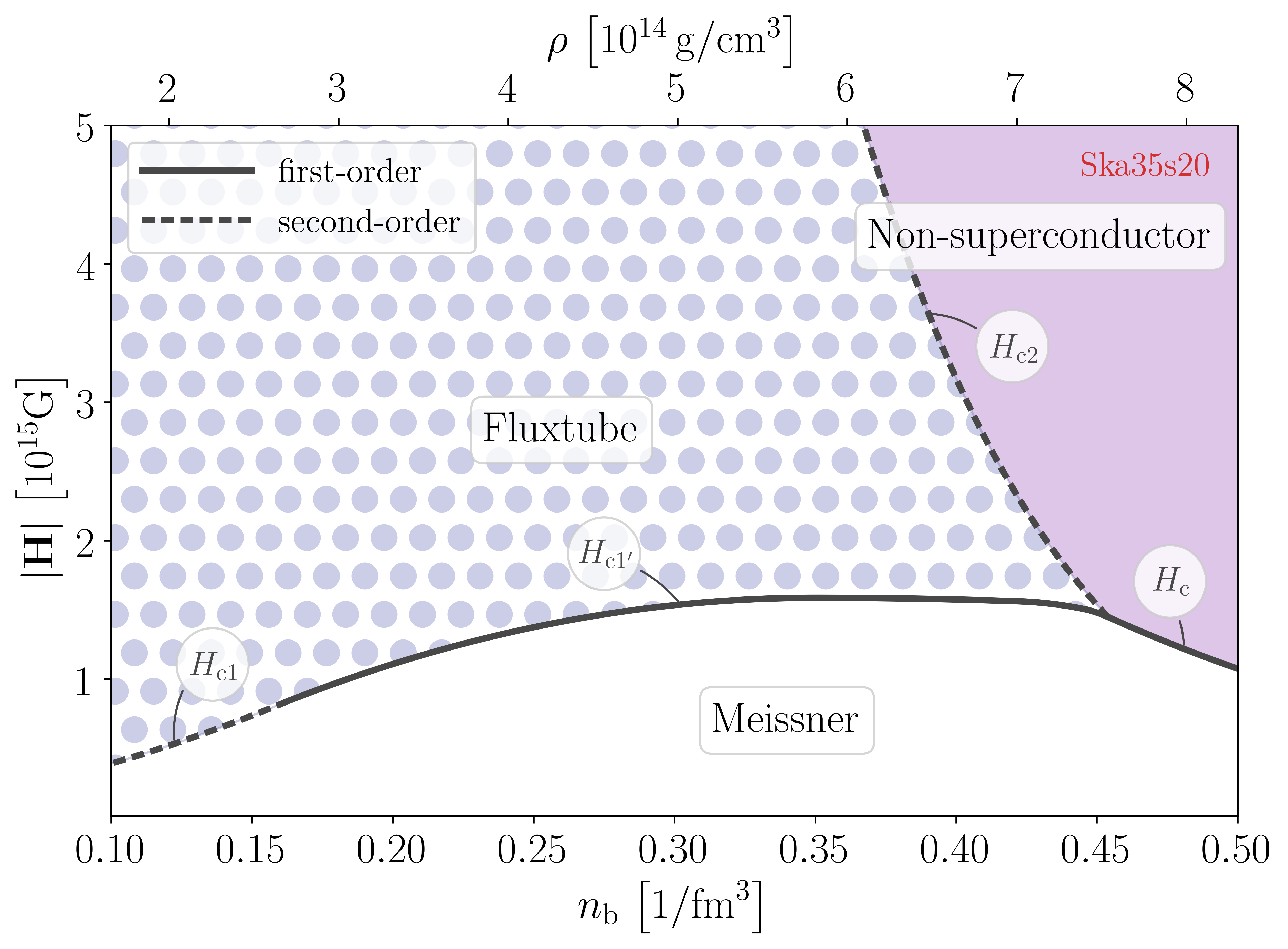}
  \hspace{0.1cm}
  \includegraphics[width=0.48\textwidth]{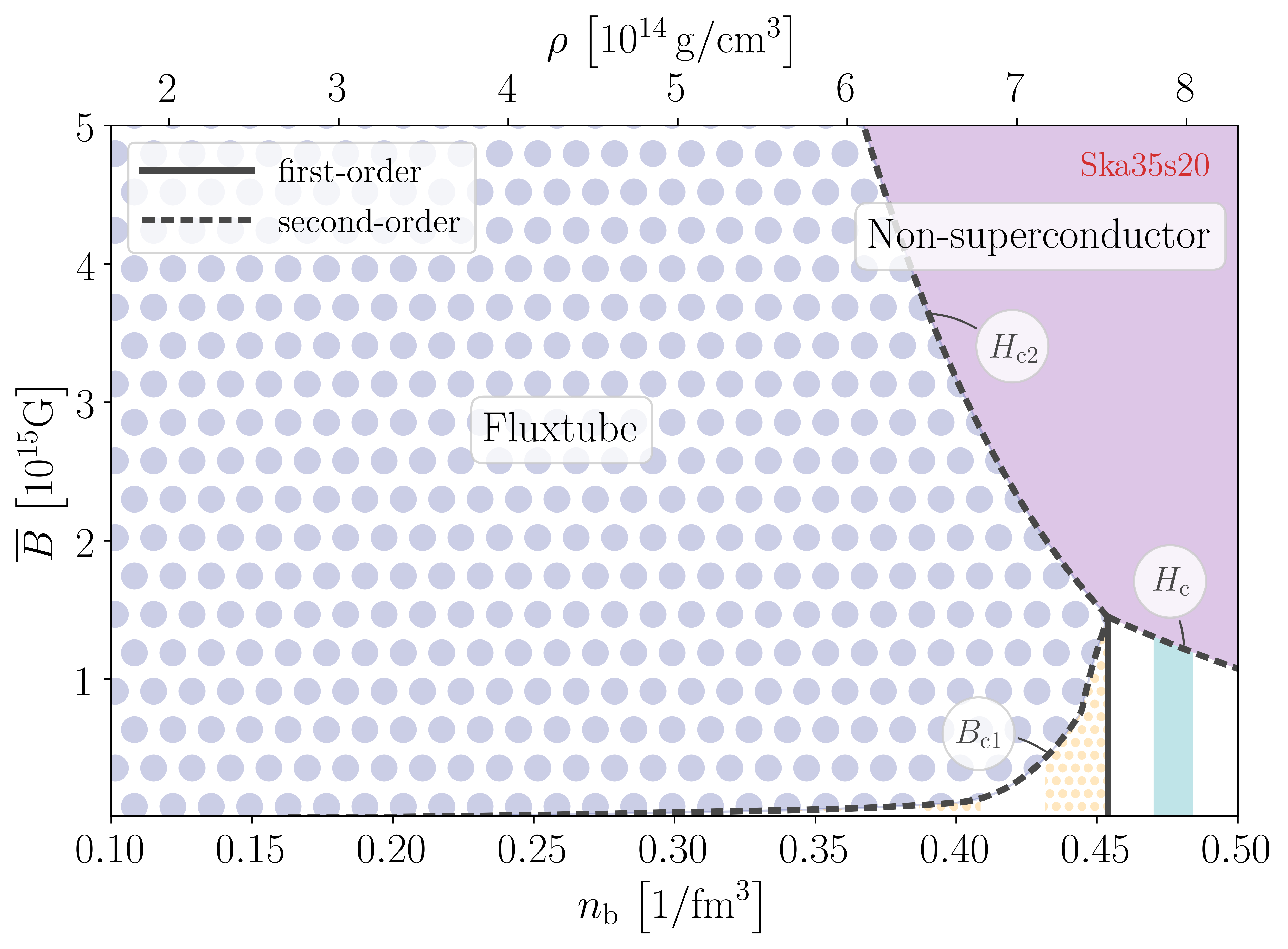}
\caption{Phase diagrams for the Ska35s20 \EoS\
with imposed external field $|\mathbf{H}|$ (left panel)
and imposed mean flux $\overline{B}$ (right panel).
Both figures are plotted in the same style as Fig.~\ref{fig:LNS}.
We observe that the upper transition is always second order,
i.e., the inhomogeneous state consisting of alternating fluxtube
and non-superconducting regions is absent.}
\label{fig:Ska35s20}
\end{figure}

For Ska35s20, whose phase diagrams are illustrated in Fig.~\ref{fig:Ska35s20},
the upper transition from the non-superconducting
to the fluxtube state is always second order.
We remind the reader that the superfluid entrainment parameter $h_1$ in this particular Skyrme model
is much weaker than for the other models (see Table~\ref{tbl:h_i}).
This likely explains
the absence of an inhomogeneous state consisting of alternating fluxtube and non-superconducting regions.
Indeed, we show in Appendix~\ref{sec:Abrikosov} that the parameters $h_1$ and $h_2$ determine how the neutron condensate responds to the presence of fluxtubes,
and that this response favors a first-order transition, if $h_1$ and $h_2$ are sufficiently large.

We note that there is considerable variation, between the different \EoSs,
in the location of the critical point,
i.e., the value of the nuclear density at which the critical fields $H_{\cc1'}$, $H_{\cc2'}$ and $H_{\cc}$ meet
and the transition to type-I superconductivity occurs.
In the case of Ska35s20,
this transition takes place at the depth at which $H_\cc = H_{\cc2}$.
Since our numerical model assumes $\gr=0$,
we can combine Eqs.~\eqref{eq:H_c} and \eqref{eq:H_c2}
to deduce that this corresponds to the point at which
\begin{equation}
  \kappa = \frac{1}{\sqrt{2}}\left(1 + \frac{h_1}{\epsilon}\right) \,.
      \label{eq:kappaI}
\end{equation}
In the absence of entrainment, i.e., for $h_1 = 0$, we recover $\kappa = 1/\sqrt{2}$,
the critical value of a classic single-component superconductor.
Using the definition of the effective London length $\lambda_\star$
in the presence of entrainment~\eqref{eq:lambda_star},
Eq.~\eqref{eq:kappaI} can be expressed as $\lambda_\star^2 = \kappa /\sqrt{2} $.
In terms of dimensional quantities (we remind the reader
that we nondimensionalized length scales with the proton coherence length $\xi_\pp$),
this last condition that determines the transition to the type-I regime can be expressed as
\begin{equation}
  \frac{\lambda_\star^2}{\lambda \xi_\pp}
    = \left(1 + \frac{h_1}{\epsilon}\right)^{-1/2} \frac{\lambda_\star}{\xi_\pp}
    = \frac{1}{\sqrt{2}}\,,
    \label{eq:triple}
\end{equation}
where $\lambda$ and $\xi_\pp$ are the bare characteristic length scales
given by Eqs.~\eqref{eq:bare_coherence} and \eqref{eq:bare_London}.
This conflicts with many previous works,\citep{Alpar-etal84, Glampedakis-etal11, Graber-etal17}
which typically assume without proper justification
that the location of this transition depends only on the ratio $\lambda_\star/\xi_\pp$
and occurs when $\lambda_\star/\xi_\pp = 1/\sqrt{2}$.
Analyzing the \gs\ of the two-component system coupled via entrainment,
and deducing the different critical fields consistently,
we have shown that this ratio needs to be adjusted by an additional factor
dependent on the entrainment coefficient $h_1$ and the parameter $\epsilon = n_\pp / n_\nn$,
when the transition to the type-I regime is to be determined.
For the other \EoSs, the transition to type-I superconductivity is not exactly given by Eq.~\eqref{eq:triple},
since $H_{\cc2'} \neq H_{\cc2}$,
but it serves as a very good approximation as long as $H_{\cc2'}$ does not differ much from $H_{\cc2}$.
\rewrite{In principle, the exact location of the critical point
can be determined semi-analytically by considering the surface energy
of the interface between Meissner and non-superconducting domains. \citep{Kobyakov20}
If we set $|\mathbf{H}|=H_\cc$ then the Gibbs energy density away from the interface is constant,
allowing the energy of the interface itself
to be calculated from a one-dimensional solution of Eqs.~\eqref{eq:ground_A}--\eqref{eq:ground_p}.
This surface energy vanishes at the point where the interface becomes unstable,
signalling the terminus of the type-I regime.
We omit the details of this calculation here,
since it is described in full by \citet{Kobyakov20}.}

\begin{figure}[t]
\centering
  % \includegraphics[width=0.48\textwidth, height=0.35\textwidth]{example-image-a}
  % \hspace{0.1cm}
  % \includegraphics[width=0.48\textwidth, height=0.35\textwidth]{example-image-a}
  \includegraphics[width=0.48\textwidth]{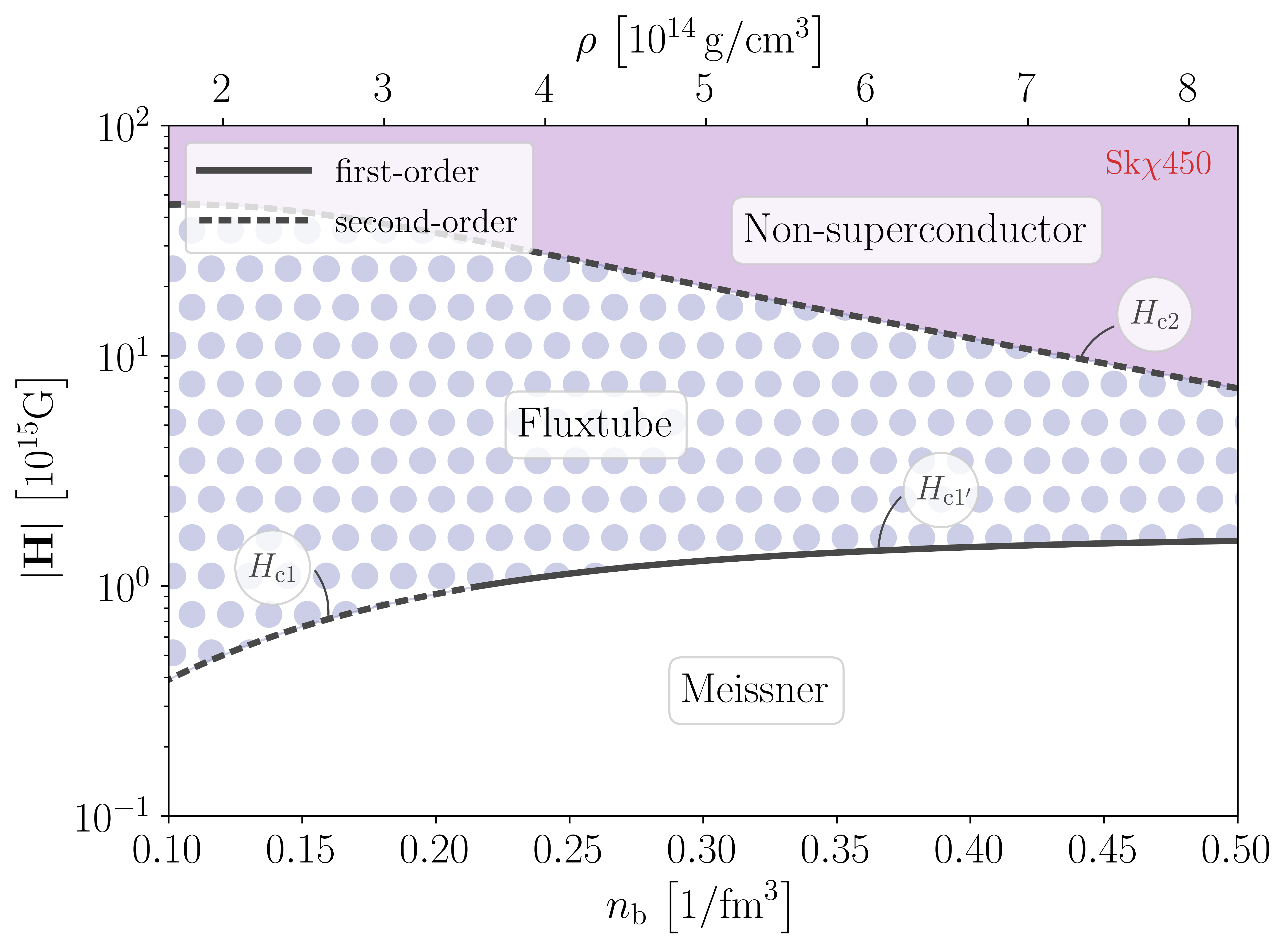}
  \hspace{0.1cm}
  \includegraphics[width=0.48\textwidth]{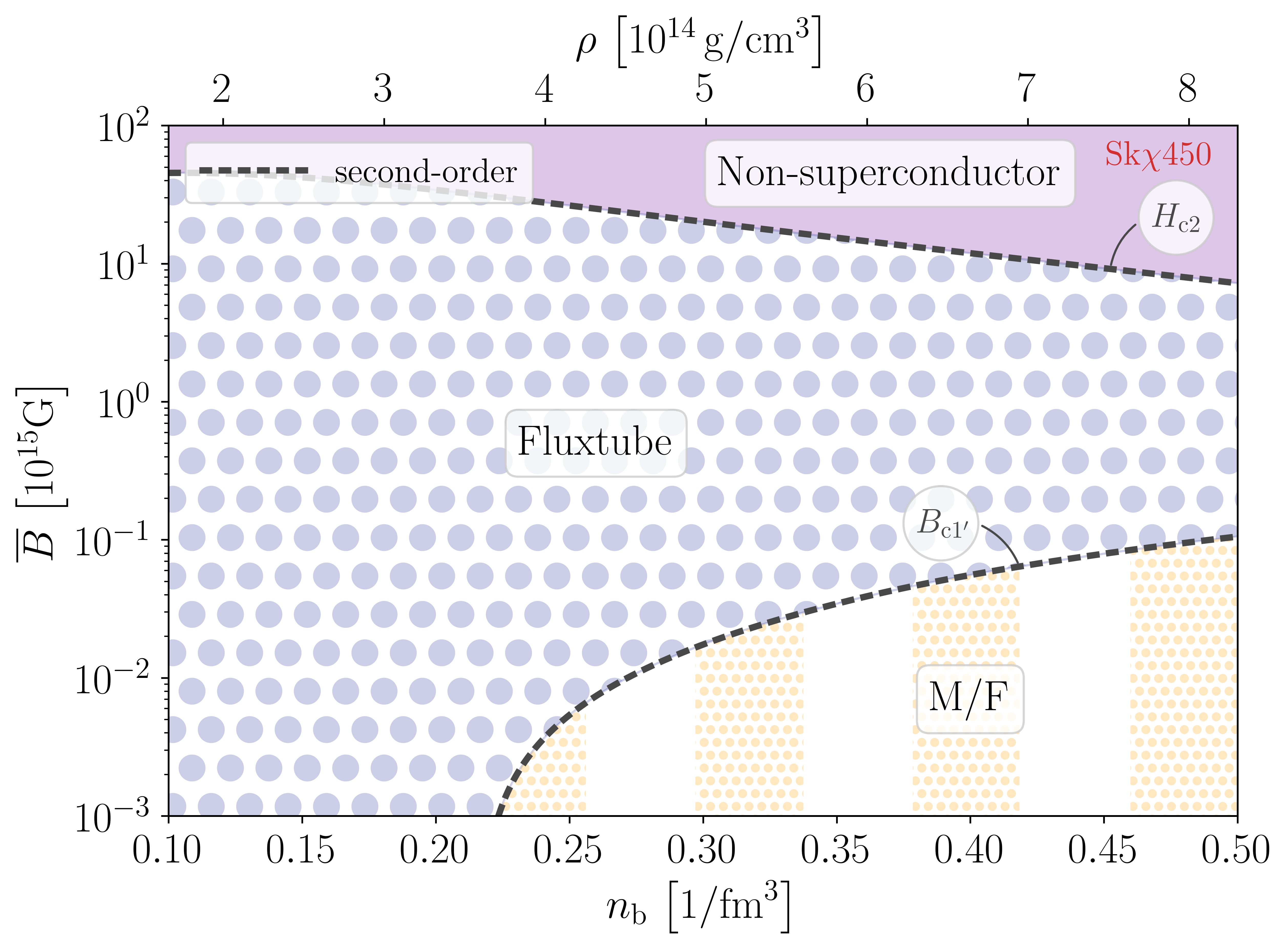}
\caption{Phase diagrams for the Sk$\chi$450 \EoS\
with imposed external field $|\mathbf{H}|$ (left panel)
and imposed mean flux $\overline{B}$ (right panel).
Both figures are plotted in the same style as Fig.~\ref{fig:LNS}.
Note that the critical point does not fall within our density range.
As a result, the intermediate type-I region,
where Meissner and non-superconducting regions alternate, is absent.
For both plots, the magnetic field axes are given on a logarithmic scale,
covering a different range.}
\label{fig:Skchi450}
\end{figure}

Finally, in Fig.~\ref{fig:Skchi450}
we present the phase diagrams for the \EoS\ Sk$\chi$450.
This case differs from the others,
because $\xi_\pp$ remains smaller than $\lambda$ (i.e., $\kappa>1$) even deep within the star,
as illustrated in the left panel of Fig.~\ref{fig:kappa_R}.
This in turn can be traced back to the large Sk$\chi$450 proton energy gap,
as shown in the right panel of Fig.~\ref{fig:gapsTc},
which is itself caused by the small proton fraction predicted by this \EoS\
(right panel of Fig.~\ref{fig:particle_fractions}).
As a result we do not see an upper transition of first order
and the critical point falls outside of our nuclear density range,
i.e., the type-I region is entirely absent.
Instead, the type-1.5 phase extends all the way to the star's center.

In the phase diagrams we have presented here,
we focused on the density range $n_\bb \gtrsim \unitfrac[0.1]{1}{fm^3}$,
i.e., we have omitted a discussion of the \gs\ close to the crust-core interface.
We neglect this region for two reasons.
First, the transition between the neutron star crust and the core is not a sharp one,
but covers a density range, which is dependent on the specific Skyrme model considered.\citep{Balliet20}
How the magnetic flux distribution changes from the innermost crustal layer,
through this extended transition into the core is not known
and these lower density regions likely impact on the \gs\ of the superconducting protons,
making our predictions less robust.
Second, the physical parameters (in particular the neutron coherence length, $\xi_\nn$)
that influence the phases of the superconductor vary significantly
in the region close to the crust-core interface,
and so the concept of a local \gs\ becomes less meaningful.
For the models Ska35s20 and Sk$\chi$450,
our model predicts a second type-1.5 phase
for $n_\bb \lesssim \unitfrac[0.09]{1}{fm^3}$,
but the preferred fluxtube separation is so large
that the distinction between type-1.5 and type-II becomes largely academic.

%%%%%%%%%%%%%%%%%%%%%%%%%%%%%%%%%%%%%%%%%%%%%%%%%%%%%%%%%%%%%%
%%%%%%%%%%%%%%%%%%%%%%%%%%%%%%%%%%%%%%%%%%%%%%%%%%%%%%%%%%%%%%

\section{Conclusion}
\label{sec:conclusion}

We have determined the \gs\ for a mixture
of two superfluid condensates,
one of which is neutral and the other electrically charged,
in the presence of a mean magnetic flux.
The condensates are coupled by density and density-gradient interactions,
and by mutual entrainment.
Our model extends the \rewrite{phenomenological} \GL\ framework used in previous works
to \rewrite{consistently} satisfy Galilean invariance \rewrite{on small scales},
and uses values for the model parameters based on a connection with the Skyrme model of nuclear matter,
allowing a realistic characterization of the stellar interior.
In addition to the three homogeneous phases (Meissner, fluxtube lattice, and non-superconducting)
that arise in simple single-component superconductors,
we have shown that inhomogeneous mixtures of any two of these phases can occur in the \gs,
due to the coupling between the condensates.
Although some features of the phase transitions can be determined analytically,
the complete phase diagrams are quite complicated,
and exact details of the transitions between different regions
need to be computed numerically.

\begin{figure}[t]
  \centering
  \includegraphics[width=10cm]{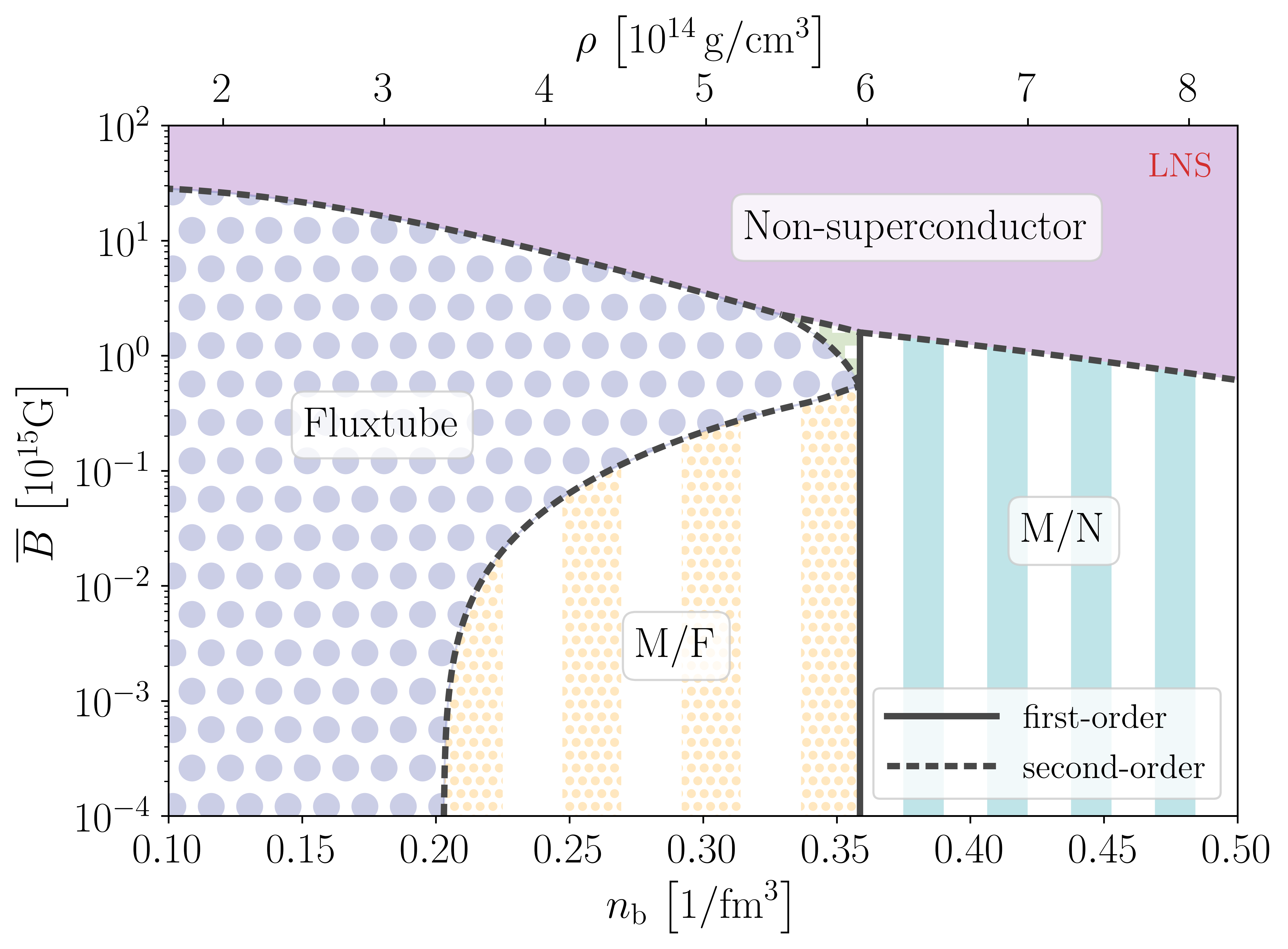}
\caption{Phase diagram for the LNS \EoS\
with imposed mean flux $\overline{B}$
(zoomed-out view of the right panel of Fig.~\ref{fig:LNS})
to illustrate the \gs s for a larger range of mean fluxes.
The magnetic field axis is given on a logarithmic scale.
For $n_\bb \lesssim \unitfrac[0.20]{1}{fm^3}$,
the protons behave like a classic type-II superconductor,
while in the range $0.20 \lesssim n_\bb \lesssim \unitfrac[0.36]{1}{fm^3}$
they exhibit type-1.5 characteristics.
At high densities, $n_\bb \gtrsim \unitfrac[0.36]{1}{fm^3}$,
we recover the behavior of a type-I superconductor.}
\label{fig:LNSzoomed}
\end{figure}

The phase diagrams vary significantly depending on the
particular set of Skyrme parameters used for the \EoS.
However, for all six Skyrme models considered in this paper,
the protons in part or all of the outer core behave as a type-1.5 superconductor,
in which fluxtubes form bundles with local hexagonal symmetry, rather than a periodic lattice.
This behavior, reminiscent of laboratory multi-band superconductors,
is a direct result of
\review{the coupling between the neutrons and the protons:
each fluxtube produces a perturbation to the neutron condensate,
which leads to a long-range attraction between fluxtubes
that gives way to electromagnetic repulsion on shorter scales.}
The fact that fluxtubes have a preferred separation distance also
leads to a first-order phase transition at the
lower critical field $H_{\cc1'} < H_{\cc1}$.
In some cases, the upper critical field is also a first-order transition, occurring at
$H_{\cc2'} > H_{\cc2}$,
\review{which is again a consequence of the coupling between the condensates.
The perturbations to the neutron condensate
produced by the fluxtubes lower the overall free energy;
if this effect is strong enough then it becomes energetically favorable to
confine the proton condensate within part of the domain,
forming a lattice with the preferred separation,
and concentrate the rest of the magnetic flux in non-superconducting regions.}

These results have important consequences for the neutron star interior.
In Fig.~\ref{fig:LNSzoomed}
we provide a zoomed-out view of the right panel of Fig.~\ref{fig:LNS},
to illustrate the \gs s for a larger range of mean fluxes, $\overline{B}$.
The majority of these compact objects are observed as radio pulsars
with inferred dipolar magnetic field strengths below $\unit[10^{14}]{G}$.
Assuming that this value is indicative of the typical field strength in the stellar interior,
we conclude that most known neutron stars fall below the transition $B_{\cc1}$
in much (if not all) of their outer cores,
and therefore contain an inhomogeneous mixture of hexagonal fluxtube lattice
and flux-free Meissner regions.
This conflicts with the assumption, generally considered in the literature,
that the protons can be treated as a type-II system,
and our results suggest that a different description is needed,
when phenomena involving the superconducting protons are modelled.
For a small fraction of neutron stars,
the so-called magnetars with field strengths $\gtrsim \unit[10^{14}]{G}$,
there may also be a narrow range of mean fluxes and densities
for which the \gs\ is a mixture of fluxtubes and non-superconductor,
although this inhomogeneous state was not observed for all \EoSs.
The majority of the outer magnetar core is permeated by a homogeneous fluxtube phase.
As field strengths are increased further, superconductivity eventually breaks down.
We point out that the upper transition to the non-superconducting phase
plateaus towards smaller densities;
for LNS the $H_{\cc2}$ curve peaks at $\sim \unit[3 \times 10^{16}]{G}$
as shown in Fig.~\ref{fig:LNSzoomed}.
We recall that the neutrons remain superfluid in our model
even when proton superconductivity disappears.

\begin{figure}[t]
  \centering
\includegraphics[width=13cm]{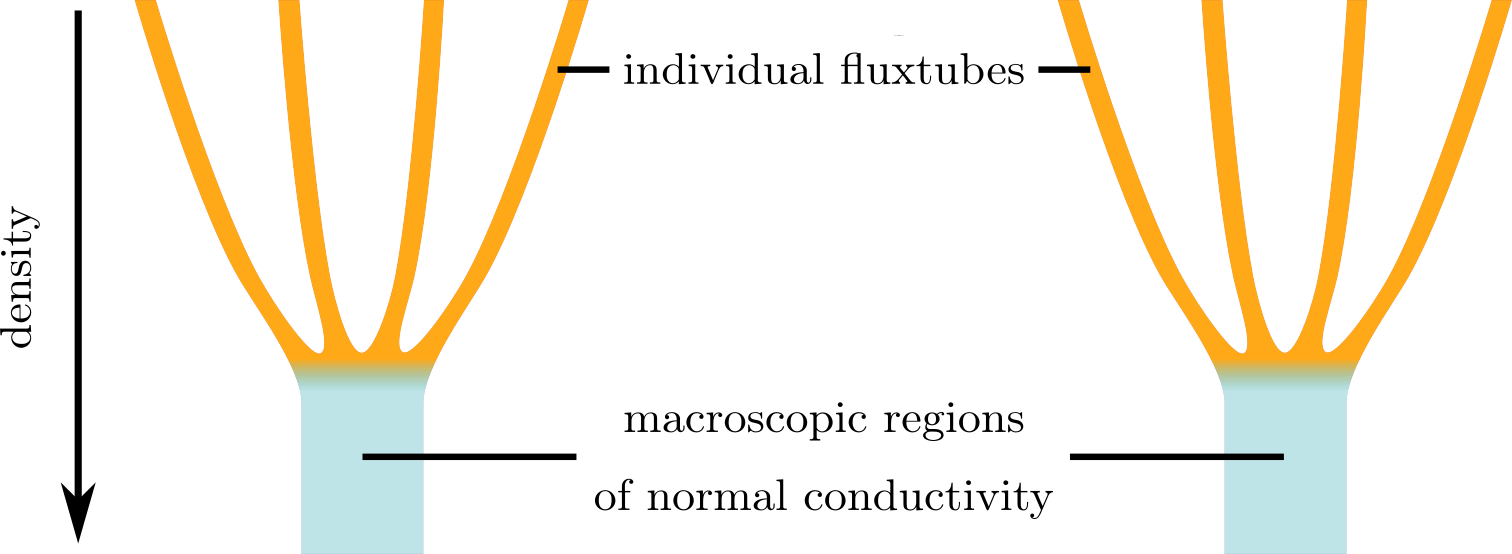}
\caption{Schematic representation of the magnetic flux
at the transition from the outer to the inner core.
At lower densities, the protons are in a type-1.5 regime,
where the flux is quantized into thin fluxtubes (orange) of preferred separation
(overall confined to a small fraction of the total stellar volume).
As the density increases towards the inner core,
which is a type-I superconductor,
magnetic flux is contained in macroscopic regions of normal conductivity (light blue).
The shading indicates the intensity of the field.
These alternate with macroscopic flux-free Meissner regions.
Note that this schematic is not to scale.}
\label{fig:schematic}
\end{figure}

In the inner core, magnetic flux is likely distributed in an intermediate type-I state,
where flux-free regions alternate with non-superconducting, i.e. normal-conducting, ones.
Although the transition between the type-1.5 regime to this intermediate phase is of first order,
it is fairly smooth and we illustrate the magnetic flux distribution schematically in Fig.~\ref{fig:schematic}.
The exact density, at which this transition from the outer to the inner core takes place,
varies significantly with the \EoS\
(for one Skyrme model it even fell outside of the density range considered)
and generally cannot be determined analytically.
However, we have provided \rewrite{an approximate} criterion for this transition, given in Eq.~\eqref{eq:triple},
that consistently incorporates entrainment
and serves as a very good approximation in all of the Skyrme models that we have considered.

We point out that our conclusions as illustrated in Fig.~\ref{fig:schematic}
rely on the assumption that the Skyrme interaction
correctly captures the physics up to the neutron star's center
and exotic particles are absent from the inner core.
If this is not satisfied, the flux distribution (especially at high densities)
could vary significantly from our results.

Overall, the phases and transitions in our diagrams closely resemble those conjectured
by \citet{HaberSchmitt17},
which were based on one-dimensional fluxtube models.
However, their study was largely focused on the role of the parameter $\alpha$,
which measures the mutual repulsion between the condensates.
We have set this parameter to zero in our numerical results,
and instead focused on the effect of
(Galilean-invariant) entrainment between the condensates.
Our two-dimensional numerical model has the advantage
that we can determine the exact positions of the transitions for different densities
and simulate the magnetic flux distribution for a given set of parameters, as demonstrated in Fig.~\ref{fig:mixed}.
Our results confirm the hypothesis of \citet{HaberSchmitt17}
that fluxtubes preferentially adopt a singly-charged, hexagonal configuration,
even in cases where a periodic lattice is not the \gs.

Until now it has generally been accepted (following \citet{Baym-etal69})
that the outer core of a neutron star is a type-II superconductor,
in which fluxtubes are stable but mutually repulsive on all scales.
In that case, fluxtubes are expected to form a lattice arrangement that expands over time,
eventually leading to a flux-free state
(albeit on a very long timescale).
Our results imply that fluxtubes have a preferred separation distance throughout much of the core,
as a result of entrainment between the protons and neutrons,
and will therefore form ``hexagonal bundles''
that can persist indefinitely.
However, to understand the time-dependent behavior of the fluxtubes in detail
will require a more complete dynamical model than that presented here.
In particular, in this work we have considered defects in the proton condensate only,
i.e., we have neglected the presence of neutron vortices
that will appear as a result of the superfluid's quantized rotation.
This is justifiable when characterizing the system's \gs,
because the fluxtubes outnumber the vortices by many orders of magnitude,
but interactions between both types of defects are crucial for the dynamics of the rotation and magnetic field on larger scales.
A fully dynamical description of the neutron star interior
must also incorporate the (non-superfluid) electrons,
whose scattering by fluxtubes is a dominant source of dissipation in the star's core.\citep{Alpar-etal84, Jones91, Gusakov19}
Finding a consistent treatment of the electrons,
whose mean free path far exceeds the typical distance between fluxtubes,
is beyond the scope of the present work
and is left for future study.

%%%%%%%%%%%%%%%%%%%%%%%%%%%%%%%%%%%%%%%%%%%%%%%%%%%%%%%%%%%%%%

\begin{acknowledgments}

The authors would like to thank the Institute for Nuclear Theory at the University of Washington
for its kind hospitality and hosting INT Program INT-19-1a during which part of this work was carried out.
We also thank Wynn Ho, John Miller, and Hayder Salman for helpful conversations related to this paper
\rewrite{and Alexander Haber for providing feedback on our manuscript.}
Part of TSW's time was funded by EPSRC Grant EP/R024952/1.
VG acknowledges partial support from a McGill Space Institute postdoctoral fellowship
and the Trottier Chair in Astrophysics and Cosmology
as well as the H2020 ERC Consolidator Grant “MAGNESIA” under grant agreement Nr.~817661 (PI: Rea).
WGN acknowledges support from NASA grant 80NSSC18K1019.
Finally, we acknowledge the use of the following software: \texttt{IPython},\citep{PerezGranger07}
\texttt{Matplotlib},\citep{Hunter07} \texttt{NumPy},\citep{Oliphant06, vanderWalt-etal11, Harris-etal20}
\texttt{Pandas},\citep{McKinney10} and \texttt{SciPy}.\citep{Jones-etal01, SciPy20}
The data used to produce the plots in this paper are available at \url{https://github.com/vanessagraber/NS_EoS} and
\url{https://data.ncl.ac.uk/}.

\end{acknowledgments}

%%%%%%%%%%%%%%%%%%%%%%%%%%%%%%%%%%%%%%%%%%%%%%%%%%%%%%%%%%%%%%
%%%%%%%%%%%%%%%%%%%%%%%%%%%%%%%%%%%%%%%%%%%%%%%%%%%%%%%%%%%%%%

\appendix

\section{Minimization of the free energy}
\label{sec:code}

The dimensionless free-energy density,
given by Eq.~\eqref{eq:F_dimless},
is approximated numerically
on a regular 2D grid,
with intervals $\delta x$ and $\delta y$ in the $x$ and $y$ directions.
The order parameters $\psip$ and $\psin$ are defined on the gridpoints as
$\psip^{i,j}$ and $\psin^{i,j}$,
where $i$ and $j$ denote the indices in $x$ and $y$, respectively.
The vector field $\mathbf{A}$ has only two components, $(A_x,A_y)$,
which are defined on the corresponding links between the gridpoints,
i.e., we have $A_x^{i+1/2,j}$ and $A_y^{i,j+1/2}$.
The gauge coupling between $\psip$ and $\mathbf{A}$ is implemented using a standard Peierls substitution,
noting that for instance
\begin{align}
    \left|\left(\frac{\partial}{\partial x}-\ii A_x\right)\psip\right|
        &= \left|\frac{\partial}{\partial x}\exp(- \textstyle\int \ii A_x\,\dd x)\psip\right| \nonumber \\
    \Rightarrow \left|\left(\frac{\partial}{\partial x}-\ii A_x\right)\psip\right|^{i+1/2,j}
        &\simeq \frac{1}{\delta x}\left|\exp(- \ii A_x^{i+1/2,j}\,\delta x)\psip^{i+1,j} - \psip^{i,j}\right|\,.
\end{align}
By including the gauge coupling in this way,
we exactly preserve the (discrete) gauge symmetry
\begin{equation}
    \psip^{i,j} \to \exp(\ii\phi^{i,j})\psip^{i,j}\,,
        \qquad
    A_x^{i+1/2,j} \to A_x^{i+1/2,j} + \frac{\phi^{i+1,j}-\phi^{i,j}}{\delta x}\,,
        \qquad
    A_y^{i,j+1/2} \to A_y^{i,j+1/2} + \frac{\phi^{i,j+1}-\phi^{i,j}}{\delta y}\,.
\end{equation}
This leads to a discrete approximation to the total free energy, $\mathcal{F}_\text{dis}[\psip^{i,j},\psin^{i,j},A_x^{i+1/2,j},A_y^{i,j+1/2}]$.
We obtain the \gs\ using a simple gradient-descent method,
in which the step size is made as large as possible while maintaining numerical stability.
Specifically, we use the iteration scheme
\begin{align}
    \psip^{i,j} &\to
        \psip^{i,j} - \left(\frac{N/4}{\delta y/\delta x + \delta x/\delta y}\right)
            \frac{\partial\mathcal{F}_\text{dis}/\partial\psip^{\star,i,j}}
            {1 + \dfrac{h_1}{\epsilon}|\psin^{i,j}|^2 + h_3|\psip^{i,j}|^2}\,, \\
    \psin^{i,j} &\to
        \psin^{i,j} - \left(\frac{N/4}{\delta y/\delta x + \delta x/\delta y}\right)
            \frac{\partial\mathcal{F}_\text{dis}/\partial\psin^{\star,i,j}}
            {1 + h_1|\psip^{i,j}|^2 + \dfrac{h_3}{\epsilon}|\psin^{i,j}|^2}\,, \\
    A_x^{i+1/2,j} &\to
        A_x^{i+1/2,j} - \left(\frac{N/4}{\delta y/\delta x + \delta x/\delta y}\right)
            \frac{\partial\mathcal{F}_\text{dis}/\partial A_x^{i+1/2,j}}{2\kappa^2}\,, \\
    A_y^{i,j+1/2} &\to
        A_x^{i,j+1/2} - \left(\frac{N/4}{\delta y/\delta x + \delta x/\delta y}\right)
        \frac{\partial\mathcal{F}_\text{dis}/\partial A_y^{i,j+1/2}}{2\kappa^2}\,.
\end{align}
We use an initial (dimensionless) resolution of $\delta x, \delta y \simeq 0.5$,
and iterate until the change in energy drops below a threshold of $10^{-7}$.
We then double the resolution in $x$ and $y$ and repeat the whole process
until the value of $\mathcal{F}_\text{dis}$ converges to at least five significant figures.

%%%%%%%%%%%%%%%%%%%%%%%%%%%%%%%%%%%%%%%%%%%%%%%%%%%%%%%%%%%%%%
%%%%%%%%%%%%%%%%%%%%%%%%%%%%%%%%%%%%%%%%%%%%%%%%%%%%%%%%%%%%%%

\section{Long-range interaction between fluxtubes}
\label{sec:Kramer}

The dimensionless free-energy density~\eqref{eq:F_dimless},
when written in terms of the real variables $f$, $g$, $\mathbf{V}$ and $\chi$
defined in Sec.~\ref{sec:lower},
takes the form
\begin{align}
    F[f,g,\mathbf{V},\chi] =\ & \frac{1}{2}(f^2 - 1)^2 + \frac{R^2}{2\epsilon}(g^2 - 1)^2
            + \frac{\gr}{\epsilon}(f^2 - 1)(g^2 - 1) \nonumber \\
        &+ \left|\nablab f\right|^2 + f^2\left|\mathbf{V}\right|^2
            + \frac{1}{\epsilon}\left|\nablab g\right|^2
            + \frac{1}{\epsilon}g^2\left|\nablab\chi\right|^2
            + \kappa^2|\nablab\times\mathbf{V}|^2 \nonumber \\
        &+ \frac{h_1}{\epsilon}\left[f^2\left|\nablab g\right|^2
            + g^2\left|\nablab f\right|^2
            + f^2g^2|\mathbf{V}-\nablab\chi|^2\right] \nonumber \\
        &+ \frac{h_2}{2\epsilon}\nablab f^2 \cdot \nablab g^2
            + \frac{h_3}{4}\left(\frac{1}{\epsilon^2}\left|\nablab g^2\right|^2
            + \left|\nablab f^2\right|^2\right)\,.
        \label{eq:F2}
\end{align}
The reason for working with these real variables will be explained shortly.
Recall that we have chosen the reference energy level so
that $F = 0$ in the absence of fluxtubes and vortices,
i.e., $f=g=1$ and $\mathbf{V}=\nablab\chi=\mathbf{0}$.
In what follows, we will refer to this as the uniform solution.

Our goal is to determine the free energy per unit length per \WS\ cell, $\mathcal{F}(a)$,
in the asymptotic limit of a widely-spaced fluxtube lattice, $a\to\infty$.
In this limit, we know that $\mathcal{F}$ converges to the value for a single fluxtube,
i.e., $\mathcal{F} \to \mathcal{F}_\infty$,
and we wish to estimate the ``interaction energy'' $\mathcal{F} - \mathcal{F}_\infty$.
If its value is positive, the lower transition is of second-order,
whereas a negative value implies a first-order transition.
To do so,
we will first generalize the method introduced by \citet{Kramer71} to the case of a free energy
in the completely general form $\mathcal{F} = \langle F[\Psi]\rangle$,
where $\Psi$ represents the complete set of independent, real variables,
and the angled brackets represent a domain integral.
We will then apply the results to the particular case given by Eq.~\eqref{eq:F2}.
We consider an infinite lattice of parallel fluxtubes that are widely separated,
in the sense that the size of each \WS\ cell is large in comparison with both
the coherence length and the London length.
Any such lattice, characterized by $\Psi$,
is a steady state, in that it is a solution of the \EL\ equations
that are obtained from the functional derivative
\begin{equation}
    \frac{\delta}{\delta\Psi}\langle F\rangle = 0\,.
        \label{eq:EL}
\end{equation}
We assume that each fluxtube is located far inside its \WS\ cell,
which allows us to make two approximations:
\begin{enumerate}
  \item within each cell, the solution can be approximated as
  a linear perturbation to the single fluxtube solution;
  \item on the boundary of the cell, the solution can be approximated as
  the uniform solution plus a superposition of
  linear, \emph{independent} perturbations produced by the fluxtubes.
\end{enumerate}
Note that we do not assume
that the fluxtube is located exactly at the center of the cell,
and we will verify later
that the exact location of the fluxtube within the cell
does not affect our result for the interaction energy.

We will use the notation $\delta^n F[\delta\Psi;\Psi]$ to represent the $n$-th variation of $F$,
i.e., the terms up to order $(\delta\Psi)^n$ in the Taylor expansion of $F[\Psi+\delta\Psi]-F[\Psi]$.
If $\Psi$ is a solution of the \EL\ equations \eqref{eq:EL},
then $\delta^1 F[\delta\Psi;\Psi]$ is an exact derivative.
This follows directly from
the definition of the functional derivative,
which dictates that
\begin{equation}
    \delta^1 F = \delta \Psi \cdot
        \underbrace{\left( \frac{\delta}{\delta\Psi} \langle F\rangle \right)}_{\text{EL equations}}
        + \nablab\cdot\mathbf{Q},
\end{equation}
for some vector field $\mathbf{Q}[\delta\Psi;\Psi]$.
This result implies that the integral of $\delta^1 F$ over any domain
can be expressed as an integral over the boundary of that domain.
Furthermore, if $\delta\Psi$ is a solution of the \emph{linearized} \EL\ equations
(i.e.~linearized about $\Psi$),
then it can be shown that $\delta^2F[\delta\Psi;\Psi]$ is also an exact derivative.
In fact, we have
\begin{equation}
    \delta^2 F = \delta \Psi \cdot
        \underbrace{\left( \frac{\delta}{\delta\Psi} \langle F\rangle \right)}_{\text{EL equations}}
        +  \frac{1}{2} \, \delta \Psi \cdot
        \underbrace{\left(\delta^1\left\{\dfrac{\delta}{\delta\Psi}\langle F\rangle\right\}\right)}_{\substack{\text{perturbed} \\ \text{EL equations} }}
        + \nablab\cdot\mathbf{Q}^{(2)}\,,
\end{equation}
where the vector field $\mathbf{Q}^{(2)}[\delta\Psi;\Psi]$
is given precisely by
the terms up to order $(\delta\Psi)^2$
in the Taylor expansion of
$\mathbf{Q}[\delta\Psi;\Psi+\tfrac{1}{2}\delta\Psi]$.
Once we identify the functional form of $\delta^1 F$,
we deduce $\mathbf{Q}$ and thence $\mathbf{Q}^{(2)}$.
We can then express the integral of $\delta^2 F$ over any domain
as an integral over the boundary of that domain.
The importance of this result in determining the interaction energy
will become more obvious shortly.
In what follows, we refer to the linear equations for $\delta\Psi$ as the Jacobi equations,
by analogy with the equations defining Jacobi fields in Riemannian geometry.\citep{Taub69}

Suppose the fluxtubes are aligned with the $z$-axis,
at locations in the $xy$-plane indexed as $\mathbf{x}_i$.
Without loss of generality, we will assume that $\mathbf{x}_0 = \mathbf{0}$,
and that $\mathbf{x}_{-i} = - \mathbf{x}_i$.
We will also label the \WS\ cells as $C_i$, such that $\mathbf{x}_i \in C_i$.
This allows us to express the interaction energy as
\begin{align}
    \mathcal{F} - \mathcal{F}_\infty &= \iint_{C_0} F^\text{(all)}\,\dd x\,\dd y
        - \iint_{\mathbb{R}^2} F^{(0)}\,\dd x\,\dd y \nonumber \\
    &= \iint_{C_0} \left[F^\text{(all)} - \sum_i F^{(i)}\right]\,\dd x\,\dd y\,,
        \label{eq:Eint}
\end{align}
where $F^\text{(all)}$ represents the free-energy density in the presence of the lattice,
and $F^{(i)}$ represents the free-energy density
in the presence of a single fluxtube at $\mathbf{x} = \mathbf{x}_i$.
To obtain the last line,
we have used the fact that,
due to the translational symmetry of the lattice,
the energy density in cell $C_i$ resulting from a single fluxtube at $\mathbf{x}=\mathbf{0}$
is equivalent to the energy density in cell $C_0$ resulting from a single fluxtube at $\mathbf{x} = \mathbf{x}_{-i}$.

We now apply assumptions 1 and 2 stated above.
Within cell $C_0$, we therefore approximate $\Psi^\text{(all)} \simeq \Psi^{(0)} + \delta\Psi^{(i\neq0)}$,
and $\Psi^{(i)} \simeq \Psi^\text{(none)} + \delta\Psi^{(i)}$ for each $i\neq0$,
where $\delta\Psi^{(i\neq0)}$ represents the perturbation produced by all fluxtubes
external to $C_0$ and $\Psi^\text{(none)}$ is the uniform solution.
Furthermore, we approximate $\delta\Psi^{(i\neq0)} \simeq \sum_{i\neq0}\delta\Psi^{(i)}$
\emph{on the boundary} of $C_0$.
We emphasize that $\delta\Psi^{(i)}$ refers to the linear perturbation
of the uniform solution in the presence of a single fluxtube.
In practice this can be calculated from the Jacobi equations,
as we have demonstrated in Sec.~\ref{sec:lower}.
Under these approximations,
we can expand $F^\text{(all)}$ in Eq.~\eqref{eq:Eint} and
cancel the zeroth-order term $F^{(0)}$ with the $i=0$ term in the sum.
Expanding the remaining $i \neq 0$ contributions
and keeping in mind that the uniform solution has $F=0$,
we are left with an expression for the interaction energy
that only contains second variations of $F$,
and can thus be written in terms of the vector field $\mathbf{Q}^{(2)}$:
\begin{align}
    \mathcal{F} - \mathcal{F}_\infty
    &\simeq \iint_{C_0}\left[\delta^2F[\delta\Psi^{(i\neq0)};\Psi^{(0)}]\right]
        - \sum_{i\neq0}\left[\delta^2F[\delta\Psi^{(i)};\Psi^\text{(none)}]\right]\,\dd x\,\dd y
            \nonumber \\
    &= \int_{\partial C_0}\left[\mathbf{Q}^{(2)}[\delta\Psi^{(i\neq0)};\Psi^{(0)}]
        - \sum_{i\neq0}\mathbf{Q}^{(2)}[\delta\Psi^{(i)};\Psi^\text{(none)}]\right]\cdot\dd\mathbf{S}
            \nonumber \\
    &\simeq \int_{\partial C_0}\left[\mathbf{Q}^{(2)}[\sum_{i\neq0}\delta\Psi^{(i)};\Psi^{(0)}]
        - \sum_{i\neq0}\mathbf{Q}^{(2)}[\delta\Psi^{(i)};\Psi^\text{(none)}]\right]\cdot\dd\mathbf{S}\,,
        \label{eq:interaction1}
\end{align}
where $\partial C_0$ represents the boundary of cell $C_0$,
and $\dd\mathbf{S}$ is the boundary element on this boundary, with outward normal.
The final step is to approximate $\Psi^{(0)} \simeq \Psi^\text{(none)} + \delta\Psi^{(0)}$
on the boundary of $C_0$,
and retain only terms up to second order in $\delta\Psi$;
this calculation is straightforward once the functional form of $\mathbf{Q}^{(2)}$ is known.
In this way, we can
express the interaction energy as an integral over the boundary of a single \WS\ cell,
and so we do not need to consider the full domain volume
to determine the nature of the phase transition at the lower critical field.

Note that our assumptions 1 and 2 generally do not apply to the proton order parameter $\psip$,
because introducing an additional fluxtube
causes a nonlinear change in the phase of the proton condensate throughout the domain.
Thus, we cannot apply the above method directly to Eq.~\eqref{eq:F_dimless}.
This is the reason for making the change of variables
to the set $\Psi \equiv (f,g,\mathbf{V},\chi)$,
for which assumptions 1 and 2 do hold.
Now taking first variations of the free energy $F$
in the form of Eq.~\eqref{eq:F2}
and using integration by parts,
we find that
\begin{align}
    \mathbf{Q} &= 2\delta f\nablab f + \frac{2}{\epsilon}\delta g\nablab g
        + \frac{2}{\epsilon}g^2\delta\chi\nablab\chi
        + 2\kappa^2\delta\mathbf{V}\times(\nablab\times\mathbf{V})
            \nonumber \\
    &+ \frac{2h_1}{\epsilon}\left(g^2\delta f\nablab f + f^2\delta g\nablab g
        + f^2g^2\delta\chi(\nablab\chi - \mathbf{V})\right)
            \nonumber \\
    &+ \frac{2h_2}{\epsilon}fg\left(\delta f\nablab g + \delta g\nablab f\right)
        + 2h_3\left(f^2\delta f\nablab f + \frac{1}{\epsilon^2}g^2\delta g\nablab g\right)\,.
        \label{eq:F1}
\end{align}
The interaction energy can now be calculated following the steps outlined above.
For brevity let us just consider the representative term
$\mathbf{Q} = 2fg\,\delta f\nablab g$.
For this term, we find that
\begin{equation}
    \mathbf{Q}^{(2)} = \mathbf{Q} + g(\delta f)^2\nablab g
        + f\,\delta f\,\delta g\nablab g + fg\,\delta f\nablab\delta g\,.
\end{equation}
Recalling that the uniform solution has $f=g=1$ and $\nablab\chi=\mathbf{V}=\mathbf{0}$,
we find that the contribution from this term to the interaction energy \eqref{eq:interaction1} is
\begin{align*}
    &\int_{\partial C_0}\sum_{i\neq0}
        \left[2\delta f^{(i)}\nablab\delta g^{(0)}
        + \sum_{j\neq0}\delta f^{(i)}\nablab\delta g^{(j)}
        - \delta f^{(i)}\nablab\delta g^{(i)}
        \right]\cdot\dd\mathbf{S} \\
    &= \int_{\partial C_0}
        \left[\sum_{i}\sum_{j\neq i}\delta f^{(i)}\nablab\delta g^{(j)}
        + \sum_{i}\left(\delta f^{(i)}\nablab\delta g^{(0)}
        - \delta f^{(0)}\nablab\delta g^{(i)}\right)
        \right]\cdot\dd\mathbf{S}\, \\
    &= \sum_i \iint_{C_0} \nablab \cdot \left[ \delta f^{(i)} \nablab \delta g^{(0)}
        - \delta f^{(0)} \nablab \delta g^{(i)} \right] \,\dd x\,\dd y \, .
\end{align*}
In deriving the last equality, we have used the divergence theorem and the fact that
the doubly-summed term vanishes, as can be shown using a similar argument
to that leading to Eq.~\eqref{eq:Eint}:
\begin{align*}
    \sum_{i} \sum_{j\neq i} \int_{\partial C_0}
        \left(\delta f^{(i)}\nablab\delta g^{(j)}\right)\cdot\dd\mathbf{S}
    &= \sum_{i}\sum_{j\neq 0}\int_{\partial C_i}
        \left(\delta f^{(0)}\nablab\delta g^{(j)}\right)\cdot\dd\mathbf{S} \\
    &= \sum_{j\neq 0}\int_{\partial\mathbb{R}^2}
        \left(\delta f^{(0)}\nablab\delta g^{(j)}\right)\cdot\dd\mathbf{S} \\
    &= 0\,.
\end{align*}
Here, $\partial\mathbb{R}^2$ is the boundary of the entire $xy$-plane,
where the integrand is exponentially small.

By applying a similar procedure to the remaining terms in Eq.~\eqref{eq:F1},
we eventually find that
\begin{equation}
    \mathcal{F} - \mathcal{F}_\infty
        \simeq \sum_i\iint_{C_0}\left[\delta\Psi^{(i)}\cdot\mathcal{L}[\delta\Psi^{(0)}]
        - \delta\Psi^{(0)}\cdot\mathcal{L}[\delta\Psi^{(i)}]\right]\,\dd x\,\dd y\,,
            \label{eq:interaction}
\end{equation}
where $\delta\Psi = (\delta f,\delta g,\delta\mathbf{V},\delta\chi)$ and
\begin{equation}
    \mathcal{L}[\delta\Psi] = \left(\begin{array}{c}
        \left(1+\dfrac{h_1}{\epsilon}+h_3\right)\nabla^2\delta f
        + \dfrac{h_2}{\epsilon}\nabla^2\delta g \\[1ex]
    \dfrac{1}{\epsilon}\left(1+h_1+\dfrac{h_3}{\epsilon}\right)\nabla^2\delta g
        + \dfrac{h_2}{\epsilon}\nabla^2\delta f \\[1ex]
    - \kappa^2\nablab\times(\nablab\times\delta\mathbf{V})
        + \dfrac{h_1}{\epsilon}\nablab\delta\chi \\
    \dfrac{1}{\epsilon}\nabla^2\delta\chi
        + \dfrac{h_1}{\epsilon}\nablab\cdot(\nablab\delta\chi - \delta\mathbf{V})
    \end{array}\right)\,.
        \label{eq:L}
\end{equation}
We note the similarity between this linear operator $\mathcal{L}[\delta\Psi]$
and the Jacobi equations~\eqref{eq:lin_f}--\eqref{eq:lin_chi}.
In fact, for an arbitrary free-energy functional $F[\Psi]$
it can be shown that the formula~\eqref{eq:interaction}
for the interaction energy still holds, provided that we define
\begin{equation}
    \mathcal{L}[\delta\Psi] \equiv
        - \frac{1}{2}\delta^1\left\{\dfrac{\delta}{\delta\Psi}\langle     F\rangle\right\}[\delta\Psi;\Psi^\text{(none)}]\,,
    \label{eq:L_general}
\end{equation}
i.e., $\mathcal{L}$ is defined by the Jacobi equations.
This implies that $\mathcal{L}[\delta\Psi^{(i)}]=0$ at all points except $\mathbf{x} = \mathbf{x}_i$
(the center of the fluxtube),
where $\delta\Psi^{(i)}$ is not differentiable.
As a result, the integrand in Eq.~\eqref{eq:interaction} vanishes at all points inside $C_0$,
except at $\mathbf{x}=\mathbf{0}$,
where it has the form of a delta function.
It is for this reason
that the exact location of the fluxtube within the \WS\ cell is immaterial in Eq.~\eqref{eq:interaction}.

We can now evaluate the integral in Eq.~\eqref{eq:interaction}
very much as for the simpler case of a single-component superconductor.\citep{Kramer71}
In the particular case given by Eq.~\eqref{eq:L},
$\delta\Psi^{(i)}$ is given by Eqs.~\eqref{eq:V0} and \eqref{eq:fg0},
after substituting $r \to |\mathbf{x} - \mathbf{x}_i|$.
The terms involving $\delta f$ and $\delta g$ can be evaluated by using the following property of the Bessel function $K_0$:
\begin{equation}
    \nabla^2 K_0(k |\mathbf{x} - \mathbf{x}_{i,0}|)
        = k^2 K_0(k |\mathbf{x} - \mathbf{x}_{i,0}|)
        - 2 \pi \delta^{(2)} (\mathbf{x} - \mathbf{x}_{i,0}),
\end{equation}
where $\delta^{(2)}$ is the two-dimensional delta function.
Using Eqs.~\eqref{eq:fi} and \eqref{eq:gi},
we then find that all of the terms cancel,
apart from those involving delta functions.
(This cancellation is expected in light of the comments below Eq.~\eqref{eq:L_general}.)
The only remaining terms are
\begin{equation}
    \left(\mathcal{F} - \mathcal{F}_\infty \right)_{\delta f, \delta g}
        \simeq - 2\pi \sum_{j=1,2} \left[
        \left(1+\dfrac{h_1}{\epsilon}+h_3\right) f_j^2 + 2 \dfrac{h_2}{\epsilon} f_j g_j
        + \dfrac{1}{\epsilon}\left(1+h_1+\dfrac{h_3}{\epsilon}\right)  g_j^2 \right]
        \sum_{i\neq0} K_0\left(\frac{\sqrt{2} |\mathbf{x}_i|}{\xi_j}\right)\,.
\end{equation}
Again using Eqs.~\eqref{eq:fi} and \eqref{eq:gi} to simplify this result,
we obtain the second contribution in Eq.~\eqref{eq:result}.
To evaluate the remaining terms in Eq.~\eqref{eq:interaction},
it is convenient to use the divergence theorem
to rewrite the area integral over $C_0$ as a contour integral over $\partial C_0$
in order to avoid the singular behavior of the Bessel functions at $r = 0$.
Remembering
that the integrand of Eq.~\eqref{eq:interaction} vanishes everywhere but at $\mathbf{x}=0$,
we can shrink the integration contour to a small circle of radius $\varepsilon$ centered around the origin.
We then have $\dd \mathbf{S} = \varepsilon \dd \theta \mathbf{e}_{r}$, and so
\begin{align}
    \left(\mathcal{F} - \mathcal{F}_\infty \right)_{\delta \mathbf{V}}
        &\simeq \kappa^2 \sum_{i\neq 0} \int_{\partial C_0}
            \left[\delta\mathbf{V}^{(i)} \times (\nablab \times \delta\mathbf{V}^{(0)})
            - \delta\mathbf{V}^{(0)} \times (\nablab \times \delta\mathbf{V}^{(i)})
            \right]\cdot\dd\mathbf{S} \nonumber \\
        &= - \kappa^2 V_0 \sum_{i\neq0} \int_0^{2 \pi}
            \varepsilon \left[ \delta\mathbf{V}^{(i)} \big|_{\mathbf{x}=0}
            K_0 \left(\frac{\varepsilon}{\lambda_\star}\right) \frac{1}{\lambda_\star}
            + K_1 \left( \frac{\varepsilon}{\lambda_\star} \right)
            (\nablab \times \delta\mathbf{V}^{(i)})_z \big|_{\mathbf{x}=0}\right] \dd \theta \,.
\end{align}
Using the asymptotic behavior of the Bessel functions,
i.e., $K_0(r) \sim - \ln(r)$ and $K_1(r) \sim 1/r$ as $r \to 0$,
we observe that in the limit $\varepsilon \to 0$ the first term vanishes,
while the second one remains finite.
More precisely, we find
\begin{equation}
    \left(\mathcal{F} - \mathcal{F}_\infty \right)_{\delta \mathbf{V}}
        \simeq - \kappa^2 V_0 \sum_{i\neq0} \int_0^{2 \pi}
            \lambda_\star (\nablab \times \delta\mathbf{V}^{(i)})_z \big|_{\mathbf{x}=0} \, \dd \theta \,
        = 2 \pi \kappa^2 V_0^2 \sum_{i\neq0} K_0 \left( \frac{|\mathbf{x}_i|}{\lambda_\star} \right)\,,
\end{equation}
the first contribution in Eq.~\eqref{eq:result}.

In principle, if we can numerically compute the nonlinear solution for a \emph{single} fluxtube,
from this we can determine the values of $V_0$, $f_1$ and $f_2$,
and then use Eq.~\eqref{eq:result} to calculate the interaction energy for any lattice of our choosing.
In practice, however, it is difficult to obtain both $f_1$ and $f_2$
to sufficiently high accuracy to achieve quantitatively reliable results.
Moreover, the assumptions made in obtaining this result
are only valid in the asymptotic limit of a widely-spaced lattice,
so caution is needed when applying this result to a lattice with finite separation between fluxtubes.
For these reasons, we would like to have a more robust method
for estimating the interaction energy.
\citet{HaberSchmitt17} have suggested a possible approach:
they followed essentially the same steps leading to Eq.~\eqref{eq:interaction},
but chose to leave the result in the form of an integral over the boundary $\partial C_0$.
They then computed this integral numerically,
approximating $\delta\Psi^{(i)}$ using the solution obtained numerically for a single fluxtube.
However, their approach has a number of shortcomings:
\begin{enumerate}
  \item Rather than computing the interaction energy for a lattice,
  they considered only a pair of fluxtubes.
  However, a pair of fluxtubes is generally not a steady state,
  i.e., it is not a solution of the \EL\ equations~\eqref{eq:EL}.
  This violates a basic assumption underlying the derivation.
  \item They chose to include some, but not all, of the higher-order terms in their calculation,
  leading to a result that lacks certain symmetries expected on physical grounds.
  By contrast, in deriving Eq.~\eqref{eq:interaction},
  we have consistently neglected all terms of higher order than $(\delta\Psi)^2$,
  and the result is antisymmetric between $\delta\Psi^{(0)}$ and $\delta\Psi^{(i)}$.
  \item Their formula (C8) for the interaction energy depends on the location of the \WS\ cell boundary,
  relative to the fluxtube lattice.
  As we have emphasized in our derivation,
  the location of the fluxtube within its cell is immaterial, and therefore
  should not change the result.
\end{enumerate}
As an alternative approach, we suggest making use of the exact result
\begin{equation}
  \frac{\dd\mathcal{F}}{\dd\ln a}
    = \iint_{C_0}\left[\frac{1}{2}(f^2 - 1)^2 + \frac{R^2}{2\epsilon}(g^2 - 1)^2
        + \frac{\gr}{\epsilon}(f^2 - 1)(g^2 - 1) - \kappa^2B_z^2\right]\,\dd x\,\dd y\,,
  \label{eq:dFda}
\end{equation}
which we derive in Appendix~\ref{sec:Abrikosov}.
Inside the integral, we can approximate the full solution
by superposing the profiles of single fluxtubes with the uniform solution:
\begin{align}
    f &\simeq 1 + \sum_i(f^{(i)} - 1) \,,
        \label{eq:f_sum} \\
    g &\simeq 1 + \sum_i(g^{(i)} - 1) \,,
        \label{eq:g_sum} \\
    B_z &\simeq \sum_i B_z^{(i)}\,.
        \label{eq:Bz_sum}
\end{align}
This formula is consistent with the rigorous result~(\ref{eq:result}) in the asymptotic limit $a\to\infty$,
and because the integrand is spatially periodic by construction,
it has none of the shortcomings described above.
The formula will be accurate as long as the approximations (\ref{eq:f_sum})--(\ref{eq:Bz_sum}) hold,
which in practice still requires that $a \gg 1$.
We have used this formula to independently verify some of the results from our 2D numerical model.

%%%%%%%%%%%%%%%%%%%%%%%%%%%%%%%%%%%%%%%%%%%%%%%%%%%%%%%%%%%%%%

\section{Weakly nonlinear lattice solution}
\label{sec:Abrikosov}

We consider a rectangular domain containing an integer number of fluxtubes, $N$, with area $aN$.
We have shown in Sec.~\ref{sec:upper} that the non-superconducting state is linearly unstable for $a > 2\pi/H_{\cc2}$,
where $H_{\cc2}$ is given by Eq.~\eqref{eq:H_c2}.
Our goal is to compute weakly nonlinear solutions for $a = 2\pi/H_{\cc2} + \delta a$,
where $\delta a \ll 1$.
To do so we will roughly follow the same procedure as \citet{Abrikosov57},
except that by working with a finite domain
and quasi-periodic boundary conditions
we avoid having to manipulate products of infinite series.

It is convenient at this point to redefine our length scale,
so that the area of the domain is normalized to unity,
and remains fixed as the parameter $a$ is varied.
At the same time, we will also rescale $\mathbf{A}$
so that the boundary conditions have no dependence on $a$.
Under the rescaling $\mathbf{x} \to (aN)^{1/2}\mathbf{x}$
and $\mathbf{A} \to (aN)^{-1/2}\mathbf{A}$,
the free energy~(\ref{eq:F_dimless}) becomes
\begin{align}
    F[\psip,\psin,\mathbf{A}]
        &= \frac{1}{2}(1 - |\psip|^2)^2 + \frac{R^2}{2\epsilon}(1 - |\psin|^2)^2
            + \frac{\gr}{\epsilon}(1 - |\psip|^2)(1 - |\psin|^2) \nonumber \\
        &+ \frac{1}{aN}\left|\left(\nablab - \ii\mathbf{A}\right)\psip\right|^2
            + \frac{1}{\epsilon aN}\left|\nablab\psin\right|^2
            + \frac{\kappa^2}{(aN)^2}|\nablab\times\mathbf{A}|^2 \nonumber \\
        &+ \frac{h_1}{\epsilon aN}\left|\left(\nablab - \ii\mathbf{A}\right)(\psin^{\star}\psip)\right|^2
            + \frac{(h_2-h_1)}{2\epsilon aN}\nablab(|\psip|^2)\cdot\nablab(|\psin|^2) \nonumber \\
        &+ \frac{h_3}{4aN}\left(\bigl|\nablab(|\psip|^2)\bigr|^2
            + \frac{1}{\epsilon^2}\bigl|\nablab(|\psin|^2)\bigr|^2\right)\,.
    \label{eq:F_rescaled}
\end{align}
Our domain now has the dimensions $\Gamma\times(1/\Gamma)$, say,
and we have the (quasi)periodic boundary conditions:
\begin{align}
    \mathbf{A}(\mathbf{x}+\mathbf{L})
        &= \mathbf{A}(\mathbf{x}) + \pi N\mathbf{e}_z\times\mathbf{L}
            \label{eq:quasi_A} \,, \\
    \psip(\mathbf{x}+\mathbf{L})
        &= \psip(\mathbf{x})\exp\left(\ii\pi N\mathbf{e}_z\times\mathbf{L}\cdot\mathbf{x}\right)
            \label{eq:quasi_p} \,, \\
    \psin(\mathbf{x}+\mathbf{L}) &= \psin(\mathbf{x})\,,
\end{align}
where $\mathbf{L}$ represents either of the translation symmetries $(\Gamma,0)$ or $(0,1/\Gamma)$.
The free energy per magnetic flux quantum (in the unscaled units) is
\begin{equation}
    \mathcal{F}(a) = a\overline{F}\,,
\end{equation}
where the overbar represents the spatial average,
which is equivalent to the area integral over the rescaled rectangular domain.
Because the quantity $a$ now appears only as a coefficient in the free energy,
we can directly compute the derivative of $\mathcal{F}(a)$,
which leads to the formula~\eqref{eq:dFda}.

In the rescaled units, the non-superconducting solution has the form
$|\psip|=0$, $|\psin| = \sqrt{1 + \alpha/R^2}$,
$\mathbf{A} = \pi N\mathbf{e}_z\times\mathbf{x}$.
At the critical point $a = 2\pi/H_{\cc2}$,
this solution becomes unstable
to perturbations $\delta\psip$
that \rewrite{lie in the kernel} of the linear operator
\begin{align}
    \mathcal{L} \equiv (\nablab - \ii\pi N\mathbf{e}_z\times\mathbf{x})^2 + 2\pi N\,.
\end{align}
These perturbations have the form
$\delta\psip = \ee^{\ii\pi Ny(x+\ii y)}\phi(x + \ii y)$,
for some function $\phi$,
which must be chosen to match the quasi-periodic boundary condition~\eqref{eq:quasi_p}.
The general solution can be expressed in terms of Jacobi theta functions:
\begin{equation}
    \phi(z) = \prod_{j=1}^{N}\exp(-2\pi\ii y_jz)\,
        \vartheta_1\left(\frac{\pi}{\Gamma}(z-z_j)\middle|\frac{\ii}{\Gamma^2}\right)\,,
\end{equation}
where the fluxtube locations, $z_j = x_j + \ii y_j$, must satisfy
$\sum_j z_j = (m+\tfrac{1}{2}N)\Gamma + (n+\tfrac{1}{2}N)\ii/\Gamma$,
for some $m,n\in\mathbb{Z}$.

Just beyond the critical point, with $a = 2\pi/H_{\cc2} + \delta a$,
we anticipate that the solutions have regular asymptotic expansions of the form
\begin{align}
    \mathbf{A} &= \mathbf{A}^{(1)} + (\delta a)\mathbf{A}^{(2)} + \ldots \,, \\
    \psin &= \psin^{(1)} + (\delta a)\psin^{(2)} + \ldots \,, \\
    \psip &= (\delta a)^{1/2}\psip^{(1)} + (\delta a)^{3/2}\psip^{(2)} + \ldots \,.
\end{align}
Substituting this ansatz into the rescaled \EL\ equations,
the leading-order contributions simply recover the non-superconducting state
for $\mathbf{A}^{(1)}$ and $\psin^{(1)}$,
as well as the linear equation $\mathcal{L}\psip^{(1)} = 0$.
Without loss of generality,
we will take $\mathbf{A}^{(1)} = \pi N\mathbf{e}_z\times\mathbf{x}$
and $\psin^{(1)} = \sqrt{1 + \gr/R^2}$.
For the moment we do not need to choose the particular form of $\psip^{(1)}$.
However, in what follows we will
make use of its quasi-periodicity,
and of the (gauge-invariant) identities
\begin{align}
    \left(\nablab - \ii\mathbf{A}^{(1)}\right)\psip^{(1)}
        &= \ii\mathbf{e}_z\times\left(\nablab - \ii\mathbf{A}^{(1)}\right)\psip^{(1)} \,, \\
    \frac{1}{2\pi}\nabla^2\ln|\psip^{(1)}|
        &= - N + \sum_j\delta^{(2)}(\mathbf{x}-\mathbf{x}_j)\,,
    \label{eq:logp}
\end{align}
where $\mathbf{x}_j$ are the fluxtube locations,
and $\delta^{(2)}$ is the two-dimensional delta function.

Now proceeding to next order in the \EL\ equations,
we eventually obtain the following:
\begin{align}
    \left(1 - \frac{\gr^2}{\epsilon R^2}\right)^{-1}\frac{H_{\cc2}^2\kappa^2}{\pi N}\nablab\times(\nablab\times\mathbf{A}^{(2)})
        &= \mathbf{e}_z\times\nablab|\psip^{(1)}|^2
            \label{eq:weak_A} \,, \\
    \left[\left(\frac{1}{|\psin^{(1)}|^2} + \frac{h_3}{\epsilon}\right)\nabla^2
        - \frac{4\pi NR^2}{H_{\cc2}}\right]\Re\left\{\psin^{(1)\star}\psin^{(2)}\right\}
        &= - \left[\frac{h_2-h_1}{2}\nabla^2
            - 2\pi N\left(h_1 + \frac{\gr}{H_{\cc2}}\right)\right]|\psip^{(1)}|^2
            \label{eq:weak_n} \,, \\
    \nabla^2\Im\left\{\psin^{(1)\star}\psin^{(2)}\right\}
        &= 0 \,, \\
    \left(1 - \frac{\gr^2}{\epsilon R^2}\right)\frac{1}{H_{\cc2}}\mathcal{L}\psip^{(2)}
        &= N\left(\frac{2\pi}{H_{\cc2}}|\psip^{(1)}|^2 - 1
            + \frac{\gr^2}{\epsilon R^2}\right)\psip^{(1)}
            - \frac{h_3}{2}\psip^{(1)}\nabla^2(|\psip^{(1)}|^2)
            \nonumber \\
        &- \psip^{(1)}\left(\frac{h_2-h_1}{\epsilon}\nabla^2 - \frac{\gr}{\epsilon}
            \frac{4\pi N}{H_{\cc2}}\right)\Re\left\{\psin^{(1)\star}\psin^{(2)}\right\} \nonumber \\
        &+ \left(1 - \frac{\gr^2}{\epsilon R^2}\right)\frac{1}{H_{\cc2}}
            \left[(\nablab-\ii\mathbf{A}^{(1)})\cdot(\ii\mathbf{A}^{(2)}\psip^{(1)})
            + \ii\mathbf{A}^{(2)}\cdot(\nablab-\ii\mathbf{A}^{(1)})\psip^{(1)}\right]
            \nonumber \\
        &- \frac{h_1}{\epsilon}\left[\psin^{(1)\star}\psin^{(2)}
            (\nablab-\ii\mathbf{A}^{(1)})^2\psip^{(1)}
            + (\nablab-\ii\mathbf{A}^{(1)})^2(\psip^{(1)}\psin^{(1)}\psin^{(2)\star})\right]
            \,.
            \label{eq:weak_p}
\end{align}
Equation~(\ref{eq:weak_A}) can be integrated once to obtain $B_z^{(2)}$.
We note that the boundary condition~\eqref{eq:quasi_A} implies
that $\mathbf{A}^{(n)}$ is spatially periodic for all $n>1$,
and so $\overline{B_z^{(n)}} = 0$.
Hence we find that
\begin{equation}
    B_z^{(2)} = \left(1 - \frac{\gr^2}{\epsilon R^2}\right)
        \frac{\pi N}{H_{\cc2}^2\kappa^2}\left(\overline{|\psip^{(1)}|^2}
        - |\psip^{(1)}|^2\right)\,.
        \label{eq:Bz2}
\end{equation}
Now, in order for Eq.~\eqref{eq:weak_p} to have regular solutions,
its right-hand side must be orthogonal to $\psip^{(1)}$.
To prove this,
we note that $\mathcal{L}$ is self-adjoint with respect to the inner product
\begin{equation}
    \langle\psi,\phi\rangle \equiv
        \int_{x=0}^{\Gamma}\int_{y=0}^{1/\Gamma}\psi^\star\phi\,\dd y\,\dd x \,,
\end{equation}
provided that \emph{both} arguments satisfy the quasi-periodic boundary condition~\eqref{eq:quasi_p}.
All of the terms $\psip^{(n)}$ satisfy these boundary conditions, and so
\begin{align*}
    \langle\psip^{(1)},\mathcal{L}\psip^{(2)}\rangle
        &= \langle\mathcal{L}\psip^{(1)},\psip^{(2)}\rangle \\
        &= 0 \,.
\end{align*}
Hence, taking the inner product of $\psip^{(1)}$ with Eq.~\eqref{eq:weak_p},
we obtain the compatibility condition
\begin{align}
    0 &= N\left(\frac{2\pi}{H_{\cc2}}\overline{|\psip^{(1)}|^4}
        - \left(1 - \frac{\gr^2}{\epsilon R^2}\right)\overline{|\psip^{(1)}|^2}\right)
        + \frac{h_3}{2}\overline{|\nablab(|\psip^{(1)}|^2)|^2}
        \nonumber \\
        &+ \frac{h_2-h_1}{\epsilon}\overline{\nablab(|\psip^{(1)}|^2)
        \cdot\nablab\Re\{\psin^{(1)\star}\psin^{(2)}\}}
        + \frac{\gr}{\epsilon}\frac{4\pi N}{H_{\cc2}}
        \overline{|\psip^{(1)}|^2\Re\{\psin^{(1)\star}\psin^{(2)}\}}
        \nonumber \\
        &+ \left(1 - \frac{\gr^2}{\epsilon R^2}\right)\frac{1}{H_{\cc2}}
        \overline{|\psip^{(1)}|^2B_z^{(2)}} + 4\pi N\frac{h_1}{\epsilon}
        \overline{|\psip^{(1)}|^2\Re\{\psin^{(1)\star}\psin^{(2)}\}}\,.
            \label{eq:compatibility}
\end{align}
Using Eq.~\eqref{eq:logp}, it can be shown that
\begin{equation}
    \overline{|\nablab(|\psip^{(1)}|^2)|^2} = 2\pi N\overline{|\psip^{(1)}|^4}\,,
\end{equation}
and combining with Eqs.~\eqref{eq:weak_n} and \eqref{eq:Bz2},
we can write the compatibility condition~\eqref{eq:compatibility} in more symmetric form:
\begin{align}
    \left(1 - \frac{\gr^2}{\epsilon R^2}\right)\frac{1}{\overline{|\psip^{(1)}|^2}}
        &= \left(\frac{2\pi}{H_{\cc2}} + \pi h_3\right)\beta
        + \left(1 - \frac{\gr^2}{\epsilon R^2}\right)^2\frac{\pi}{H_{\cc2}^3\kappa^2}
            \left(1 - \beta\right)
        \nonumber \\
        &- \frac{R^2}{\epsilon}\left[\frac{1}{2N}\left(\frac{1}{R^2 + \gr}
        + \frac{h_3}{\epsilon R^2}\right)\overline{\left|\nablab\gamma\right|^2}
        + \frac{2\pi}{H_{\cc2}}\overline{\gamma^2}\right] \,,
            \label{eq:compatibility2}
\end{align}
where we have defined
\begin{align}
    \beta \equiv \frac{\overline{|\psip^{(1)}|^4}}{\left(\overline{|\psip^{(1)}|^2}\right)^2}
    \qquad \mbox{and} \qquad
    \gamma(\mathbf{x}) \equiv \frac{2\Re\left\{\psin^{(1)\star}\psin^{(2)}\right\}}
        {\overline{|\psip^{(1)}|^2}}\,.
\end{align}
If we now expand the free-energy density~\eqref{eq:F_rescaled} up to $O(\delta a^2)$,
and use the identities derived above, we eventually find
\begin{align}
    \mathcal{F} &= a\overline{F}
        = \frac{a}{2}\left(1 - \frac{\gr^2}{\epsilon R^2}\right)
        \left[1 - \frac{H_{\cc2}(\delta a)^2}{2\pi}\overline{|\psip^{(1)}|^2}\right]
            + \frac{(2\pi\kappa)^2}{a} + O(\delta a^3)\,.
            \label{eq:weak}
\end{align}
The transition to the non-superconducting state is second order if (and only if)
$\mathcal{F}(a)$ is convex in a neighbourhood of the critical point $a = 2\pi/H_{\cc2}$,
which requires that $\mathcal{F}' < 0$ and $\mathcal{F}'' > 0$.
Using Eq.~\eqref{eq:weak}, and the compatibility condition~\eqref{eq:compatibility2},
these criteria become
\begin{align}
    (H_{\cc2}\kappa)^2 &> \frac{1}{2}\left(1 - \frac{\gr^2}{\epsilon R^2}\right) \,, \\
    \left[\frac{2}{H_{\cc2}} + h_3 - \left(1 - \frac{\gr^2}{\epsilon R^2}\right)^2
        \frac{1}{H_{\cc2}^3\kappa^2}\right]\pi\beta
    &> \frac{R^2}{\epsilon}\left[\frac{1}{2N}\left(\frac{1}{R^2 + \gr}
        + \frac{h_3}{\epsilon R^2}\right)\overline{\left|\nablab\gamma\right|^2}
        + \frac{2\pi}{H_{\cc2}}\overline{\gamma^2}\right]\,.
    \label{eq:criterion}
\end{align}
In a simple, single-component \GL\ superconductor
these criteria both reduce to $\kappa > 1/\sqrt{2}$,
and so the upper transition in a type-II superconductor is always second-order.\citep{Abrikosov57}
Moreover, minimising the free energy is equivalent to minimising $\beta$,
and hence the hexagonal lattice is energetically preferred.\citep{Kleiner-etal64}
In our more complicated two-component system,
the second criterion cannot be evaluated analytically.
However, for any particular fluxtube arrangement,
we can in principle compute $\psip^{(1)}$,
then solve Eq.~\eqref{eq:weak_n} to obtain $\gamma$,
and thereby test this criterion numerically.
In particular, for the case of a square or hexagonal lattice,
Eq.~\eqref{eq:weak_n} can be solved in Fourier space,
which leads eventually to the result~\eqref{eq:result2}.
We note that the perturbation to the neutron condensate produced by the fluxtubes,
which is represented by $\gamma$ in Eq.~\eqref{eq:criterion},
always acts to reduce the overall free energy,
making a first-order transition more likely.
The magnitude of $\gamma$ depends, via Eq.~\eqref{eq:weak_n},
on the coupling parameters $h_1$ and $h_2$,
and so a first-order transition is guaranteed if these parameters are sufficiently large.

\begin{figure}[t]
    \centering
    \includegraphics[width=10cm]{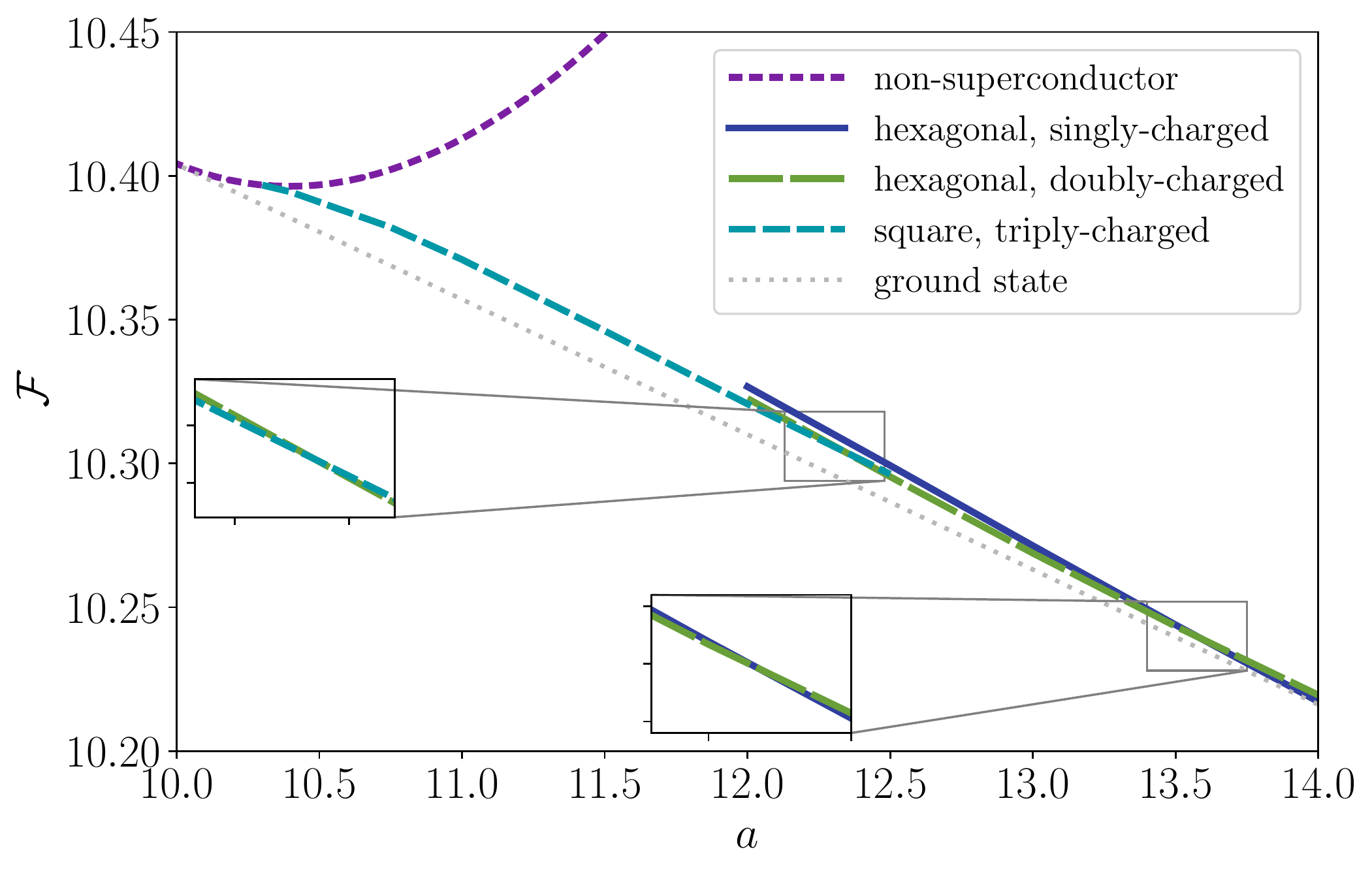}
    \caption{The Helmholtz free energy per flux quantum per unit length,
    $\mathcal{F}(a)$, for various lattice states,
    with (dimensionless) parameters
    $h_1 \simeq 0.061$, $h_2 \simeq 0.422$, $h_3 \simeq 0.062$,
    $R \simeq 0.412$, $\epsilon \simeq 0.096$, and $\kappa = 1.17$.
    The true \gs\ (dotted, gray line) arises as a subcritical bifurcation
    from the non-superconducting state (short-dashed, purple line).
    The two insets show more detail of the cross-over regions between
    two lattice configurations of different charge.}
    \label{fig:crossover}
\end{figure}

Interestingly, there are certain combinations of the parameters
for which Eq.~\eqref{eq:weak_n} can be solved analytically.
In particular, if $\gr=0$ and
\begin{equation}
    \frac{h_2}{h_1} - 1 = \frac{1}{R^2} \,
        \frac{1 + \frac{h_3}{\epsilon}} {1 + \frac{h_1}{\epsilon}} \,,
            \label{eq:analytic}
\end{equation}
then we find that $\gamma \propto |\psip^{(1)}|^2$.
In that case, the $O(\delta a^2)$ term in the free energy~\eqref{eq:weak} has the form
\begin{align}
    \left[2\pi\left(1 + \frac{h_1}{\epsilon}\right) + \pi h_3
        - \frac{\pi}{\kappa^2}\left(1 + \frac{h_1}{\epsilon}\right)^3
        - \frac{1}{\epsilon R^2}\frac{h_1^2}{1 + \frac{h_1}{\epsilon}}
            \left(1 + \frac{1}{2R^2}\frac{1 + \frac{h_3}{\epsilon}}
            {1 + \frac{h_1}{\epsilon}}\right)\right]\beta \,.
        \label{eq:compatibility_special}
\end{align}
If one of the parameters $\epsilon$, $R$ or $\kappa$ is sufficiently small,
then the quantity in square brackets will be negative.
Taken at face value, this result seems to suggest that the free energy can be made arbitrarily small,
because $\beta$ can be made arbitrarily large in an unbounded domain.
However, this singularity actually just reflects the breakdown
of our weakly nonlinear analysis in the limit of an infinite domain.
To illustrate how this breakdown occurs,
we have calculated the free energy of various multiply-charged
hexagonal and square lattice states for one particular set of parameters.
We found that condition~\eqref{eq:analytic} is satisfied for the LNS \EoS\
at the depth where $n_\bb = \unitfrac[0.325]{1}{fm^3}$,
and by reducing $\kappa$ from its true value there of $\simeq 1.47$ to $1.17$
the quantity in Eq.~\eqref{eq:compatibility_special} was made slightly negative.
In Fig.~\ref{fig:crossover}, we plot the resulting free energy as a function of $a$,
for several different lattice types.
As the value of $a$ is reduced from $14$ towards the critical value $2\pi/H_{\cc2} \simeq 10.3$,
we find that the singly-charged hexagonal lattice is replaced by the doubly-charged hexagonal lattice,
and then by the triply-charged square lattice,
as the energetically preferred lattice state.
As $a$ is further reduced,
we expect that even more highly-charged lattice states
(with higher values of $\beta$) will become energetically preferred.
However, throughout this whole range of $a$,
the true \gs\ for an unbounded domain is not a lattice at all,
and is instead a mixture of the non-superconducting state and a singly-charged hexagonal lattice.
Our weakly-nonlinear analysis does not apply to the true \gs, which develops as a nonlinear instability
for $a > 2\pi/H_{\cc2'} \simeq 9.9$.
In general, we find that the singly-charged hexagonal lattice is the only permitted lattice configuration
in a domain free from geometrical constraints.

%%%%%%%%%%%%%%%%%%%%%%%%%%%%%%%%%%%%%%%%%%%%%%%%%%%%%%%%%%%%%%

\bibliography{bibliography}

\end{document}